\documentclass[12pt]{article}
 
\usepackage{axodraw}

\def\ltwid{\mathrel{\raise.3ex\hbox{$<$\kern-.75em\lower1ex\hbox{$\sim$}}}}
\def \be{\begin{equation}}
\def \ee{\end{equation}}
\def \bea{\begin{eqnarray}}
\def \eea{\end{eqnarray}}

\def \del{\partial}

\def \f{\frac}

\def \g{\gamma}

\def \ka{\kappa}
\def \d{\delta}
\def \D{\Delta}
\def \e{\eta}

\begin{document}

\begin{titlepage}
 
\begin{flushright}
gr-qc/0511140 \\ UFIFT-QG-05-06
\end{flushright}

\vspace{2cm}

\begin{center}
{\bf The Fermion Self-Energy during Inflation}
\end{center}

\vspace{.5cm}

\begin{center}
S. P. Miao$^{\dagger}$ and R. P. Woodard$^{\ddagger}$
\end{center}

\vspace{.5cm}

\begin{center}
\it{Department of Physics \\ 
University of Florida \\
Gainesville, FL 32611}
\end{center}

\vspace{1cm}

\begin{center}
ABSTRACT
\end{center}
We compute the one loop fermion self-energy for massless Dirac 
$+$ Einstein in the presence of a locally de Sitter background.
We employ dimensional regularization and obtain a fully renormalized
result by absorbing all divergences with BPHZ counterterms. An 
interesting technical aspect of this computation is the need for a
noninvariant counterterm owing to the breaking of de Sitter invariance
by our gauge condition. Our result can be used in the quantum-corrected
Dirac equation to search for inflation-enhanced quantum effects from 
gravitons, analogous to those which have been found for massless,
minimally coupled scalars.

\vspace{2cm}

\begin{flushleft}
\end{flushleft}

\begin{flushleft}
$^{\dagger}$ e-mail: miao@phys.ufl.edu \\
$^{\ddagger}$ e-mail: woodard@phys.ufl.edu
\end{flushleft}

\end{titlepage}

\section{Introduction}

In this paper we compute and renormalize the one loop quantum gravitational
corrections to the self-energy of massless fermions in a locally de Sitter
background. The physical motivation for this exercise is to check for
graviton analogues of the enhanced quantum effects seen in this background
for interactions which involve one or more undifferentiated, massless, 
minimally coupled (MMC) scalars. Those effects are driven by the fact that
inflation tends to rip virtual, long wavelength scalars out of the vacuum 
and thereby lengthens the time during which they can interact with themselves
or other particles. Gravitons possess the same crucial property of 
masslessness without classical conformal invariance that is responsible for 
the inflationary production of MMC scalars. One might therefore expect 
a corresponding strengthening of quantum gravitational effects during 
inflation.

Of particular interest to us is what happens when a MMC scalar is Yukawa 
coupled to a massless Dirac fermion for non-dynamical gravity. The one 
loop fermion self-energy has been computed for this model and used to 
solve the quantum-corrected Dirac equation \cite{PW},
\begin{equation}
\sqrt{-g} \, i\hspace{-.1cm}\not{\hspace{-.15cm} \mathcal{D}}_{ij}
\psi_{j}(x) - \int d^4x' \, \Bigl[\mbox{}_i \Sigma_j \Bigr](x;x') \, 
\psi_{j}(x') = 0 \; . \label{Diraceq} 
\end{equation}
Powers of the inflationary scale factor $a = e^{Ht}$ play a crucial role
in understanding this equation for the Yukawa model and also for what we
expect from quantum gravity. The Yukawa result for the self-energy consists
\cite{PW} of terms which were originally ultraviolet divergent and which 
end up, after renormalization, carrying the same number of scale factors 
as the classical term. Had the scalar been conformally coupled these would
be the only contributions to the one loop self-energy. However, minimally
coupled scalars also give contributions due to inflationary particle 
production. These are ultraviolet finite from the beginning and possesses 
an extra factor of $a \ln(a)$ relative to the classical term. Higher loops 
can bring more factors of $\ln(a)$, but no more powers of $a$, so it is 
consistent to solve the equation with only the one loop corrections. The 
result is a declining oscillatory behavior in the loop-corrected mode 
functions that seems to betoken the evolution of a nonzero fermion mass. 
A recent one loop computation of the Yukawa scalar self-mass-squared 
indicates that the scalar which catalyzes this process cannot develop a 
large enough mass quickly enough to prevent the super-horizon fermion 
modes from fully experiencing this effect \cite{DW}.

Analogous graviton effects should be suppressed by the fact that the
$h_{\mu\nu} \overline{\psi} \psi$ interaction carries a derivative as 
opposed to the undifferentiated Yukawa interaction. What we expect is that 
the corresponding quantum gravitational self-energy will consist of two terms.
The most ultraviolet singular one will require higher derivative counterterms
and will end up, after renormalization, possessing {\it one less} factor of
$a$ than the classical term. The less singular term due to inflationary
particle production should require only lower derivative counterterms and 
will be enhanced from the classical term by a factor of $\ln(a)$. This would
give a much weaker effect than the analogous term in the Yukawa model, but it
would still be interesting. And note that {\it any such effect from gravitons
would be universal}, independent of assumptions about the existence or 
couplings of unnaturally light scalars.

Dirac $+$ Einstein is not perturbatively renormalizable \cite{DVN}, however,
ultraviolet divergences can always be absorbed in the BPHZ sense 
\cite{BP,H,Z1,Z2}. A widespread misconception exists that no valid quantum 
predictions can be extracted from such an exercise. This is false: while 
nonrenormalizability does preclude being able to compute {\it everything}, 
that is not the same thing as being able to compute {\it nothing}. The problem
with a nonrenormalizable theory is that no physical principle fixes the finite 
parts of the escalating series of BPHZ counterterms needed to absorb 
ultraviolet divergences, order-by-order in perturbation theory. Hence any 
prediction of the theory that can be changed by adjusting the finite parts of 
these counterterms is essentially arbitrary. However, loops of massless 
particles make nonlocal contributions to the effective action that can never 
be affected by local counterterms. These nonlocal contributions typically 
dominate the infrared. Further, they cannot be affected by whatever 
modification of ultraviolet physics ultimately results in a completely 
consistent formalism. As long as the eventual fix introduces no new massless 
particles, and does not disturb the low energy couplings of the existing 
ones, the far infrared predictions of a BPHZ-renormalized quantum theory will 
agree with those of its fully consistent descendant.

It is worthwhile to review the vast body of distinguished work that has
exploited this fact. The oldest example is the solution of the infrared
problem in quantum electrodynamics by Bloch and Nordsieck \cite{BN}, long 
before that theory's renormalizability was suspected. Weinberg \cite{SW} 
was able to achieve a similar resolution for quantum gravity with zero 
cosmological constant. The same principle was at work in the Fermi theory 
computation of the long range force due to loops of massless neutrinos by 
Feinberg and Sucher \cite{FS,HS}. Matter which is not supersymmetric 
generates nonrenormalizable corrections to the graviton propagator at one 
loop, but this did not prevent the computation of photon, massless neutrino
and massless, conformally coupled scalar loop corrections to the long range 
gravitational force \cite{CDH,CD,DMC1,DL}. More recently, Donoghue 
\cite{JFD1,JFD2} has touched off a minor industry \cite{MV,HL,ABS,KK1,KK2} 
by applying the principles of low energy effective field theory to compute 
graviton corrections to the long range gravitational force. Our analysis 
exploits the power of low energy effective field theory in the same way, 
differing from the previous examples only in the detail that our background 
geometry is locally de Sitter rather than flat.\footnote{For another recent
example in a nontrivial cosmology see \cite{EMV}.}

That summarizes why the exercise we have undertaken is both valid and
interesting. In the next section we work out the fermionic part of the Feynman 
rules for Dirac $+$ Einstein. Section 3 is devoted to the issues associated 
with the graviton propagator. A major complication concerns the impossibility 
of employing a de Sitter invariant gauge condition \cite{AM,TW1}. We give a 
short review of the complex literature on this issue. Then we introduce a 
noninvariant gauge fixing term, isolate the subgroup of de Sitter 
transformations that it respects, and present the gauge-fixed graviton 
propagator. The section closes with a discussion of the BPHZ counterterms 
necessary for our computation. In section 4 we evaluate the contributions 
from diagrams involving a single 4-point interaction. In section 5 we evaluate 
the more difficult contributions which involve two 3-point interactions. 
Renormalization is accomplished in section 6, and our conclusions are given 
in section 7.

\section{Fermions in Quantum Gravity}

The coupling of gravity to particles with half integer spin is usually
accomplished by shifting the fundamental gravitational field variable 
from the metric $g_{\mu\nu}(x)$ to the vierbein $e_{\mu m}(x)$.\footnote{
For another approach see \cite{HAW}.} Greek letters stand 
for coordinate indices and Latin letters denote Lorentz indices, and 
both kinds of indices take values in the set $\{0,1,2,\dots,(D\!-\!1)\}$. 
One recovers the metric by contracting two vierbeins into the Lorentz 
metric $\eta^{bc}$,
\begin{equation}
g_{\mu\nu}(x) = e_{\mu b}(x) e_{\nu c}(x) \eta^{bc} \; .
\end{equation}
The coordinate index is raised and lowered with the metric ($e^{\mu}\,_{b} 
= g^{\mu\nu} e_{\nu b}$), while the Lorentz index is raised and lowered with 
the Lorentz metric ($e_{\mu}\,^{b} = \eta^{bc} e_{\mu c}$). We employ the
usual metric-compatible and vierbein compatible connections,
\begin{eqnarray}
g_{\rho\sigma ; \mu} = 0 & \Longrightarrow & \Gamma^{\rho}_{~\mu\nu} =
\frac12 g^{\rho\sigma} \Bigl(g_{\sigma \mu , \nu} + g_{\nu \sigma , \mu}
- g_{\mu\nu , \sigma}\Bigr) \; , \\
e_{\beta b ; \mu} = 0 & \Longrightarrow & A_{\mu c d} = e^{\nu}_{~c} \Bigl(
e_{\nu d, \mu} - \Gamma^{\rho}_{~\mu\nu} e_{\rho d}\Bigr) \; . \label{spin}
\end{eqnarray}

Fermions also require gamma matrices, $\gamma^b_{ij}$. The anti-commutation
relations,
\begin{equation}
\Bigl\{\gamma^b , \gamma^c\Bigr\} \equiv \Bigl(\gamma^b \gamma^c + 
\gamma^c \gamma^b\Bigr) = -2 \eta^{bc} I \; ,
\end{equation}
imply that only fully anti-symmetric products of gamma matrices are 
actually independent. The Dirac Lorentz representation matrices are such
an anti-symmetric product,
\begin{equation}
J^{bc} \equiv \frac{i}4 \Bigl(\gamma^b \gamma^c - \gamma^c \gamma^b\Bigr) 
\equiv \frac{i}2 \gamma^{[b} \gamma^{c]} \; .
\end{equation}
They can be combined with the spin connection (\ref{spin}) to form the Dirac
covariant derivative operator,
\begin{equation}
\mathcal{D}_{\mu} \equiv \partial_{\mu} + \frac{i}2 A_{\mu cd} J^{cd} \; .
\end{equation}
Other identities we shall often employ involve anti-symmetric products,
\begin{eqnarray}
\gamma^b \gamma^c \gamma^d & = & \gamma^{[b} \gamma^c \gamma^{d]} - \eta^{bc}
\gamma^d + \eta^{db} \gamma^c - \eta^{cd} \gamma^b \; , \\
\gamma^b J^{cd} & = & \frac{i}2 \gamma^{[b} \gamma^c \gamma^{d]} + \frac{i}2
\eta^{bd} \gamma^c - \frac{i}2 \eta^{bc} \gamma^d \; . \label{Jred}
\end{eqnarray}
We shall also encounter cases in which one gamma matrix is contracted into
another through some other combination of gamma matrices,
\begin{eqnarray}
\gamma^b \gamma_b & = & -D I \; , \\
\gamma^b \gamma^c \gamma_b & = & (D\!-\!2) \gamma^c \; , \\
\gamma^b \gamma^c \gamma^d \gamma_b & = & 4 \eta^{cd} I -(D\!-\!4) \gamma^c 
\gamma^d \; , \\
\gamma^b \gamma^c \gamma^d \gamma^e \gamma_b & = & 2 \gamma^e \gamma^d
\gamma^c + (D\!-\!4) \gamma^c \gamma^d \gamma^e \; .
\end{eqnarray}

The Lagrangian of massless fermions is,
\begin{equation}
\mathcal{L}_{\rm Dirac} \equiv \overline{\psi} e^{\mu}_{~b} \gamma^{b} i 
\mathcal{D}_{\mu} \psi \sqrt{-g} \; . \label{Dirac}
\end{equation}
Because our locally de Sitter background is conformally flat it is useful
to rescale the vierbein by an arbitrary function of spacetime $a(x)$,
\begin{equation}
e_{\beta b} \equiv a \, \widetilde{e}_{\beta b} \qquad \Longrightarrow \qquad
e^{\beta b} = a^{-1} \, \widetilde{e}^{\beta b} \; .
\end{equation}
Of course this implies a rescaled metric $\widetilde{g}_{\mu\nu}$,
\begin{equation}
g_{\mu\nu} = a^2 \, \widetilde{g}_{\mu\nu} \qquad \Longrightarrow \qquad
g^{\mu\nu} = a^{-2} \, \widetilde{g}^{\mu\nu} \;. \label{confg}
\end{equation}
The old connections can be expressed as follows in terms of the ones
formed from the rescaled fields,
\begin{eqnarray}
\Gamma^{\rho}_{~\mu\nu} & = & a^{-1} \Bigl(\delta^{\rho}_{~\mu} \, a_{,\nu} 
\!+\!  \delta^{\rho}_{~\nu} \, a_{,\mu} \!-\! \widetilde{g}^{\rho\sigma} \, 
a_{,\sigma} \, \widetilde{g}_{\mu\nu}\Bigr) + \widetilde{\Gamma}^{\rho}_{
~\mu\nu} \label{confG} \; \\
A_{\mu cd} & = &-a^{-1} \Bigl(\widetilde{e}^{\nu}_{~c} \, \widetilde{e}_{\mu d} 
\!-\!  \widetilde{e}^{\nu}_{~d} \, \widetilde{e}_{\mu c} \Bigr) a_{,\nu} + 
\widetilde{A}_{\mu cd} \; .
\end{eqnarray}
We define rescaled fermion fields as follows,
\begin{equation}
\Psi \equiv a^{\frac{D-1}2} \psi \qquad {\rm and} \qquad
\overline{\Psi} \equiv a^{\frac{D-1}2} \overline{\psi} \; .
\end{equation}
The utility of these definitions stems from the conformal invariance of the
Dirac Lagrangian,
\begin{equation}
\mathcal{L}_{\rm Dirac} = \overline{\Psi} \, \widetilde{e}^{\mu}_{~b} \, 
\gamma^b \, i \widetilde{\mathcal{D}}_{\mu} \Psi \sqrt{-\widetilde{g}} \; , 
\label{Diract}
\end{equation}
where $\widetilde{\mathcal{D}}_{\mu} \equiv \partial_{\mu} \!+\! \frac{i}2 
\widetilde{A}_{\mu cd} J^{cd}$. 

One could follow early computations about flat space background \cite{BG1,BG2} 
in defining the graviton field as a first order perturbation of the 
(conformally rescaled) vierbein. However, so much of gravity involves the 
vierbein only through the metric that it is simpler to instead take the 
graviton field to be a first order perturbation of the conformally rescaled 
metric,
\begin{equation}
\widetilde{g}_{\mu\nu} \equiv \eta_{\mu\nu} + \kappa h_{\mu\nu} \qquad 
{\rm with} \qquad \kappa^2 = 16 \pi G \; . 
\end{equation}
We then impose symmetric gauge ($e_{\beta b} = e_{b \beta}$) to fix the
local Lorentz gauge freedom, and solve for the vierbein in terms of the
graviton,
\begin{equation}
\widetilde{e}[\widetilde{g}]_{\beta b} \equiv \Bigl(\sqrt{\widetilde{g}
\eta^{-1}} \, \Bigr)_{\!\beta}^{~\gamma} \, \eta_{\gamma b} = \eta_{\beta b} 
+ \frac12 \kappa h_{\beta b} - \frac18 \kappa^2 h_{\beta}^{~\gamma} 
h_{\gamma b} + \dots 
\end{equation}
It can be shown that the local Lorentz ghosts decouple in this gauge and
one can treat the model, at least perturbatively, as if the fundamental 
variable were the metric and the only symmetry were diffeomorphism invariance 
\cite{RPW1}. At this stage there is no more point in distinguishing between
Latin letters for local Lorentz indices and Greek letters for vector 
indices. Other conventions are that graviton indices are raised and lowered 
with the Lorentz metric ($h^{\mu}_{~\nu} \equiv \eta^{\mu\rho} h_{\rho\nu}$,
$h^{\mu\nu} \equiv \eta^{\mu\rho} \eta^{\nu\sigma} h_{\rho\sigma}$) and that
the trace of the graviton field is $h \equiv \eta^{\mu\nu} h_{\mu\nu}$. We
also employ the usual Dirac ``slash'' notation,
\begin{equation}
\hspace{-.1cm}\not{\hspace{-.05cm} V}_{ij} \equiv V_{\mu} 
\gamma^{\mu}_{ij} \; .
\end{equation}

It is straightforward to expand all familiar operators in powers of the
graviton field,
\begin{eqnarray}
\widetilde{e}^{\mu}_{~b} & = & \delta^{\mu}_{~b} - \frac12 \kappa h^{\mu}_{~b}
+ \frac38 \kappa^2 h^{\mu\rho} h_{\rho b} + \dots \; , \\
\widetilde{g}^{\mu\nu} & = & \eta^{\mu\nu} - \kappa h^{\mu\nu} + \kappa^2 
h^{\mu}_{~\rho} h^{\rho\nu} - \dots \; , \\
\sqrt{-\widetilde{g}} & = & 1 + \frac12 \kappa h + \frac18 \kappa^2 h^2 - 
\frac14 \kappa^2 h^{\rho\sigma} h_{\rho\sigma} + \dots
\end{eqnarray}
Applying these identities to the conformally rescaled Dirac Lagrangian gives,
\begin{eqnarray}
\lefteqn{\mathcal{L}_{\rm Dirac} = \overline{\Psi} i \hspace{-.1cm}\not{
\hspace{-.1cm} \partial} \Psi + \frac{\kappa}2 \Biggl\{h \overline{\Psi} 
i \hspace{-.1cm}\not{\hspace{-.1cm} \partial} \Psi \!-\! h^{\mu\nu} 
\overline{\Psi} \gamma_{\mu} i \partial_{\nu} \Psi \!-\!
h_{\mu\rho , \sigma} \overline{\Psi} \gamma^{\mu} J^{\rho\sigma} \Psi
\Biggr\} } \nonumber \\
& & + \kappa^2 \Biggl\{ \Bigl[\frac18 h^2 \!-\! \frac14 h^{\rho\sigma} 
h_{\rho\sigma}\Bigr] \overline{\Psi} i \hspace{-.1cm}\not{\hspace{-.1cm} 
\partial} \Psi \!+\! \Bigl[-\frac14 h h^{\mu\nu} \!+\! \frac38 h^{\mu\rho}
h_{\rho}^{~\nu}\Bigr] \overline{\Psi} \gamma_{\mu} i \partial_{\nu} \Psi
+ \Biggl[-\frac14 h h_{\mu \rho , \sigma} \nonumber \\
& & \hspace{1.5cm} + \frac18 h^{\nu}_{~\rho} h_{\nu \sigma , \mu} + \frac14 
(h^{\nu}_{~\mu} h_{\nu\rho})_{,\sigma} \!+\! \frac14 h^{\nu}_{~ \sigma} 
h_{\mu\rho ,\nu}\Biggr] \overline{\Psi} \gamma^{\mu} J^{\rho\sigma} \Psi 
\Biggr\} + O(\kappa^3) \; . \qquad \label{Dexp}
\end{eqnarray}
From the first term we see that the rescaled fermion propagator is the same 
as for flat space,
\begin{equation}
i\Bigl[\mbox{}_i S_j \Bigr](x;x') = \frac{\Gamma(\frac{D}2\!-\!1)}{4\pi^{
\frac{D}2}} \, i \hspace{-.1cm}\not{\hspace{-.1cm} \partial}_{ij}
\Big(\frac1{\Delta x^2}\Big)^{\frac{D}2-1} , \label{fprop}
\end{equation}
where the coordinate interval is $\Delta x^2(x;x') \equiv \Vert \vec{x} \!-\! 
\vec{x}' \Vert^2 - (\vert \eta\!-\!\e'\vert -i \delta)^2$.

We now represent the various interaction terms in (\ref{Dexp}) as vertex 
operators acting on the fields. At order $\kappa$ the interactions involve 
fields, $\overline{\Psi}_i$, $\Psi_j$ and $h_{\alpha\beta}$, which we number 
``1'', ``2'' and ``3'', respectively. Each of the three interactions can be 
written as some combination $V_{I ij}^{ \alpha\beta}$ of tensors, spinors 
and a derivative operator acting on these fields. For example, the first 
interaction is,
\begin{equation}
\frac{\kappa}2 h \overline{\Psi} i \hspace{-.1cm}\not{\hspace{-.1cm} \partial} 
\Psi = \frac{\kappa}2 \eta^{\alpha \beta} i \hspace{-.1cm}\not{\hspace{-.1cm} 
\partial}_{2 ij} \times \overline{\Psi}_i \Psi_j h_{\alpha\beta} \equiv
V_{1ij}^{\alpha\beta} \times \overline{\Psi}_i \Psi_j h_{\alpha\beta} \; .
\end{equation}
Hence the 3-point vertex operators are,
\begin{equation}
V_{1ij}^{\alpha\beta} = \frac{\kappa}2 \eta^{\alpha \beta} i \hspace{-.1cm}
\not{\hspace{-.1cm} \partial}_{2 ij} \quad , \quad V_{2ij}^{\alpha\beta} = 
-\frac{\kappa}2 \gamma^{(\alpha}_{ij} i\partial_2^{\beta)} \quad , \quad 
V_{3ij}^{\alpha\beta} = -\frac{\kappa}2 \Bigl(\gamma^{(\alpha} J^{\beta)\mu}
\Bigr)_{ij} \partial_{3 \mu} \; . \label{3VO}
\end{equation}
The order $\kappa^2$ interactions define 4-point vertex operators $U_{I ij}^{
\alpha\beta\rho\sigma}$ similarly, for example,
\begin{equation}
\frac18 \kappa^2 h^2 \overline{\Psi} i\hspace{-.1cm}\not{\hspace{-.1cm} 
\partial} \Psi = \frac18 \kappa^2 \eta^{\alpha \beta} \eta^{\rho\sigma} i 
\hspace{-.1cm} \not{\hspace{-.1cm} \partial}_{2 ij} \times \overline{\Psi}_i 
\Psi_j h_{\alpha\beta} h_{\rho\sigma} \equiv U_{1ij}^{\alpha\beta\rho\sigma} 
\times \overline{\Psi}_i \Psi_j h_{\alpha\beta} h_{\rho\sigma} \; .
\end{equation}
The eight 4-point vertex operators are given in Table \ref{v4ops}. Note that 
we do not bother to symmetrize upon the identical graviton fields.

\begin{table}

\vbox{\tabskip=0pt \offinterlineskip
\def\tablerule{\noalign{\hrule}}
\halign to390pt {\strut#& \vrule#\tabskip=1em plus2em& 
\hfil#& \vrule#& \hfil#\hfil& \vrule#& \hfil#& \vrule#& \hfil#\hfil& 
\vrule#\tabskip=0pt\cr
\tablerule
\omit&height4pt&\omit&&\omit&&\omit&&\omit&\cr
&&\omit\hidewidth \# 
&&\omit\hidewidth {\rm Vertex Operator}\hidewidth&& 
\omit\hidewidth \#\hidewidth&& 
\omit\hidewidth {\rm Vertex Operator}
\hidewidth&\cr
\omit&height4pt&\omit&&\omit&&\omit&&\omit&\cr
\tablerule
\omit&height2pt&\omit&&\omit&&\omit&&\omit&\cr
&& 1 && $\frac18 \kappa^2 \eta^{\alpha\beta} \eta^{\rho\sigma}
i \hspace{-.1cm} \not{\hspace{-.1cm} \partial}_{2 ij}$
&& 5 && $-\frac14 \kappa^2 \eta^{\alpha\beta} (\gamma^{\rho} 
J^{\sigma\mu})_{ij} \partial_{4\mu}$ &\cr
\omit&height2pt&\omit&&\omit&&\omit&&\omit&\cr
\tablerule
\omit&height2pt&\omit&&\omit&&\omit&&\omit&\cr
&& 2 && $-\frac14 \kappa^2 \eta^{\alpha\rho} \eta^{\sigma\beta}
i \hspace{-.1cm} \not{\hspace{-.1cm} \partial}_{2 ij}$
&& 6 && $\frac18 \kappa^2 \eta^{\alpha\rho} (\gamma^{\mu} J^{\beta\sigma})_{ij}
\partial_{4\mu}$ &\cr
\omit&height2pt&\omit&&\omit&&\omit&&\omit&\cr
\tablerule
\omit&height2pt&\omit&&\omit&&\omit&&\omit&\cr
&& 3 && $-\frac14 \kappa^2 \eta^{\alpha\beta}\gamma^{\rho}_{ij} 
i\partial^{\sigma}_2$
&& 7 && $\frac14 \kappa^2 \eta^{\alpha\rho} (\gamma^{\beta} J^{\sigma\mu})_{ij}
(\partial_3 + \partial_4)_{\mu}$ &\cr
\omit&height2pt&\omit&&\omit&&\omit&&\omit&\cr
\tablerule
\omit&height2pt&\omit&&\omit&&\omit&&\omit&\cr
&& 4 && $\frac38 \kappa^2 \eta^{\alpha\rho} \gamma^{\beta}_{ij}
i\partial^{\sigma}_2$
&& 8 && $\frac14 \kappa^2 (\gamma^{\rho} J^{\sigma\alpha})_{ij} 
\partial_4^{\beta}$ &\cr
\omit&height2pt&\omit&&\omit&&\omit&&\omit&\cr
\tablerule}}

\caption{Vertex operators $U_{I ij}^{\alpha\beta\rho\sigma}$ contracted into 
$\overline{\Psi}_i \Psi_j h_{\alpha\beta} h_{\rho\sigma}$.}

\label{v4ops}

\end{table}

\section{Graviton Propagator and Counterterms}

The gravitational Lagrangian of low energy effective field theory is,
\begin{equation}
\mathcal{L}_{\rm Einstein} \equiv \frac1{16\pi G} \Bigl( R - (D\!-\!2) 
\Lambda\Bigr) \sqrt{-g} \; . \label{Einstein}
\end{equation}
The symbols $G$ and $\Lambda$ stand for Newton's constant and the cosmological
constant, respectively. The unfamiliar factor of $D\!-\!2$ multiplying 
$\Lambda$ makes the pure gravity field equations imply $R_{\mu\nu} = \Lambda 
g_{\mu\nu}$ in any dimension. The symbol $R$ stands for the Ricci scalar 
where our metric is spacelike and our curvature convention is,
\begin{equation}
R \equiv g^{\mu\nu} R_{\mu\nu} \equiv g^{\mu\nu} \Bigl(\Gamma^{\rho}_{~\nu\mu ,
\rho} - \Gamma^{\rho}_{~\rho \mu , \nu} + \Gamma^{\rho}_{~\rho \sigma}
\Gamma^{\sigma}_{~\nu\mu} - \Gamma^{\rho}_{~\nu\sigma} \Gamma^{\sigma}_{~\rho
\mu} \Bigr) \; .
\end{equation}
Unlike massless fermions, gravity is not conformally invariant. However,
it is still useful to express it in terms of the rescaled metric (\ref{confg})
and connection (\ref{confG}),
\begin{eqnarray}
\lefteqn{\mathcal{L}_{\rm Einstein} = \frac1{16 \pi G} \Biggl\{ a^{D-2} 
\widetilde{R} \!-\! 2 (D\!-\!1) a^{D-3} \widetilde{g}^{\mu\nu} 
\Bigl(a_{,\mu\nu} \!-\! \widetilde{\Gamma}^{\rho}_{~\mu\nu} a_{,\rho}\Bigr) } 
\nonumber \\
& & \hspace{3cm} - (D\!-\!4) (D\!-\! 1) a^{D-4} \widetilde{g}^{\mu\nu} a_{,\mu}
a_{,\nu} \!-\! (D\!-\!2) \Lambda a^D \Biggr\} \sqrt{-\widetilde{g}} \; .
\qquad \label{Econf}
\end{eqnarray}
The factors of $a$ which complicate this expression are the ultimate reason
there is interesting physics in this model!

None of the fermionic Feynman rules depended upon the functional form of 
the scale factor $a$ because the Dirac Lagrangian is conformally 
invariant. However, we shall need to fix $a$ in order to work out the 
graviton propagator from the Einstein Lagrangian (\ref{Econf}). The 
unique, maximally symmetric solution for positive $\Lambda$ is known as
de Sitter space. In order to regard this as a paradigm for inflation we 
work on a portion of the full de Sitter manifold known as the open conformal 
coordinate patch. The invariant element for this is,
\begin{equation}
ds^2 = a^2 \Bigl( -d\eta^2 + d\vec{x} \!\cdot\! d\vec{x}\Bigr) \qquad
{\rm where} \qquad a(\eta) = -\frac1{H\eta} \; ,
\end{equation}
and the $D$-dimensional Hubble constant is $H \equiv \sqrt{\Lambda/(D\!-\!1)}$.
Note that the conformal time $\eta$ runs from $-\infty$ to zero. For this
choice of scale factor we can extract a surface term from the invariant 
Lagrangian and write it in the form \cite{TW1},
\begin{eqnarray}
\lefteqn{\mathcal{L}_{\rm Einstein} \!-\! {\rm Surface} = {\scriptstyle 
(\frac{D}2 - 1)} H a^{D-1} \sqrt{-\widetilde{g}} \widetilde{g}^{\rho\sigma} 
\widetilde{g}^{\mu \nu} h_{\rho\sigma ,\mu} h_{\nu 0} + a^{D-2} 
\sqrt{-\widetilde{g}} \widetilde{g}^{\alpha\beta} \widetilde{g}^{\rho\sigma} 
\widetilde{g}^{\mu\nu} } \nonumber \\
& & \hspace{2cm} \times \Bigl\{{\scriptstyle \frac12} h_{\alpha\rho ,\mu} 
h_{\beta\sigma ,\nu} \!-\! {\scriptstyle \frac12} h_{\alpha\beta ,\rho} 
h_{\sigma\mu ,\nu} \!+\! {\scriptstyle \frac14} h_{\alpha\beta ,\rho} 
h_{\mu\nu ,\sigma} \!-\!  {\scriptstyle \frac14} h_{\alpha\rho ,\mu} 
h_{\beta\sigma ,\nu} \Bigr\} . \quad \label{Linv}
\end{eqnarray}

Gauge fixing is accomplished as usual by adding a gauge fixing term.
However, it turns out not to be possible to employ a de Sitter invariant
gauge for reasons that are not yet completely understood. One can add
such a gauge fixing term and then use the well-known formalism of Allen
and Jacobson \cite{AJ} to solve for the a fully de Sitter invariant propagator
\cite{AT,AM,HHT,HK,HW}. However, a curious thing happens when one uses the 
imaginary part of any such propagator to infer what ought to be the retarded 
Green's function of classical general relativity on a de Sitter background. 
The resulting Green's function gives a divergent response for a point mass
which also fails to obey the linearized invariant Einstein equation \cite{AM}!
We stress that the various propagators really do solve the gauge-fixed,
linearized equations with a point source. It is the physics which is wrong, 
not the math. There must be some obstacle to adding a de Sitter invariant 
gauge fixing term in gravity.

The problem seems to be related to combining constraint equations with 
the causal structure of the de Sitter geometry. Before gauge fixing the
constraint equations are elliptic, and they typically generate a nonzero
response throughout the de Sitter manifold, even in regions which are not
future-related to the source. Imposing a de Sitter invariant gauge results
in hyperbolic equations for which the response is zero in any region that
is not future-related to the source. This feature of gauge theories on de
Sitter space was first noted by Penrose in 1963 \cite{RP} and has since 
been studied for gravity \cite{TW1} and electromagnetism \cite{BK}. 

One consequence of the causality obstacle is that no completely de Sitter 
invariant gauge field propagator can correctly describe even classical 
physics over the entire de Sitter manifold. The confusing point is the
extent of the region over which the original, gauge invariant field 
equations are violated. For electromagnetism it turns out that a de Sitter 
invariant gauge can respect the gauge invariant equations on the submanifold 
which is future-directed from the source \cite{RPW2}. For gravity there seem 
to be violations of the Einstein equations everywhere \cite{AM}. The reason 
for this difference is not understood.

Quantum corrections bring new problems when using de Sitter invariant 
gauges. The one loop scalar self-mass-squared has recently been computed
in two different gauges for scalar quantum electrodynamics \cite{KW}. With
each gauge the computation was made for charged scalars which are massless,
minimally coupled and for charged scalars which are massless, conformally 
coupled. What goes wrong is clearest for the conformally coupled scalar, 
which should experience no large de Sitter enhancement over the flat space 
result on account of the conformal flatness of the de Sitter geometry. This 
is indeed the case when one employs the de Sitter breaking gauge that takes 
maximum account of the conformal invariance of electromagnetism in $D\!=\!3
\!+\!1$ spacetime dimensions. However, when the computation was done in the 
de Sitter invariant analogue of Feynman gauge the result was on-shell 
singularities! Off shell one-particle-irreducible functions need not agree 
in different gauges \cite{RJ} but they should agree on shell \cite{Lam}. In 
view of its on-shell singularities the result in the de Sitter invariant 
gauge is clearly wrong. 

The nature of the problem may be the apparent inconsistency between de
Sitter invariance and the manifold's linearization instability. Any
propagator gives the response (with a certain boundary condition) to a 
single point source. If the propagator is also de Sitter invariant then
this response must be valid throughout the full de Sitter manifold. But
the linearization instability precludes solving the invariant field 
equations for a single point source on the full manifold! This feature 
of the invariant theory is lost when a de Sitter invariant gauge fixing 
term is simply added to the action so it must be that the process of
adding it was not legitimate. In striving to attain a propagator which 
is valid everywhere, one invariably obtains a propagator that is not 
valid anywhere!

Although the pathology has not be identified as well as we should like, 
the procedure for dealing with it does seem to be clear. One can avoid 
the problem either by working on the full manifold with a noncovariant 
gauge condition that preserves the elliptic character of the constraint
equations, or else by employing a covariant, but not de Sitter invariant
gauge on an open submanifold \cite{TW1}. We choose the latter course and
employ the following analogue of the de Donder gauge fixing term of flat
space,
\begin{equation}
\mathcal{L}_{GF} = -\frac12 a^{D-2} \eta^{\mu\nu} F_{\mu} F_{\nu} \; , \;
F_{\mu} \equiv \eta^{\rho\sigma} \Bigl(h_{\mu\rho , \sigma} 
- \frac12 h_{\rho \sigma , \mu} + (D \!-\! 2) H a h_{\mu \rho}
\delta^0_{\sigma} \Bigr) . \label{GR}
\end{equation}

Because our gauge condition breaks de Sitter invariance it will be necessary
to contemplate noninvariant counterterms. It is therefore appropriate to
digress at this point with a description of the various de Sitter symmetries
and their effect upon (\ref{GR}). In our $D$-dimensional conformal coordinate
system the $\frac12 D(D\!+\!1)$ de Sitter transformations take the following
form:
\begin{enumerate}
\item{Spatial translations --- comprising $(D\!-\!1)$ transformations.}
\begin{eqnarray}
\eta' & = & \eta \label{homot} \; , \\
x^{\prime i} & = & x^i + \epsilon^i \label{homox} \; .
\end{eqnarray}
\item{Rotations --- comprising $\frac12 (D\!-\!1) (D\!-\!2)$ transformations.}
\begin{eqnarray}
\eta' & = & \eta \; , \label{isot} \\
x^{\prime i} & = & R^{ij} x^j \label{isox} \; .
\end{eqnarray}
\item{Dilatation --- comprising $1$ transformation.}
\begin{eqnarray}
\eta' & = & k \, \eta \; , \label{dilt} \\
x^{\prime i} & = & k \, x^i \label{dilx} \; .
\end{eqnarray}
\item{Spatial special conformal transformations --- comprising $(D\!-\!1)$
transformations.}
\begin{eqnarray}
\eta' & = & \frac{\eta}{1 \!-\! 2 \vec{\theta} \!\cdot\! \vec{x}
\!+\! \Vert \vec{\theta} \Vert^2 x\!\cdot\! x} \; , \label{ssct} \\
x^{\prime i} & = & \frac{x^i - \theta^i x\!\cdot\! x}{1 \!-\! 2 \vec{\theta} 
\!\cdot\! \vec{x} \!+\! \Vert \vec{\theta} \Vert^2 x\!\cdot\! x} \; . 
\label{sscx}
\end{eqnarray}
\end{enumerate}
It is easy to check that our gauge condition respects all of these but the 
spatial special conformal transformations. We will see that the other
symmetries impose important restrictions upon the BPHZ counterterms which 
are allowed.

It is now time to solve for the graviton propagator. Because its space and 
time components are treated differently in our coordinate system and gauge,
it is useful to have an expression for the purely spatial parts of the 
Lorentz metric and the Kronecker delta,
\begin{equation}
\overline{\eta}_{\mu\nu} \equiv \eta_{\mu\nu} + \delta^0_{\mu} \delta^0_{\nu}
\qquad {\rm and} \qquad \overline{\delta}^{\mu}_{\nu} \equiv \delta^{\mu}_{\nu}
- \delta_0^{\mu} \delta^0_{\nu} \; .
\end{equation}
The quadratic part of $\mathcal{L}_{\rm Einstein} + \mathcal{L}_{GF}$ can be 
partially integrated to take the form $\frac12 h^{\mu\nu} D_{\mu\nu}^{~~\rho
\sigma} h_{\rho\sigma}$, where the kinetic operator is,
\begin{eqnarray}
\lefteqn{D_{\mu\nu}^{~~\rho\sigma} \equiv \left\{ \frac12 \overline{\delta}_{
\mu}^{~(\rho} \overline{\delta}_{\nu}^{~\sigma)} - \frac14 \eta_{\mu\nu} 
\eta^{\rho\sigma} - \frac1{2(D\!-\!3)} \delta_{\mu}^0 \delta_{\nu}^0
\delta_0^{\rho} \delta_0^{\sigma} \right\} D_A } \nonumber \\
& & \hspace{3cm} + \delta^0_{(\mu} \overline{\delta}_{\nu)}^{(\rho}
\delta_0^{\sigma)} \, D_B + \frac12 \Bigl(\frac{D\!-\!2}{D\!-\!3}\Bigr) 
\delta_{\mu}^0 \delta_{\nu}^0 \delta_0^{\rho} \delta_0^{\sigma} \, D_C 
\; , \qquad
\end{eqnarray}
and the three scalar differential operators are,
\begin{eqnarray}
D_A & \equiv & \partial_{\mu} \Bigl(\sqrt{-g} g^{\mu\nu} \partial_{\nu}\Bigr)
\; , \\
D_B & \equiv & \partial_{\mu} \Bigl(\sqrt{-g} g^{\mu\nu} \partial_{\nu}\Bigr)
- \frac1{D} \Bigl(\frac{D\!-\!2}{D\!-\!1}\Bigr) R \sqrt{-g} \; , \\
D_C & \equiv & \partial_{\mu} \Bigl(\sqrt{-g} g^{\mu\nu} \partial_{\nu}\Bigr)
- \frac2{D} \Bigl(\frac{D\!-\!3}{D\!-\!1}\Bigr) R \sqrt{-g} \; .
\end{eqnarray}

The graviton propagator in this gauge takes the form of a sum of constant 
index factors times scalar propagators,
\begin{equation}
i\Bigl[{}_{\mu\nu} \Delta_{\rho\sigma}\Bigr](x;x') = \sum_{I=A,B,C}
\Bigl[{}_{\mu\nu} T^I_{\rho\sigma}\Bigr] i\Delta_I(x;x') \; . \label{gprop}
\end{equation}
The three scalar propagators invert the various scalar kinetic operators,
\begin{equation}
D_I \times i\Delta_I(x;x') = i \delta^D(x - x') \qquad {\rm for} \qquad
I = A,B,C \; , \label{sprops}
\end{equation}
and we will presently give explicit expressions for them. The index factors 
are,
\begin{eqnarray}
\Bigl[{}_{\mu\nu} T^A_{\rho\sigma}\Bigr] & = & 2 \, \overline{\eta}_{\mu (\rho}
\overline{\eta}_{\sigma) \nu} - \frac2{D\!-\! 3} \overline{\eta}_{\mu\nu}
\overline{\eta}_{\rho \sigma} \; , \\
\Bigl[{}_{\mu\nu} T^B_{\rho\sigma}\Bigr] & = & -4 \delta^0_{(\mu} 
\overline{\eta}_{\nu) (\rho} \delta^0_{\sigma)} \; , \\
\Bigl[{}_{\mu\nu} T^C_{\rho\sigma}\Bigr] & = & \frac2{(D \!-\!2) (D \!-\!3)}
\Bigl[(D \!-\!3) \delta^0_{\mu} \delta^0_{\nu} + \overline{\eta}_{\mu\nu}\Bigr]
\Bigl[(D \!-\!3) \delta^0_{\rho} \delta^0_{\sigma} + \overline{\eta}_{\rho
\sigma}\Bigr] \; .
\end{eqnarray}
With these definitions and equation (\ref{sprops}) for the scalar propagators
it is straightforward to verify that the graviton propagator (\ref{gprop})
indeed inverts the gauge-fixed kinetic operator,
\begin{equation}
D_{\mu\nu}^{~~\rho\sigma} \times i\Bigl[{}_{\rho\sigma} \Delta^{\alpha\beta}
\Bigr](x;x') = \delta_{\mu}^{(\alpha} \delta_{\nu}^{\beta)} i \delta^D(x-x')
\; .
\end{equation}

The scalar propagators can be expressed in terms of the following function 
of the invariant length $\ell(x;x')$ between $x^{\mu}$ and $x^{\prime \mu}$,
\begin{eqnarray}
y(x;x') & \equiv & 4 \sin^2\Bigl(\frac12 H \ell(x;x')\Bigr) 
= a a' H^2 {\Delta x }^2(x;x') \; , \\
& = & a a' H^2 \Bigl( \Vert \vec{x} - \vec{x}'\Vert^2 - (\vert \eta \!-\! 
\eta'\vert \!-\! i\delta)^2 \Bigr) \; . \label{fully}
\end{eqnarray}
The most singular term for each case is the propagator for a massless,
conformally coupled scalar \cite{BD},
\begin{equation}
{i\Delta}_{\rm cf}(x;x') = \frac{H^{D-2}}{(4\pi)^{\frac{D}2}} \Gamma\Bigl(
\frac{D}2 \!-\! 1\Bigr) \Bigl(\frac4{y}\Bigr)^{\frac{D}2-1} \; .
\end{equation}
The $A$-type propagator obeys the same equation as that of a massless,
minimally coupled scalar. It has long been known that no de Sitter invariant
solution exists \cite{AF}. If one elects to break de Sitter invariance 
while preserving homogeneity (\ref{homot}-\ref{homox}) and isotropy 
(\ref{isot}-\ref{isox}) --- this is known as the ``E(3)'' vacuum \cite{BA}
--- the minimal solution is \cite{OW1,OW2},
\begin{eqnarray}
\lefteqn{i \Delta_A(x;x') =  i \Delta_{\rm cf}(x;x') } \nonumber \\
& & + \frac{H^{D-2}}{(4\pi)^{\frac{D}2}} \frac{\Gamma(D \!-\! 1)}{\Gamma(
\frac{D}2)} \left\{\! \frac{D}{D\!-\! 4} \frac{\Gamma^2(\frac{D}2)}{\Gamma(D
\!-\! 1)} \Bigl(\frac4{y}\Bigr)^{\frac{D}2 -2} \!\!\!\!\!\! - \pi 
\cot\Bigl(\frac{\pi}2 D\Bigr) + \ln(a a') \!\right\} \nonumber \\
& & + \frac{H^{D-2}}{(4\pi)^{\frac{D}2}} \! \sum_{n=1}^{\infty}\! \left\{\!
\frac1{n} \frac{\Gamma(n \!+\! D \!-\! 1)}{\Gamma(n \!+\! \frac{D}2)} 
\Bigl(\frac{y}4 \Bigr)^n \!\!\!\! - \frac1{n \!-\! \frac{D}2 \!+\! 2} 
\frac{\Gamma(n \!+\!  \frac{D}2 \!+\! 1)}{\Gamma(n \!+\! 2)} \Bigl(\frac{y}4
\Bigr)^{n - \frac{D}2 +2} \!\right\} \! . \quad \label{DeltaA}
\end{eqnarray}
Note that this solution breaks dilatation invariance (\ref{dilt}-\ref{dilx})
in addition to the spatial special conformal invariance (\ref{ssct}-\ref{sscx})
broken by the gauge condition. By convoluting naive de Sitter transformations
with the compensating diffeomorphisms necessary to restore our gauge condition 
(\ref{GR}) one can show that the breaking of dilatation invariance is physical 
whereas the apparent breaking of spatial special conformal invariance is a 
gauge artifact \cite{GK}.

The B-type and $C$-type propagators possess de Sitter invariant (and also
unique) solutions,
\begin{eqnarray}
\lefteqn{i \Delta_B(x;x') =  i \Delta_{\rm cf}(x;x') - \frac{H^{D-2}}{(4
\pi)^{\frac{D}2}} \! \sum_{n=0}^{\infty}\! \left\{\!  \frac{\Gamma(n \!+\! D 
\!-\! 2)}{\Gamma(n \!+\! \frac{D}2)} \Bigl(\frac{y}4 \Bigr)^n \right. } 
\nonumber \\
& & \hspace{6.5cm} \left. - \frac{\Gamma(n \!+\!  \frac{D}2)}{\Gamma(n \!+\! 
2)} \Bigl( \frac{y}4 \Bigr)^{n - \frac{D}2 +2} \!\right\} \! , \qquad 
\label{DeltaB} \\
\lefteqn{i \Delta_C(x;x') =  i \Delta_{\rm cf}(x;x') + 
\frac{H^{D-2}}{(4\pi)^{\frac{D}2}} \! \sum_{n=0}^{\infty} \left\{\!
(n\!+\!1) \frac{\Gamma(n \!+\! D \!-\! 3)}{\Gamma(n \!+\! \frac{D}2)} 
\Bigl(\frac{y}4 \Bigr)^n \right. } \nonumber \\
& & \hspace{4.5cm} \left. - \Bigl(n \!-\! \frac{D}2 \!+\!  3\Bigr) \frac{
\Gamma(n \!+\! \frac{D}2 \!-\! 1)}{\Gamma(n \!+\! 2)} \Bigl(\frac{y}4 
\Bigr)^{n - \frac{D}2 +2} \!\right\} \! . \qquad \label{DeltaC}
\end{eqnarray}
They can be more compactly, but less usefully, expressed as hypergeometric 
functions \cite{CR,DC},
\begin{eqnarray}
i\Delta_B(x;x') & = & \frac{H^{D-2}}{(4\pi)^{\frac{D}2}} \frac{\Gamma(D\!-\!2)
\Gamma(1)}{\Gamma(\frac{D}2)} \, \mbox{}_2F_1\Bigl(D\!-\!2,1;\frac{D}2;1 \!-\!
\frac{y}4\Bigr) \; , \label{FDB} \\
i\Delta_C(x;x') & = & \frac{H^{D-2}}{(4\pi)^{\frac{D}2}} \frac{\Gamma(D\!-\!3)
\Gamma(2)}{\Gamma(\frac{D}2)} \, \mbox{}_2F_1\Bigl(D\!-\!3,2;\frac{D}2;1 \!-\!
\frac{y}4\Bigr) \; . \label{FDC}
\end{eqnarray}
These expressions might seem daunting but they are actually simple to use
because the infinite sums vanish in $D=4$, and each term in these sums
goes like a positive power of $y(x;x')$. This means the infinite
sums can only contribute when multiplied by a divergent term, and even
then only a small number of terms can contribute. Note also that the $B$-type
and $C$-type propagators agree with the conformal propagator in $D=4$.

In view of the subtle problems associated with the graviton propagator in
what seemed to be perfectly valid, de Sitter invariant gauges \cite{AM,TW1}, 
it is well to review the extensive checks that have been made on the 
consistency of this noninvariant propagator. On the classical level it has 
been checked that the response to a point mass is in perfect agreement with 
the linearized, de Sitter-Schwarzchild geometry \cite{TW1}. The linearized 
diffeomorphisms which enforce the gauge condition have also been explicitly
constructed \cite{TW2}. Although a tractable, $D$-dimensional form for the 
various scalar propagators $i\Delta_I(x;x')$ was not originally known, 
some simple identities obeyed by the mode functions in their Fourier 
expansions sufficed to verify the tree order Ward identity \cite{TW2}. The 
full, $D$-dimensional formalism has been used recently to compute the 
graviton 1-point function at one loop order \cite{TW3}. The result seems to 
be in qualitative agreement with canonical computations in other gauges 
\cite{LHF,FMVV}. A $D\!=\!3\!+\!1$ version of the formalism --- with 
regularization accomplished by keeping the parameter $\delta \neq 0$ in the 
de Sitter length function $y(x;x')$ (\ref{fully}) --- was used to evaluate 
the leading late time correction to the 2-loop 1-point function \cite{TW4,TW5}. 
The same technique was used to compute the unrenormalized graviton self-energy 
at one loop order \cite{TW6}. An explicit check was made that the flat space 
limit of this quantity agrees with Capper's result \cite{DMC2} for the 
graviton self-energy in the same gauge. The one loop Ward identity was also
checked in de Sitter background \cite{TW6}. Finally, the $D\!=\!4$ formalism 
was used to compute the two loop contribution from a massless, minimally 
coupled scalar to the 1-graviton function \cite{TW7}. The result was shown to 
obey an important bound imposed by global conformal invariance on the maximum 
possible late time effect.

It remains to deal with the local counterterms we must add, order-by-order
in perturbation theory, to absorb divergences in the sense of BPHZ 
renormalization. The particular counterterms which renormalize the ferm\-i\-on
self-energy must obviously involve a single $\overline{\psi}$ and a single
$\psi$.\footnote{Although the Dirac Lagrangian is conformally invariant,
the counterterms required to renormalize the fermion self-energy will not 
possess this symmetry because quantum gravity does not. We must therefore
work with the original fields rather than the conformally rescaled ones.}
At one loop order the superficial degree of divergence of quantum gravitational
contributions to the fermion self-energy is three, so the necessary 
counterterms can involve zero, one, two or three derivatives. These derivatives
can either act upon the fermi fields or upon the metric, in which case they 
must be organized into curvatures or derivatives of curvatures. We will first 
exhaust the possible invariant counterterms for a general renormalized fermion 
mass and a general background geometry, and then specialize to the case of zero 
mass in de Sitter background. We close with a discussion of possible 
noninvariant counterterms.

All one loop corrections from quantum gravity must carry a factor of $\kappa^2
\sim {\rm mass}^{-2}$. There will be additional dimensions associated with 
derivatives and with the various fields, and the balance must be struck using 
the renormalized fermion mass, $m$. Hence the only invariant counterterm with 
no derivatives has the form,
\begin{equation}
\kappa^2 m^3 \overline{\psi} \psi \sqrt{-g} \; . \label{zero}
\end{equation}
With one derivative we can always partially integrate to act upon the $\psi$
field, so the only invariant counterterm is,
\begin{equation}
\kappa^2 m^2 \overline{\psi} i \hspace{-.1cm} \not{\hspace{-.15cm} \mathcal{D}}
\psi \sqrt{-g} \; . \label{one}
\end{equation}
Two derivatives can either act upon the fermions or else on the metric to
produce curvatures. We can organize the various possibilities as follows,
\begin{equation}
\kappa^2 m \overline{\psi} (i \hspace{-.1cm} \not{\hspace{-.15cm}
\mathcal{D}})^2 \psi \sqrt{-g} \quad , \quad \kappa^2 m R \overline{\psi} 
\psi \sqrt{-g} \; . \label{two}
\end{equation}
Three derivatives can be all acted on the fermions, or one on the fermions 
and two in the form of curvatures, or there can be a differentiated curvature,
\begin{eqnarray}
\kappa^2 \overline{\psi} \Bigl( (i \hspace{-.1cm} \not{\hspace{-.15cm} 
\mathcal{D}})^2 \!+\! \frac{R}{D(D\!-\!1)} \Bigr) i \hspace{-.1cm} 
\not{\hspace{-.15cm} \mathcal{D}} \psi \sqrt{-g} 
& , & \kappa^2 R \, \overline{\psi} \, i \hspace{-.1cm} \not{\hspace{-.15cm} 
\mathcal{D}} \psi \sqrt{-g} \; , \nonumber \\
\kappa^2 e_{\mu m} \Bigl(R^{\mu\nu} - \frac1{D} g^{\mu\nu} R\Bigr)
\overline{\psi} \gamma^m i \mathcal{D}_{\nu} \psi \sqrt{-g} & , & \kappa^2 
e^{\mu}_{~m} R_{,\mu} \overline{\psi} \gamma^m \psi \sqrt{-g} \; .\label{three}
\end{eqnarray}

Because mass is multiplicatively renormalized in dimensional regularization,
and because we are dealing with zero mass fermions, counterterms (\ref{zero}),
(\ref{one}) and (\ref{two}) are all unnecessary for our calculation. Although
all four counterterms (\ref{three}) are nonzero and distinct for a general
metric background, they only affect our fermion self-energy for the special
case of de Sitter background. For that case $R_{\mu\nu} = (D\!-\!1) H^2 
g_{\mu\nu}$, so the last two counterterms vanish. The specialization of the 
invariant counter-Lagrangian we require to de Sitter background is therefore,
\begin{eqnarray}
\lefteqn{\Delta \mathcal{L}_{\rm inv} = \alpha_1 \kappa^2 \overline{\psi} 
\Bigl( (i \hspace{-.1cm} \not{\hspace{-.15cm} \mathcal{D}})^2 \!+\! \frac{R}{D
(D\!-\!1)}\Bigr) i \hspace{-.1cm} \not{\hspace{-.15cm} \mathcal{D}}
 \psi \sqrt{-g} + \alpha_2 \kappa^2 R \, \overline{\psi} \, i \hspace{-.1cm} 
\not{\hspace{-.15cm} \mathcal{D}} \psi \sqrt{-g} \; , \label{invctms} } \\
& & \longrightarrow \alpha_1 \kappa^2 \overline{\Psi} \Bigl(i \hspace{-.1cm} 
\not{\hspace{-.1cm} \partial} a^{-1} i \hspace{-.1cm} \not{\hspace{-.1cm} 
\partial} a^{-1} \!+\! \frac{R}{D(D\!-\!1)} \Bigr) i \hspace{-.1cm} 
\not{\hspace{-.1cm} \partial} \Psi + \alpha_2 (D\!-\!1) D \kappa^2 H^2 
\overline{\Psi} i \hspace{-.1cm} \not{\hspace{-.1cm} \partial} \Psi \; . \qquad
\end{eqnarray}
Here $\alpha_1$ and $\alpha_2$ are $D$-dependent constants which are 
dimensionless for $D\!=\!4$. The associated vertex operators are,
\begin{eqnarray}
C_{1 ij} & \equiv & \alpha_1 \kappa^2 \Bigl(i \hspace{-.1cm} 
\not{\hspace{-.1cm} \partial} a^{-1} i \hspace{-.1cm} \not{\hspace{-.1cm} 
\partial} a^{-1} i \hspace{-.1cm} \not{\hspace{-.1cm} \partial} \!+\! H^2 
i \hspace{-.1cm} \not{\hspace{-.1cm} \partial}\Bigr)_{ij} = \alpha_1 \kappa^2 
\Bigl(a^{-1} i \hspace{-.1cm} \not{\hspace{-.1cm} \partial} \partial^2 a^{-1}
\Bigr)_{ij} \; , \label{C1} \\
C_{2 ij} & \equiv & \alpha_2 (D\!-\!1) D \kappa^2 H^2 i \hspace{-.1cm} 
\not{\hspace{-.1cm} \partial}_{ij} \; . \label{C2}
\end{eqnarray}
Of course $C_1$ is the higher derivative counterterm mentioned in section 1.
It will renormalize the most singular terms --- coming from the $i\Delta_{\rm 
cf}$ part of the graviton propagator --- which are unimportant because they 
are suppressed by powers of the scale factor. The other vertex operator, 
$C_2$, is a sort of dimensionful field strength renormalization in de Sitter 
background. It will renormalize the less singular contributions which derive 
physically from inflationary particle production.

The one loop fermion self-energy would require no additional counterterms 
had it been possible to use the background field technique in background 
field gauge \cite{BSD1,BSD2,BSD3,LFA}. However, the obstacle to using a de 
Sitter invariant gauge obviously precludes this. We must therefore come to 
terms with the possibility that divergences may arise which require 
noninvariant counterterms. What form can these counterterms take? Applying 
the BPHZ theorem \cite{BP,H,Z1,Z2} to the gauge-fixed theory in de Sitter 
background implies that the relevant counterterms must still consist of 
$\kappa^2$ times a spinor differential operator with the dimension of 
mass-cubed, involving no more than three derivatives and acting between 
$\overline{\Psi}$ and $\Psi$. As the only dimensionful constant in our 
problem, powers of $H$ must be used to make up whatever dimensions are not 
supplied by derivatives.

Because dimensional regularization respects diffeomorphism invariance, it is 
only the gauge fixing term (\ref{GR}) that permits noninvariant 
counterterms.\footnote{One might think that they could come as well from
the fact that the vacuum breaks de Sitter invariance, but symmetries broken 
by the vacuum do not introduce new counterterms \cite{SRC}. Highly relevant, 
explicit examples are provided by recent computations for a massless,
minimally coupled scalar with a quartic self-interaction in the same locally
de Sitter background used here. The vacuum in this theory also breaks de
Sitter invariance but noninvariant counterterms fail to arise even at {\it two
loop} order in either the expectation value of the stress tensor \cite{OW1,OW2} 
or the self-mass-squared \cite{BOW}. It is also relevant that the one loop
vacuum polarization from (massless, minimally coupled) scalar quantum 
electrodynamics is free of noninvariant counterterms in the same background
\cite{PTW}.} Conversely, noninvariant counterterms must respect the residual 
symmetries of the gauge condition. Homogeneity (\ref{homot}-\ref{homox}) 
implies that the spinor differential operator cannot depend upon the spatial 
coordinate $x^i$. Similarly, isotropy (\ref{isot}-\ref{isox}) requires that 
any spatial derivative operators $\partial_i$ must either be contracted into 
$\gamma^i$ or another spatial derivative. Owing to the identity,
\begin{equation}
(\gamma^i \partial_i)^2 = - \nabla^2 \; , 
\end{equation}
we can think of all spatial derivatives as contracted into $\gamma^i$.
Although the temporal derivative is not required to be multiplied by $\gamma^0$
we lose nothing by doing so provided additional dependence upon $\gamma^0$ is
allowed.

The final residual symmetry is invariance under dilatations 
(\ref{dilt}-\ref{dilx}). It has the crucial consequence that derivative 
operators can only appear in the form $a^{-1} \partial_{\mu}$. In addition 
the entire counterterm must have an overall factor of $a$, and there can 
be no other dependence upon $\eta$. So the most general counterterm 
consistent with our gauge condition takes the form,
\begin{equation}
\Delta \mathcal{L}_{\rm non} = \kappa^2 H^3 a \overline{\Psi} 
\mathcal{S}\Bigl((H a)^{-1} \gamma^0 \partial_0, (H a)^{-1} \gamma^i 
\partial_i\Bigr) \Psi \; , \label{noninv}
\end{equation}
where the spinor function $\mathcal{S}(b,c)$ is at most a third order 
polynomial function of its arguments, and it may involve $\gamma^0$ in an 
arbitrary way.

Three more principles constrain noninvariant counterterms. The first of
these principles is that the fermion self-enery involves only odd powers 
of gamma matrices. This follows from the masslessness of our fermion and the
consequent fact that the fermion propagator and each interaction vertex
involves only odd numbers of gamma matrices. This principle fixes the
dependence upon $\gamma^0$ and allows us to express the spinor differential
operator in terms of just ten constants $\beta_i$,
\begin{eqnarray}
\lefteqn{\kappa^2 H^3 a \mathcal{S}\Bigl((H a)^{-1} \gamma^0 \partial_0, 
(H a)^{-1} \gamma^i \partial_i\Bigr) = \kappa^2 a \Biggl\{ \beta_1 (a^{-1}
\gamma^0 \partial_0)^3 } \nonumber \\
& & \hspace{.5cm} + \beta_2 \Bigl[(a^{-1} \gamma^0 \partial_0)^2 (a^{-1} 
\gamma^i \partial_i)\Bigr] + \beta_3 \Bigl[(a^{-1} \gamma^0 \partial_0) 
(a^{-1} \gamma^i \partial_i)^2\Bigr] + \beta_4 (a^{-1} \gamma^i \partial_i)^3 
\nonumber \\
& & \hspace{1.5cm} + H \gamma^0 \Biggl( \beta_5 (a^{-1} \gamma^0 \partial_0)^2 
+ \beta_6 \Bigl[(a^{-1} \gamma^0 \partial_0) (a^{-1} \gamma^i \partial_i) 
\Bigr] + \beta_7 (a^{-1} \gamma^i \partial_i)^2 \Biggr) \nonumber \\
& & \hspace{3cm} + H^2 \Biggl( \beta_8 (a^{-1} \gamma^0 \partial_0) + \beta_9
(a^{-1} \gamma^i \partial_i) \Biggr)+ H^3 \gamma^0 \beta_{10} \Biggr\} . \qquad
\label{expS}
\end{eqnarray}
In this expansion, but for the rest of this section only, we define 
noncommuting factors within square brackets to be symmetrically ordered, for 
example,
\begin{eqnarray}
\lefteqn{\Bigl[(a^{-1} \gamma^0 \partial_0)^2 (a^{-1} \gamma^i \partial_i)
\Bigr] \equiv \frac13 (a^{-1} \gamma^0 \partial_0)^2 (a^{-1} \gamma^i 
\partial_i)} \nonumber \\
& & \hspace{1.5cm} + \frac13 (a^{-1} \gamma^0 \partial_0) (a^{-1} \gamma^i 
\partial_i) (a^{-1} \gamma^0 \partial_0) + \frac13 (a^{-1} \gamma^i \partial_i) 
(a^{-1} \gamma^0 \partial_0)^2 \; . \qquad
\end{eqnarray}

The second principle is that our gauge condition (\ref{GR}) becomes 
Poincar\'e invariant in the flat space limit of $H \rightarrow 0$, where the 
conformal time is $\eta = -e^{-Ht}/H$ with $t$ held fixed. In that limit 
only the four cubic terms of (\ref{expS}) survive,
\begin{eqnarray}
\lefteqn{\lim_{H \rightarrow 0} \kappa^2 H^3 a \mathcal{S}\Bigl((H a)^{-1} 
\gamma^0 \partial_0, (H a)^{-1} \gamma^i \partial_i\Bigr) = \kappa^2 \Biggl\{ 
\beta_1 (\gamma^0 \partial_0)^3 } \nonumber \\
& & \hspace{1.5cm} + \beta_2 \Bigl[(\gamma^0 \partial_0)^2 (\gamma^i \partial_i)
\Bigr] + \beta_3 \Bigl[(\gamma^0 \partial_0) (\gamma^i \partial_i)^2\Bigr] 
+ \beta_4 (\gamma^i \partial_i)^3 \Biggr\} . \qquad
\end{eqnarray}
Because the entire theory is Poincar\'e invariant in that limit, these four
terms must sum to a term proportional to $(\gamma^{\mu} \partial_{\mu})^3$,
which implies,
\begin{equation}
\beta_1 = \frac13 \beta_2 = \frac13 \beta_3 = \beta_4 \; .
\end{equation}
But in that case the four cubic terms sum to give a linear combination of
the invariant counterterms (\ref{C1}) and (\ref{C2}),
\begin{eqnarray}
\lefteqn{\kappa^2 a \Biggl\{ (a^{-1} \gamma^0 \partial_0)^3 + 3 \Bigl[(a^{-1} 
\gamma^0 \partial_0)^2 (a^{-1} \gamma^i \partial_i)\Bigr] } \nonumber \\
& & \hspace{1.5cm} + 3 \Bigl[(a^{-1} \gamma^0 \partial_0) (a^{-1} \gamma^i 
\partial_i)^2\Bigr] + (a^{-1} \gamma^i \partial_i)^3 \Biggr\} = \kappa^2 
\hspace{-.1cm} \not{\hspace{-.1cm} \partial} \, a^{-1} \hspace{-.1cm} 
\not{\hspace{-.1cm} \partial} a^{-1} \hspace{-.1cm} \not{\hspace{-.1cm} 
\partial} \; . \qquad 
\end{eqnarray}
Because we have already counted this combination among the invariant 
counterterms it need not be included in $\mathcal{S}$.

The final simplifying principle is that the fermion self-energy is odd under
interchange of $x^{\mu}$ and $x^{\prime \mu}$,
\begin{equation}
-i \Bigl[\mbox{}_i \Sigma_j\Bigr](x;x') = + i \Bigl[\mbox{}_i \Sigma_j
\Bigr](x';x) \; . \label{refl}
\end{equation}
This symmetry is trivial at tree order, but not easy to show generally. 
Moreover, it isn't a property of individual terms, many of which violate 
(\ref{refl}). However, when everything is summed up the result must obey
(\ref{refl}), hence so too must the counterterms. This has the immediate
consequence of eliminating the counterterms with an even number of derivatives:
those proportional to $\beta_{5-7}$ and to $\beta_{10}$. We have already 
dispensed with $\beta_{1-4}$, which leaves only the linear terms, 
$\beta_{8-9}$. Because one linear combination of these already appears in 
the invariant (\ref{C2}) the sole noninvariant counterterm we require is,
\begin{equation}
\Delta \mathcal{L}_{\rm non} = \overline{\Psi} C_3 \Psi \qquad {\rm where}
\qquad C_{3ij} \equiv \alpha_3 \kappa^2 H^2 i \,
\hspace{-.1cm} \overline{\not{\hspace{-.1cm} \partial}}_{ij} \; . \label{nictm}
\end{equation}

\section{Contributions from the 4-Point Vertices}

In this section we evaluate the contributions from 4-point vertex operators
of Table \ref{v4ops}. The generic diagram topology is depicted in Fig.~1. 
The analytic form is,
\begin{equation}
-i\Bigl[\mbox{}_i \Sigma_j^{\rm 4pt}\Bigr](x;x') = \sum_{I=1}^8 i U^{\alpha\beta
\rho\sigma}_{Iij} \, i\Bigl[\mbox{}_{\alpha\beta} \Delta_{\rho\sigma}
\Bigr](x;x') \, \delta^D(x\!-\!x') \; . \label{4ptloop}
\end{equation}
\begin{center}
\begin{picture}(300,80)(0,0) 
\ArrowLine(150,20)(90,20) 
\ArrowLine(210,20)(150,20)
\Vertex(150,20){3}
\Text(150,10)[b]{$x$}
\GlueArc(150,45)(23,-90,270){3}{8}
\end{picture}
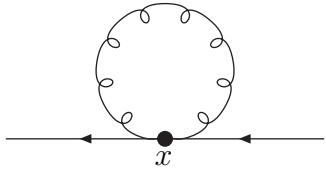
\\ {\rm Fig.~1: Contribution from 4-point vertices.}
\end{center}
And the generic contraction for each of the vertex operators in 
Table~\ref{v4ops} is given in Table~\ref{4con}.
\begin{table}

\vbox{\tabskip=0pt \offinterlineskip
\def\tablerule{\noalign{\hrule}}
\halign to390pt {\strut#& \vrule#\tabskip=1em plus2em& 
\hfil#\hfil& \vrule#& \hfil#\hfil& \vrule#\tabskip=0pt\cr
\tablerule
\omit&height4pt&\omit&&\omit&\cr
&&$\!\!\!\!{\rm I}\!\!\!\!$ && $\!\!\!\! i [\mbox{}_{\alpha\beta} 
\Delta_{\rho\sigma}](x;x') \, i U_I^{\alpha\beta\rho\sigma} \, 
\delta^D(x\!-\!x') \!\!\!\!$ & \cr
\omit&height4pt&\omit&&\omit&\cr
\tablerule
\omit&height2pt&\omit&&\omit&\cr
&& 1 && $-\frac18 \kappa^2 \, i [\mbox{}^{\alpha}_{~\alpha}
\Delta^{\rho}_{~\rho}](x;x) \, \hspace{-.1cm} \not{\hspace{-.1cm} \partial} 
\, \delta^D(x\!-\!x')$ & \cr
\omit&height2pt&\omit&&\omit&\cr
\tablerule
\omit&height2pt&\omit&&\omit&\cr
&& 2 && $\frac14 \kappa^2 \, i [\mbox{}^{\alpha\beta}
\Delta_{\alpha\beta}](x;x) \, \hspace{-.1cm} \not{\hspace{-.1cm} \partial} 
\, \delta^D(x\!-\!x')$ & \cr
\omit&height2pt&\omit&&\omit&\cr
\tablerule
\omit&height2pt&\omit&&\omit&\cr
&& 3 && $\frac14 \kappa^2 \, i [\mbox{}^{\alpha}_{~\alpha}
\Delta_{\rho\sigma}](x;x) \, \gamma^{\rho} \partial^{\sigma}
\, \delta^D(x\!-\!x')$ & \cr
\omit&height2pt&\omit&&\omit&\cr
\tablerule
\omit&height2pt&\omit&&\omit&\cr
&& 4 && $-\frac38 \kappa^2 \, i [\mbox{}^{\alpha}_{~\beta}
\Delta_{\alpha\sigma}](x;x) \, \gamma^{\beta} \partial^{\sigma}
\, \delta^D(x\!-\!x')$ & \cr
\omit&height2pt&\omit&&\omit&\cr
\tablerule
\omit&height2pt&\omit&&\omit&\cr
&& 5 && $-\frac{i}4 \kappa^2 \, \partial_{\mu}' i [\mbox{}^{\alpha}_{~\alpha}
\Delta_{\rho\sigma}](x;x') \, \gamma^{\rho} J^{\sigma \mu} \, 
\delta^D(x\!-\!x')$ & \cr
\omit&height2pt&\omit&&\omit&\cr
\tablerule
\omit&height2pt&\omit&&\omit&\cr
&& 6 && $\frac{i}8 \kappa^2 \, \partial_{\mu}' i [\mbox{}^{\alpha}_{~\beta}
\Delta_{\alpha\sigma}](x;x') \, \gamma^{\mu} J^{\beta\sigma} \, 
\delta^D(x\!-\!x')$ & \cr
\omit&height2pt&\omit&&\omit&\cr
\tablerule
\omit&height2pt&\omit&&\omit&\cr
&& 7 && $\frac{i}4 \kappa^2 \, \partial_{\mu} i [\mbox{}^{\alpha}_{~\beta}
\Delta_{\alpha\sigma}](x;x) \, \gamma^{\beta} J^{\sigma\mu} \, 
\delta^D(x\!-\!x')$ & \cr
\omit&height2pt&\omit&&\omit&\cr
\tablerule
\omit&height2pt&\omit&&\omit&\cr
&& 8 && $\frac{i}4 \kappa^2 \, \partial^{\prime \beta} i [\mbox{}_{\alpha
\beta} \Delta_{\rho\sigma}](x;x') \, \gamma^{\rho} J^{\sigma \alpha} \, 
\delta^D(x\!-\!x')$ & \cr
\omit&height2pt&\omit&&\omit&\cr
\tablerule}}

\caption{Generic 4-point contractions}

\label{4con}

\end{table}

From an examination of the generic contractions in Table~\ref{4con} it is 
apparent that we must work out how the three index factors $[\mbox{}_{\alpha
\beta} T^I_{\rho \sigma}]$ which make up the graviton propagator contract into 
$\eta^{\alpha\beta}$ and $\eta^{\alpha\rho}$. For the $A$-type and $B$-type 
index factors the various contractions give,
\begin{eqnarray}
\eta^{\alpha\beta} \, \Bigl[{}_{\alpha\beta} T^A_{\rho\sigma}\Bigr]
= - \Bigl(\frac4{D\!-\!3}\Bigr) \, \overline{\eta}_{\rho\sigma} & , &
\eta^{\alpha\rho} \, \Bigl[{}_{\alpha\beta} T^A_{\rho\sigma}\Bigr] 
= \Bigl(D\!-\!\frac2{D\!-\!3}\Bigr) \,\overline{\eta}_{\beta\sigma} \; , \\
\eta^{\alpha\beta} \, \Bigl[{}_{\alpha\beta} T^B_{\rho\sigma}\Bigr] 
= 0 & , & \eta^{\alpha\rho} \, \Bigl[{}_{\alpha\beta} T^B_{\rho\sigma} \Bigr] 
= -(D \!-\! 1) \, \delta^0_{\beta} \delta^0_{\sigma} + \overline{\eta}_{
\beta\sigma} \; , 
\end{eqnarray}
For the $C$-type index factor they are,
\begin{eqnarray}
\eta^{\alpha\beta} \, \Bigl[{}_{\alpha\beta} T^C_{\rho\sigma}\Bigr]
& = & \Bigl(\frac4{D-2}\Bigr) \, \delta^0_{\rho} \delta^0_{\sigma} +
\frac4{(D\!-\!2)(D\!-\!3)} \, \overline{\eta}_{\rho\sigma} \; , \nonumber \\
\eta^{\alpha\rho} \, \Bigl[{}_{\alpha\beta} T^C_{\rho\sigma}\Bigr] 
& = & -2 \Bigl(\frac{D\!-\!3}{D\!-\!2}\Bigr) \, \delta^0_{\beta} 
\delta^0_{\sigma} \!+\! \frac2{(D\!-\!2) (D\!-\!3)} \,
\overline{\eta}_{\beta\sigma} \; . \qquad
\end{eqnarray}
On occasion we also require double contractions. For the $A$-type index 
factor these are,
\begin{eqnarray}
\eta^{\alpha\beta} \eta^{\rho\sigma} \, \Bigl[{}_{\alpha\beta} 
T^A_{\rho\sigma}\Bigr] & = & -4 \Bigl(\frac{D\!-\!1}{D\!-\!3}\Bigr) \; , 
\nonumber \\
\eta^{\alpha\rho} \eta^{\beta\sigma} \, \Bigl[{}_{\alpha\beta} 
T^A_{\rho\sigma}\Bigr] & = & D (D\!-\!1) - 2
\Bigl(\frac{D\!-\!1}{D\!-\!3}\Bigr) \; .
\end{eqnarray}
The double contractions of the $B$-type and $C$-type index factors are,
\begin{eqnarray}
\eta^{\alpha\beta} \eta^{\rho\sigma} \, \Bigl[{}_{\alpha\beta} 
T^B_{\rho\sigma}\Bigr] = 0 & , & \eta^{\alpha\rho} \eta^{\beta\sigma} \, 
\Bigl[{}_{\alpha\beta} T^B_{\rho\sigma} \Bigr] = 2 (D \!-\! 1) \; , \\
\eta^{\alpha\beta} \eta^{\rho\sigma} \, \Bigl[{}_{\alpha\beta} 
T^C_{\rho\sigma}\Bigr] = \frac8{(D\!-\!2)(D\!-\!3)} & , & 
\eta^{\alpha\rho} \eta^{\beta\sigma} \, \Bigl[{}_{\alpha\beta} 
T^C_{\rho\sigma}\Bigr] = 2 \frac{(D^2 \!-\! 5D \!+\! 8)}{(D\!-\!2)(D\!-\!3)} 
\; . \qquad
\end{eqnarray}

Table~\ref{4props} was generated from Table~\ref{4con} by expanding the
graviton propagator in terms of index factors,
\begin{equation}
i\Bigl[{}_{\alpha\beta} \Delta_{\rho\sigma}\Bigr](x;x') =
\Bigl[{}_{\alpha\beta} T^A_{\rho\sigma}\Bigr] i\Delta_A(x;x') +
\Bigl[{}_{\alpha\beta} T^B_{\rho\sigma}\Bigr] i\Delta_B(x;x') +
\Bigl[{}_{\alpha\beta} T^C_{\rho\sigma}\Bigr] i\Delta_C(x;x') \; .
\end{equation}
We then perform the relevant contractions using the previous identities.
Relation (\ref{Jred}) was also exploited to simplify the gamma matrix 
structure.
\begin{table}

\vbox{\tabskip=0pt \offinterlineskip
\def\tablerule{\noalign{\hrule}}
\halign to390pt {\strut#& \vrule#\tabskip=1em plus2em& 
\hfil#\hfil& \vrule#& \hfil#\hfil& \vrule#& \hfil#\hfil& \vrule#\tabskip=0pt\cr
\tablerule
\omit&height4pt&\omit&&\omit&&\omit&\cr
&&$\!\!\!\!{\rm I}\!\!\!\!$ && $\!\!\!\!{\rm J}\!\!\!\!$ &&
$\!\!\!\! i [\mbox{}_{\alpha\beta} T^J_{\rho\sigma}]\, i\Delta_J(x;x') 
\, i U_I^{\alpha\beta\rho\sigma} \, \delta^D(x\!-\!x') \!\!\!\!$ & \cr
\omit&height4pt&\omit&&\omit&&\omit&\cr
\tablerule
\omit&height2pt&\omit&&\omit&&\omit&\cr
&& 1 && A && $\frac12 (\frac{D-1}{D-3}) \kappa^2 \, i \Delta_A(x;x) \,
\hspace{-.1cm} \not{\hspace{-.1cm} \partial} \, \delta^D(x\!-\!x')$ & \cr
\omit&height2pt&\omit&&\omit&&\omit&\cr
\tablerule
\omit&height2pt&\omit&&\omit&&\omit&\cr
&& 1 && B && $0$ & \cr
\omit&height2pt&\omit&&\omit&&\omit&\cr
\tablerule
\omit&height2pt&\omit&&\omit&&\omit&\cr
&& 1 && C && $\!\!\!\!-\frac1{(D-2)(D-3)} \kappa^2 \, i \Delta_C(x;x) \, 
\hspace{-.1cm} \not{\hspace{-.1cm} \partial} \, \delta^D(x\!-\!x')\!\!\!\!$ 
& \cr
\omit&height2pt&\omit&&\omit&&\omit&\cr
\tablerule
\omit&height2pt&\omit&&\omit&&\omit&\cr
&& 2 && A && $\!\!\!\!(\frac{D-1}4) (\frac{D^2 -3D -2}{D-3}) \kappa^2 \, i 
\Delta_A(x;x) \, \hspace{-.1cm} \not{\hspace{-.1cm} \partial} \, 
\delta^D(x\!-\!x')\!\!\!\!$ & \cr
\omit&height2pt&\omit&&\omit&&\omit&\cr
\tablerule
\omit&height2pt&\omit&&\omit&&\omit&\cr
&& 2 && B && $(\frac{D-1}2) \kappa^2 \, i \Delta_B(x;x) \, \hspace{-.1cm} 
\not{\hspace{-.1cm} \partial} \, \delta^D(x\!-\!x')$ & \cr
\omit&height2pt&\omit&&\omit&&\omit&\cr
\tablerule
\omit&height2pt&\omit&&\omit&&\omit&\cr
&& 2 && C && $\frac12 \frac{(D^2 -5D + 8)}{(D-2)(D-3)} \kappa^2 \, i 
\Delta_C(x;x) \, \hspace{-.1cm} \not{\hspace{-.1cm} \partial} \, 
\delta^D(x\!-\!x')$ & \cr
\omit&height2pt&\omit&&\omit&&\omit&\cr
\tablerule
\omit&height2pt&\omit&&\omit&&\omit&\cr
&& 3 && A && $-\frac1{D-3} \kappa^2 \, i \Delta_A(x;x) \, \hspace{-.1cm} 
\overline{\not{\hspace{-.1cm} \partial}} \, \delta^D(x\!-\!x')$ & \cr
\omit&height2pt&\omit&&\omit&&\omit&\cr
\tablerule
\omit&height2pt&\omit&&\omit&&\omit&\cr
&& 3 && B && $0$ & \cr
\omit&height2pt&\omit&&\omit&&\omit&\cr
\tablerule
\omit&height2pt&\omit&&\omit&&\omit&\cr
&& 3 && C && $\!\!\!\!\frac1{(D-2)(D-3)} \kappa^2 \, i \Delta_C(x;x) [\; 
\hspace{-.1cm} \overline{\not{\hspace{-.1cm} \partial}} \!-\! {\scriptstyle 
(D-3)} \gamma^0 \partial_0] \delta^D(x\!-\!x') \!\!\!\!$ & \cr
\omit&height2pt&\omit&&\omit&&\omit&\cr
\tablerule
\omit&height2pt&\omit&&\omit&&\omit&\cr
&& 4 && A && $-\frac38 (\frac{D^2 -3D -2}{D-3}) \kappa^2 \, i \Delta_A(x;x) \,
\hspace{-.1cm} \overline{\not{\hspace{-.1cm} \partial}} \, \delta^D(x\!-\!x')$ 
& \cr
\omit&height2pt&\omit&&\omit&&\omit&\cr
\tablerule
\omit&height2pt&\omit&&\omit&&\omit&\cr
&& 4 && B && $\!\!\!\!-\frac38 \kappa^2 \, i \Delta_B(x;x) [\; \hspace{-.1cm} 
\overline{\not{\hspace{-.1cm} \partial}} \!+\! {\scriptstyle (D-1)} 
\gamma^0 \partial_0] \delta^D(x\!-\!x')\!\!\!\!$ & \cr
\omit&height2pt&\omit&&\omit&&\omit&\cr
\tablerule
\omit&height2pt&\omit&&\omit&&\omit&\cr
&& 4 && C && $\!\!\!\!-\frac34 \frac1{(D-2)(D-3)} \kappa^2 \, i 
\Delta_C(x;x) [\; \hspace{-.1cm} \overline{\not{\hspace{-.1cm} \partial}}
\!+\! {\scriptstyle (D-3)}^2 \gamma^0 \partial_0] \delta^D(x\!-\!x')
\!\!\!\!$ & \cr
\omit&height2pt&\omit&&\omit&&\omit&\cr
\tablerule
\omit&height2pt&\omit&&\omit&&\omit&\cr
&& 5 && A && $\!\!\!\!\kappa^2 [-\frac1{2(D-3)} \; \hspace{-.1cm} \overline{
\not{\hspace{-.1cm} \partial}}' \!+\! \frac12 (\frac{D-1}{D-3})
\hspace{-.1cm} \not{\hspace{-.1cm} \partial}' ] \, i\Delta_A(x;x') 
\, \delta^D(x\!-\!x') \!\!\!\!$ & \cr
\omit&height2pt&\omit&&\omit&&\omit&\cr
\tablerule
\omit&height2pt&\omit&&\omit&&\omit&\cr
&& 5 && B && $0$ & \cr
\omit&height2pt&\omit&&\omit&&\omit&\cr
\tablerule
\omit&height2pt&\omit&&\omit&&\omit&\cr
&& 5 && C && $\!\!\!\!- \frac1{(D-2)(D-3)} \kappa^2 \, [\frac12\; 
\hspace{-.1cm} \overline{\not{\hspace{-.1cm} \partial}}' \!+\! (\frac{D-1}2)
\gamma^0 \partial_0'] \, i\Delta_C(x;x') \, \delta^D(x\!-\!x')\!\!\!\!$ & \cr
\omit&height2pt&\omit&&\omit&&\omit&\cr
\tablerule
\omit&height2pt&\omit&&\omit&&\omit&\cr
&& 6 && A && $0$ & \cr
\omit&height2pt&\omit&&\omit&&\omit&\cr
\tablerule
\omit&height2pt&\omit&&\omit&&\omit&\cr
&& 6 && B && $0$ & \cr
\omit&height2pt&\omit&&\omit&&\omit&\cr
\tablerule
\omit&height2pt&\omit&&\omit&&\omit&\cr
&& 6 && C && $0$ & \cr
\omit&height2pt&\omit&&\omit&&\omit&\cr
\tablerule
\omit&height2pt&\omit&&\omit&&\omit&\cr
&& 7 && A && $\!\!\!\! (\frac{D^2-3D-2}{D-3}) \kappa^2 [-\frac18\; 
\hspace{-.1cm} \overline{\not{\hspace{-.1cm} \partial}} \!+\! 
(\frac{D-1}8) \hspace{-.1cm} \not{\hspace{-.1cm} \partial} ] 
\, i\Delta_A(x;x) \, \delta^D(x\!-\!x') \!\!\!\!$ & \cr
\omit&height2pt&\omit&&\omit&&\omit&\cr
\tablerule
\omit&height2pt&\omit&&\omit&&\omit&\cr
&& 7 && B && $\!\!\!\! \kappa^2 [(\frac{D-2}8) \; \hspace{-.1cm} \overline{
\not{\hspace{-.1cm} \partial}} \!+\! (\frac{D-1}8) \hspace{-.1cm} 
\not{\hspace{-.1cm} \partial} ] \, i\Delta_B(x;x) \, \delta^D(x\!-\!x') 
\!\!\!\!$ & \cr
\omit&height2pt&\omit&&\omit&&\omit&\cr
\tablerule
\omit&height2pt&\omit&&\omit&&\omit&\cr
&& 7 && C && $\!\!\!\! \frac14 \kappa^2 [\frac{(D^2-6D+8)}{(D-2)(D-3)} \;
\hspace{-.1cm} \overline{\not{\hspace{-.1cm} \partial}} \!+\! \frac{(D-1)}{
(D-2)(D-3)} \hspace{-.1cm} \not{\hspace{-.1cm} \partial} ] \, i\Delta_C(x;x) 
\, \delta^D(x\!-\!x') \!\!\!\!$ & \cr
\omit&height2pt&\omit&&\omit&&\omit&\cr
\tablerule
\omit&height2pt&\omit&&\omit&&\omit&\cr
&& 8 && A && $\!\!\!\! - \kappa^2 \frac{(D-2)(D-1)}{8(D-3)} \; \hspace{-.1cm} 
\overline{\not{\hspace{-.1cm} \partial}}' \, i\Delta_A(x;x') \, 
\delta^D(x\!-\!x') \!\!\!\!$ & \cr
\omit&height2pt&\omit&&\omit&&\omit&\cr
\tablerule
\omit&height2pt&\omit&&\omit&&\omit&\cr
&& 8 && B && $\!\!\!\! -\kappa^2 [\frac18 \; \hspace{-.1cm} \overline{
\not{\hspace{-.1cm} \partial}}' \!+\! (\frac{D-1}8) \gamma^0 \partial_0' ] \, 
i\Delta_B(x;x') \, \delta^D(x\!-\!x') \!\!\!\!$ & \cr
\omit&height2pt&\omit&&\omit&&\omit&\cr
\tablerule
\omit&height2pt&\omit&&\omit&&\omit&\cr
&& 8 && C && $\!\!\!\! \frac14 \kappa^2 [\frac1{(D-2)(D-3)} \; \hspace{-.1cm} 
\overline{\not{\hspace{-.1cm} \partial}}' \!-\! (\frac{D-1}{D-2}) \gamma^0
\partial_0' ] \, i\Delta_C(x;x') \, \delta^D(x\!-\!x') \!\!\!\!$ & \cr
\omit&height2pt&\omit&&\omit&&\omit&\cr
\tablerule}}

\caption{4-point contribution from each part of the graviton propagator.}

\label{4props}

\end{table}

From Table~\ref{4props} it is apparent that we require the coincidence 
limits of zero or one derivatives acting on each of the scalar propagators. 
For the $A$-type propagator these are,
\begin{eqnarray}
\lim_{x' \rightarrow x} \, {i\Delta}_A(x;x') & = & \frac{H^{D-2}}{(4\pi)^{
\frac{D}2}} \frac{\Gamma(D-1)}{\Gamma(\frac{D}2)} \left\{-\pi \cot\Bigl(
\frac{\pi}2 D \Bigr) + 2 \ln(a) \right\} , \label{Acoin} \\
\lim_{x' \rightarrow x} \, \partial_{\mu} {i\Delta}_A(x;x') & = & 
\frac{H^{D-2}}{(4\pi)^{\frac{D}2}} \frac{\Gamma(D-1)}{\Gamma(\frac{D}2)} 
\times H a \delta^0_{\mu} \; . \label{Acoin'}
\end{eqnarray}
The analogous coincidence limits for the $B$-type propagator are actually 
finite in $D=4$ dimensions,
\begin{eqnarray}
\lim_{x' \rightarrow x} \, {i\Delta}_B(x;x') & = & \frac{H^{D-2}}{(4\pi)^{
\frac{D}2}} \frac{\Gamma(D-1)}{\Gamma(\frac{D}2)}\times -\frac1{D\!-\!2} \; ,\\
\lim_{x' \rightarrow x} \, \partial_{\mu} {i\Delta}_B(x;x') & = & 0 \; .
\end{eqnarray}
The same is true for the coincidence limits of the $C$-type propagator,
\begin{eqnarray}
\lim_{x' \rightarrow x} \, {i\Delta}_C(x;x') & = & \frac{H^{D-2}}{(4\pi)^{
\frac{D}2}} \frac{\Gamma(D-1)}{\Gamma(\frac{D}2)}\times \frac1{(D\!-\!2)
(D\!-\!3)} \; ,\\
\lim_{x' \rightarrow x} \, \partial_{\mu} {i\Delta}_C(x;x') & = & 0 \; .
\label{Ccoin'}
\end{eqnarray}

Our final result for the 4-point contributions is given in Table~\ref{4fin}.
It was obtained from Table~\ref{4props} by using the previous coincidence 
limits. We have also always chosen to re-express conformal time derivatives
thusly,
\begin{equation}
\gamma^0 \partial_0 = \;\; \hspace{-.1cm} \not{\hspace{-.1cm} \partial} - \;
\hspace{-.1cm} \overline{\not{\hspace{-.1cm} \partial}} \; .
\end{equation}
A final point concerns the fact that the terms in the final column of
Table~\ref{4fin} do not obey the reflection symmetry. In the next section
we will find the terms which exactly cancel these.

\begin{table}

\vbox{\tabskip=0pt \offinterlineskip
\def\tablerule{\noalign{\hrule}}
\halign to390pt {\strut#& \vrule#\tabskip=1em plus2em& 
\hfil#\hfil& \vrule#& \hfil#\hfil& \vrule#& \hfil#\hfil& \vrule#& 
\hfil#\hfil& \vrule#& \hfil#\hfil& \vrule#\tabskip=0pt\cr
\tablerule
\omit&height4pt&\omit&&\omit&&\omit&&\omit&&\omit&\cr
&&$\!\!\!\!{\rm I}\!\!\!\!$ && $\!\!\!\!{\rm J}\!\!\!\!$ &&
$\!\!\!\! \hspace{-.1cm} \not{\hspace{-.1cm} \partial} \,
\delta^D(x\!-\!x')\!\!\!\!$ && $\!\!\!\! \hspace{-.1cm} \overline{\not{
\hspace{-.1cm} \partial}} \, \delta^D(x\!-\!x')\!\!\!\!$ && $\!\!\!\!a H
\gamma^0 \, \delta^D(x\!-\!x')\!\!\!\!$ & \cr
\omit&height4pt&\omit&&\omit&&\omit&&\omit&&\omit&\cr
\tablerule
\omit&height2pt&\omit&&\omit&&\omit&&\omit&&\omit&\cr
&& 1 && A && $-(\frac{D-1}{D-3}) A$ && $0$ && $0$ & \cr
\omit&height2pt&\omit&&\omit&&\omit&&\omit&&\omit&\cr
\tablerule
\omit&height2pt&\omit&&\omit&&\omit&&\omit&&\omit&\cr
&& 1 && B && $0$ && $0$ && $0$ & \cr
\omit&height2pt&\omit&&\omit&&\omit&&\omit&&\omit&\cr
\tablerule
\omit&height2pt&\omit&&\omit&&\omit&&\omit&&\omit&\cr
&& 1 && C && $\!\!\!\!-\frac1{(D-2)^2 (D-3)^2}\!\!\!\!$ && $0$ && $0$ & \cr
\omit&height2pt&\omit&&\omit&&\omit&&\omit&&\omit&\cr
\tablerule
\omit&height2pt&\omit&&\omit&&\omit&&\omit&&\omit&\cr
&& 2 && A && $\!\!\!\! [-\frac{D(D-1)}2 \!+\! (\frac{D-1}{D-3})] A \!\!\!\!$ &&
$0$ && $0$ & \cr
\omit&height2pt&\omit&&\omit&&\omit&&\omit&&\omit&\cr
\tablerule
\omit&height2pt&\omit&&\omit&&\omit&&\omit&&\omit&\cr
&& 2 && B && $-\frac12 (\frac{D-1}{D-2})$ && $0$ && $0$ & \cr
\omit&height2pt&\omit&&\omit&&\omit&&\omit&&\omit&\cr
\tablerule
\omit&height2pt&\omit&&\omit&&\omit&&\omit&&\omit&\cr
&& 2 && C && $\frac12 \frac{(D^2 -5D + 8)}{(D-2)^2(D-3)^2}$ && $0$ && $0$ & \cr
\omit&height2pt&\omit&&\omit&&\omit&&\omit&&\omit&\cr
\tablerule
\omit&height2pt&\omit&&\omit&&\omit&&\omit&&\omit&\cr
&& 3 && A && $0$ && $\frac2{D-3} A$ && $0$ & \cr
\omit&height2pt&\omit&&\omit&&\omit&&\omit&&\omit&\cr
\tablerule
\omit&height2pt&\omit&&\omit&&\omit&&\omit&&\omit&\cr
&& 3 && B && $0$ && $0$ && $0$ & \cr
\omit&height2pt&\omit&&\omit&&\omit&&\omit&&\omit&\cr
\tablerule
\omit&height2pt&\omit&&\omit&&\omit&&\omit&&\omit&\cr
&& 3 && C && $\!\!\!\!-\frac1{(D-2)^2(D-3)}\!\!\!\!$ &&
$\!\!\!\!\frac1{(D-2)(D-3)^2}\!\!\!\!$ && $0$ & \cr
\omit&height2pt&\omit&&\omit&&\omit&&\omit&&\omit&\cr
\tablerule
\omit&height2pt&\omit&&\omit&&\omit&&\omit&&\omit&\cr
&& 4 && A && $0$ && $\!\!\!\![\frac{3D}4 \!-\! \frac{3}{2 (D-3)}] A\!\!\!\!$ 
&& $0$ & \cr
\omit&height2pt&\omit&&\omit&&\omit&&\omit&&\omit&\cr
\tablerule
\omit&height2pt&\omit&&\omit&&\omit&&\omit&&\omit&\cr
&& 4 && B && $\frac38 (\frac{D-1}{D-2})$ && $-\frac38$ && $0$ & \cr
\omit&height2pt&\omit&&\omit&&\omit&&\omit&&\omit&\cr
\tablerule
\omit&height2pt&\omit&&\omit&&\omit&&\omit&&\omit&\cr
&& 4 && C && $-\frac3{4(D-2)^2}$ && $\!\!\!\!\frac34 \frac{(D^2-6D+8)}{(D-2)^2
(D-3)^2} \!\!\!\!$ && $0$ & \cr
\omit&height2pt&\omit&&\omit&&\omit&&\omit&&\omit&\cr
\tablerule
\omit&height2pt&\omit&&\omit&&\omit&&\omit&&\omit&\cr
&& 5 && A && $0$ && $0$ && $\!\!\!\!\frac12 (\frac{D-1}{D-3})\!\!\!\!$ & \cr
\omit&height2pt&\omit&&\omit&&\omit&&\omit&&\omit&\cr
\tablerule
\omit&height2pt&\omit&&\omit&&\omit&&\omit&&\omit&\cr
&& 5 && B && $0$ && $0$ && $0$ & \cr
\omit&height2pt&\omit&&\omit&&\omit&&\omit&&\omit&\cr
\tablerule
\omit&height2pt&\omit&&\omit&&\omit&&\omit&&\omit&\cr
&& 5 && C && $0$ && $0$ && $0$ & \cr
\omit&height2pt&\omit&&\omit&&\omit&&\omit&&\omit&\cr
\tablerule
\omit&height2pt&\omit&&\omit&&\omit&&\omit&&\omit&\cr
&& 6 && A && $0$ && $0$ && $0$ & \cr
\omit&height2pt&\omit&&\omit&&\omit&&\omit&&\omit&\cr
\tablerule
\omit&height2pt&\omit&&\omit&&\omit&&\omit&&\omit&\cr
&& 6 && B && $0$ && $0$ && $0$ & \cr
\omit&height2pt&\omit&&\omit&&\omit&&\omit&&\omit&\cr
\tablerule
\omit&height2pt&\omit&&\omit&&\omit&&\omit&&\omit&\cr
&& 6 && C && $0$ && $0$ && $0$ & \cr
\omit&height2pt&\omit&&\omit&&\omit&&\omit&&\omit&\cr
\tablerule
\omit&height2pt&\omit&&\omit&&\omit&&\omit&&\omit&\cr
&& 7 && A && $0$ && $0$ && $\!\!\!\! \frac{D (D-1)}4 \!-\! \frac12 
(\frac{D-1}{D-3})\!\!\!\!$ & \cr
\omit&height2pt&\omit&&\omit&&\omit&&\omit&&\omit&\cr
\tablerule
\omit&height2pt&\omit&&\omit&&\omit&&\omit&&\omit&\cr
&& 7 && B && $0$ && $0$ && $0$ & \cr
\omit&height2pt&\omit&&\omit&&\omit&&\omit&&\omit&\cr
\tablerule
\omit&height2pt&\omit&&\omit&&\omit&&\omit&&\omit&\cr
&& 7 && C && $0$ && $0$ && $0$ & \cr
\omit&height2pt&\omit&&\omit&&\omit&&\omit&&\omit&\cr
\tablerule
\omit&height2pt&\omit&&\omit&&\omit&&\omit&&\omit&\cr
&& 8 && A && $0$ && $0$ && $0$ & \cr
\omit&height2pt&\omit&&\omit&&\omit&&\omit&&\omit&\cr
\tablerule
\omit&height2pt&\omit&&\omit&&\omit&&\omit&&\omit&\cr
&& 8 && B && $0$ && $0$ && $0$ & \cr
\omit&height2pt&\omit&&\omit&&\omit&&\omit&&\omit&\cr
\tablerule
\omit&height2pt&\omit&&\omit&&\omit&&\omit&&\omit&\cr
&& 8 && C && $0$ && $0$ && $0$ & \cr
\omit&height2pt&\omit&&\omit&&\omit&&\omit&&\omit&\cr
\tablerule}}

\caption{Final 4-point contributions. All contributions are
multiplied by $\frac{\kappa^2 H^{D-2}}{(4 \pi)^{\frac{D}2}} \frac{\Gamma(D-1)}{
\Gamma(\frac{D}2)}$. We define $A \equiv \frac{\pi}2 \cot(\frac{\pi D}2) \!-\!
\ln(a)$.}

\label{4fin}

\end{table}

\section{Contributions from the 3-Point Vertices}

In this section we evaluate the contributions from two 3-point vertex 
operators. The generic diagram topology is depicted in Fig.~2. The analytic 
form is,
\begin{equation}
-i\Bigl[\mbox{}_i \Sigma_j^{\rm 3pt}\Bigr](x;x') = \sum_{I=1}^3 iV^{\alpha\beta
}_{Iik}(x) \, i\Bigl[\mbox{}_k S_{\ell}\Bigr](x;x') \sum_{J=1}^3 iV^{\rho\sigma
}_{J\ell j}(x') \, i\Bigl[\mbox{}_{\alpha\beta} \Delta_{\rho\sigma}\Bigr](x;x') 
\; . \label{3ptloop}
\end{equation}
\begin{center} 
\begin{picture}(300,70)(0,0) 
\GlueArc (150,20)(40,0,180){5}{8}
\ArrowLine(190,20)(110,20) 
\ArrowLine(110,20)(50,20)
\Vertex(110,20){3}
\Text(110,10)[b]{$x$}
\ArrowLine(250,20)(190,20) 
\Vertex(190,20){3}
\Text(191,10)[b]{$x'$}
\end{picture}
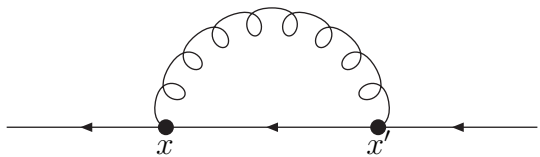
\\ {\rm Fig.~2: Contribution from two 3-point vertices.}
\end{center}

\begin{table}

\vbox{\tabskip=0pt \offinterlineskip
\def\tablerule{\noalign{\hrule}}
\halign to390pt {\strut#& \vrule#\tabskip=1em plus2em& 
\hfil#& \vrule#& \hfil#& \vrule#& \hfil#\hfil& 
\vrule#\tabskip=0pt\cr
\tablerule
\omit&height4pt&\omit&&\omit&&\omit&\cr
&&\omit\hidewidth {\rm I} &&\omit\hidewidth {\rm J} \hidewidth&& 
\omit\hidewidth $iV_I^{\alpha\beta}(x) \, i[S](x;x') \, i V_J^{\rho\sigma}(x') 
\, i[\mbox{}_{\alpha\beta} \Delta_{\rho\sigma}](x;x')$ \hidewidth&\cr
\omit&height4pt&\omit&&\omit&&\omit&\cr
\tablerule
\omit&height2pt&\omit&&\omit&&\omit&\cr
&& 1 && 1 && $\frac14 \kappa^2 \hspace{-.1cm} \not{\hspace{-.1cm} \partial} 
\delta^D(x\!-\!x') \, i[^{\alpha}_{~\alpha}\Delta^{\rho}_{~\rho}](x;x)$ & \cr
\omit&height2pt&\omit&&\omit&&\omit&\cr
\tablerule
\omit&height2pt&\omit&&\omit&&\omit&\cr
&& 1 && 2 && $-\frac14 \kappa^2 \gamma^{\rho} \partial^{\sigma} 
\delta^D(x\!-\!x') \, i[^{\alpha}_{~\alpha}\Delta_{\rho\sigma}](x;x)$ & \cr
\omit&height2pt&\omit&&\omit&&\omit&\cr
\tablerule
\omit&height2pt&\omit&&\omit&&\omit&\cr
&& 1 && 3 && $\frac14 i \kappa^2 \gamma^{\rho} J^{\sigma \mu} 
\delta^D(x\!-\!x') \, \partial^{\prime}_{\mu} i[^{\alpha}_{~\alpha}
\Delta_{\rho\sigma}](x;x')$ & \cr
\omit&height2pt&\omit&&\omit&&\omit&\cr
\tablerule
\omit&height2pt&\omit&&\omit&&\omit&\cr
&& 2 && 1 && $\frac14 \kappa^2 \partial^{\prime}_{\mu} \{ \gamma^{\alpha} 
\partial^{\beta} \, i[S](x;x') \, \gamma^{\mu} \, i[\mbox{}_{\alpha\beta}
\Delta^{\rho}_{~\rho}](x;x') \}$ & \cr
\omit&height2pt&\omit&&\omit&&\omit&\cr
\tablerule
\omit&height2pt&\omit&&\omit&&\omit&\cr
&& 2 && 2 && $-\frac14 \kappa^2 \partial^{\prime \rho} \{ \gamma^{\alpha} 
\partial^{\beta} \, i[S](x;x') \, \gamma^{\sigma} \, i[\mbox{}_{\alpha\beta}
\Delta_{\rho\sigma}](x;x') \}$ & \cr
\omit&height2pt&\omit&&\omit&&\omit&\cr
\tablerule
\omit&height2pt&\omit&&\omit&&\omit&\cr
&& 2 && 3 && $-\frac14 i\kappa^2 \, \gamma^{\alpha} \partial^{\beta} \, 
i[S](x;x') \, \gamma^{\rho} J^{\sigma\mu} \partial^{\prime}_{\mu} \, 
i[\mbox{}_{\alpha\beta} \Delta_{\rho\sigma}](x;x')$ & \cr
\omit&height2pt&\omit&&\omit&&\omit&\cr
\tablerule
\omit&height2pt&\omit&&\omit&&\omit&\cr
&& 3 && 1 && $-\frac14 i \kappa^2 \partial^{\prime}_{\nu} \{ \gamma^{\alpha}
J^{\beta\mu} \, i[S](x;x') \, \gamma^{\nu} \partial_{\mu} \, i[\mbox{}_{\alpha
\beta} \Delta^{\rho}_{~\rho}](x;x') \}$ & \cr
\omit&height2pt&\omit&&\omit&&\omit&\cr
\tablerule
\omit&height2pt&\omit&&\omit&&\omit&\cr
&& 3 && 2 && $\frac14 i \kappa^2 \partial^{\prime \rho} \{ \gamma^{\alpha}
J^{\beta\mu} \, i[S](x;x') \, \gamma^{\sigma} \partial_{\mu} \, 
i[\mbox{}_{\alpha\beta} \Delta_{\rho\sigma}](x;x') \}$ & \cr
\omit&height2pt&\omit&&\omit&&\omit&\cr
\tablerule
\omit&height2pt&\omit&&\omit&&\omit&\cr
&& 3 && 3 && $-\frac14 \kappa^2 \, \gamma^{\alpha} J^{\beta\mu} \, i[S](x;x') 
\, \gamma^{\rho} J^{\sigma\nu} \partial_{\mu} \partial^{\prime}_{\nu} \, 
i[\mbox{}_{\alpha\beta} \Delta_{\rho\sigma}](x;x')$ & \cr
\omit&height2pt&\omit&&\omit&&\omit&\cr
\tablerule}}

\caption{Generic Contributions from the 3-Point Vertices.}

\label{gen3}

\end{table}

Because there are three 3-point vertex operators (\ref{3VO}), there are
nine vertex products in (\ref{3ptloop}). We label each contribution by the
numbers on its vertex pair, for example, 
\begin{equation}
\Bigl[I\!\!-\!\!J\Bigr] \equiv iV_I^{\alpha\beta}(x) \times 
i\Bigl[S\Bigr](x;x') \times i V_J^{\rho\sigma}(x') \times i\Bigl[\mbox{}_{
\alpha\beta} \Delta_{\rho\sigma}\Bigr](x;x') \; .
\end{equation}
Table \ref{gen3}  gives the generic reductions, before decomposing the graviton
propagator. Most of these reductions are straightforward but two subtleties 
deserve mention. First, the Dirac slash of the fermion propagator gives a
delta function,
\begin{equation}
i \hspace{-.1cm} \not{\hspace{-.1cm} \partial} i \Bigl[S\Bigr](x;x') = 
i \delta^D(x-x') \; . \label{fpeqn}
\end{equation}
This occurs whenever the first vertex is $I\!=\!1$, for example,
\begin{eqnarray}
\Bigl[1\!\!-\!\!3\Bigr] & \equiv & \frac{i\kappa}2 \eta^{\alpha\beta} i 
\hspace{-.1cm} \not{\hspace{-.1cm} \partial} \times i \Bigl[S\Bigr](x;x') 
\times -\frac{i\kappa}2 \gamma^{\rho} J^{\sigma\mu} \partial^{\prime}_{\mu} 
\times i\Bigl[\mbox{}_{\alpha\beta} \Delta_{\rho\sigma}\Bigr](x;x') \; , 
\qquad \\
& = & \frac{i \kappa^2}4 \gamma^{\rho} J^{\sigma \mu} \delta^D(x\!-\!x') \, 
\partial^{\prime}_{\mu} i \Bigl[ \mbox{}^{\alpha}_{~\alpha} \Delta_{\rho\sigma}
\Bigr](x;x') \; .
\end{eqnarray}
The second subtlety is that derivatives on external lines must be partially
integrated back on the entire diagram. This happens whenever the second
vertex is $J\!=\!1$ or $J\!=\!2$, for example,
\begin{eqnarray}
\Bigl[2\!\!-\!\!2\Bigr] & \equiv & -\frac{i\kappa}2 \gamma^{\alpha} i 
\partial^{\beta} \times i \Bigl[S\Bigr](x;x') \times -\frac{i\kappa}2 
\gamma^{\rho} i \partial^{\prime \sigma}_{\rm ext} \times i\Bigl[\mbox{}_{
\alpha\beta} \Delta_{\rho\sigma}\Bigr](x;x') \; , \qquad \\
& = & -\frac{\kappa^2}4 \partial^{\prime \sigma} \Biggl\{ \gamma^{\alpha} 
\partial^{\beta} \, i\Bigl[S\Bigr](x;x') \, \gamma^{\rho} \, i \Bigl[ 
\mbox{}_{\alpha\beta} \Delta_{\rho\sigma} \Bigr](x;x') \Biggr\} .
\end{eqnarray}

In comparing Table~\ref{gen3} and Table~\ref{4con} it will be seen that
the 3-point contributions with $I\!=\!1$ are closely related to three
of the 4-point contributions. In fact the $[1\!\!-\!\!1]$ contribution
is $-2$ times the 4-point contribution with $I \!=\! 1$; while $[1\!\!-\!\!2]$
and $[1\!\!-\!\!3]$ cancel the 4-point contributions with $I\!=\!3$ 
and $I\!=\!5$, respectively. Because of this it is convenient to add the 
3-point contributions with $I\!=\!1$ to the 4-point contributions
from Table~\ref{4fin},
\begin{eqnarray}
\lefteqn{-i \Bigl[\Sigma^{\rm 4pt} + \Sigma^{\rm 3pt}_{I=1}\Bigr](x;x') 
= \frac{\kappa^2 H^{D-2}}{(4 \pi)^{\frac{D}2}} \frac{\Gamma(D\!-\!
1)}{\Gamma(\frac{D}2)} 
\Biggl\{ \Bigl[- \frac{(D\!+\!1) (D\!-\!1) (D\!-\!4)}{2 (D\!-\!3)} A }
\nonumber \\
& & - \frac{(D\!-\!1)(D^3 \!-\! 8 D^2 \!+\! 23 D \!-\! 32)}{8 (D\!-\!2)^2 
(D\!-\!3)^2} \Bigr] \hspace{-.1cm} \not{\hspace{-.1cm} \partial} + 
\Bigl[ \frac34 \Bigl(D \!-\! \frac2{D\!-\!3}\Bigr) A \nonumber \\
& & \hspace{.2cm} + \frac{3 (D^2 \!-\! 6 D \!+\! 8)}{4 (D\!-\!2)^2 (D\!-\!3)^2} 
\!-\! \frac38 \Bigr] \; \hspace{-.1cm} \overline{\not{\hspace{-.1cm} \partial}}
+ \Bigl(\frac{D\!-\!1}4\Bigr) \Bigl(D \!-\! \frac2{D\!-\!3}\Bigr) a H \gamma^0 
\Biggr\} \delta^D(x\!-\!x') . \qquad \label{1stcon}
\end{eqnarray}
In what follows we will focus on the 3-point contributions with $I\!=\!2$ 
and $I\!=\!3$.

\subsection{Conformal Contributions}

The key to achieving a tractable reduction of the diagrams of Fig.~2 is
that the first term of each of the scalar propagators $i\Delta_I(x;x')$
is the conformal propagator $i\Delta_{\rm cf}(x;x)$. The sum of the three
index factors also gives a simple tensor, so it is very efficient to
write the graviton propagator in the form,
\begin{eqnarray}
\lefteqn{i\Bigl[{}_{\mu\nu} \Delta_{\rho\sigma}\Bigr](x;x') = \Bigl[2 
\eta_{\mu (\rho} \eta_{\sigma) \nu} - \frac2{D\!-\!2} \eta_{\mu\nu} 
\eta_{\rho\sigma}\Bigr] i\Delta_{\rm cf}(x;x') } \nonumber \\
& & \hspace{6cm} + \sum_{I=A,B,C} \Bigl[\mbox{}_{\mu\nu} T^I_{\rho\sigma}
\Bigr] \, i{\delta \! \Delta}_I(x;x') \; , \qquad
\end{eqnarray}
where $i{\delta \! \Delta}_I(x;x') \equiv i\Delta_I(x;x') -
i\Delta_{\rm cf}(x;x')$.
In this subsection we evaluate the contribution to (\ref{3ptloop}) using
the 3-point vertex operators (\ref{3VO}) and the fermion propagator
(\ref{fprop}) but only the conformal part of the graviton propagator,
\begin{equation}
i\Bigl[{}_{\mu\nu} \Delta_{\rho\sigma}\Bigr](x;x') \longrightarrow \Bigl[2 
\eta_{\mu (\rho} \eta_{\sigma) \nu} - \frac2{D\!-\!2} \eta_{\mu\nu} 
\eta_{\rho\sigma}\Bigr] i\Delta_{\rm cf}(x;x') \equiv \Bigl[\mbox{}_{\alpha
\beta} T^{\rm cf}_{\rho\sigma}\Bigr] i\Delta_{\rm cf}(x;x) \; . \label{cfpart}
\end{equation}

We carry out the reduction in three stages. In the first stage the conformal 
part (\ref{cfpart}) of the graviton propagator is substituted into the generic
results from Table \ref{gen3} and the contractions are performed. We also make 
use of gamma matrix identities such as (\ref{Jred}) and,
\begin{equation}
\gamma^{\mu} i\Bigl[S\Bigr](x;x') \gamma_{\mu} = (D\!-\!2) i\Bigl[S\Bigr](x;x')
\qquad {\rm and} \qquad \gamma_{\alpha} J^{\alpha \mu} = -\frac{i}2 (D\!-\!1)
\gamma^{\mu} \; .
\end{equation}
Finally, we employ relation (\ref{fpeqn}) whenever $\;\hspace{-.1cm} \not{
\hspace{-.1cm} \partial}$ acts upon the fermion propagator. However, we do
not at this stage act any other derivatives. The results of these reductions
are summarized in Table \ref{Dcfcon}. Because the conformal tensor factor
$[{}_{\alpha\beta} T^{\rm cf}_{\rho\sigma}]$ contains three distinct terms,
and because the factors of $\gamma^{\alpha} J^{\beta \mu}$ in 
Table~\ref{gen3} can contribute different terms with a distinct structure,
we have sometimes broken up the result for a given vertex pair into parts.
These parts are distinguished in Table~\ref{Dcfcon} and subsequently by
subscripts taken from the lower case Latin letters.

\begin{table}

\vbox{\tabskip=0pt \offinterlineskip
\def\tablerule{\noalign{\hrule}}
\halign to390pt {\strut#& \vrule#\tabskip=1em plus2em& 
\hfil#\hfil& \vrule#& \hfil#\hfil& \vrule#& \hfil#\hfil& \vrule#& \hfil#\hfil& 
\vrule#\tabskip=0pt\cr
\tablerule
\omit&height4pt&\omit&&\omit&&\omit&&\omit&\cr
&&$\!\!\!\!{\rm I}\!\!\!\!$ && $\!\!\!\!{\rm J}\!\!\!\!$ && $\!\!\!\!{\rm sub}
\!\!\!\!$ && $\!\!\!\!iV_I^{\alpha\beta}(x) \, i[S](x;x') \, i V_J^{\rho
\sigma}(x') \, [\mbox{}_{\alpha\beta} T^{\rm cf}_{\rho\sigma}] \, 
i\Delta_{\rm cf}(x;x') \!\!\!\!$ & \cr
\omit&height4pt&\omit&&\omit&&\omit&&\omit&\cr
\tablerule
\omit&height2pt&\omit&&\omit&&\omit&&\omit&\cr
&& 2 && 1 && \omit && $-\frac1{D-2} \kappa^2 \hspace{-.1cm} \not{\hspace{-.1cm} 
\partial}' \{ \delta^D(x\!-\!x') \, i\Delta_{\rm cf}(x;x) \}$ & \cr
\omit&height2pt&\omit&&\omit&&\omit&&\omit&\cr
\tablerule
\omit&height2pt&\omit&&\omit&&\omit&&\omit&\cr
&& 2 && 2 && a && $-\frac14 (\frac{D-4}{D-2}) \kappa^2 \hspace{-.1cm} 
\not{\hspace{-.1cm} \partial}' \{ \delta^D(x\!-\!x') \, i\Delta_{\rm cf}(x;x) 
\}$ & \cr
\omit&height2pt&\omit&&\omit&&\omit&&\omit&\cr
\tablerule
\omit&height2pt&\omit&&\omit&&\omit&&\omit&\cr
&& 2 && 2 && b && $- (\frac{D-2}4) \kappa^2 \partial_{\mu}^{\prime} \{ 
\partial^{\mu} i[S](x;x') \, i\Delta_{\rm cf}(x;x') \}$ & \cr
\omit&height2pt&\omit&&\omit&&\omit&&\omit&\cr
\tablerule
\omit&height2pt&\omit&&\omit&&\omit&&\omit&\cr
&& 2 && 3 && a && $\frac18 (\frac{D}{D-2}) \kappa^2 \delta^D(x\!-\!x')
\hspace{-.1cm} \not{\hspace{-.1cm} \partial}' \, i\Delta_{\rm cf}(x;x)$ & \cr
\omit&height2pt&\omit&&\omit&&\omit&&\omit&\cr
\tablerule
\omit&height2pt&\omit&&\omit&&\omit&&\omit&\cr
&& 2 && 3 && b && $+ (\frac{D-2}8) \kappa^2 \partial_{\mu} \, i[S](x;x')
\partial^{\prime \mu} \, i\Delta_{\rm cf}(x;x')$ & \cr
\omit&height2pt&\omit&&\omit&&\omit&&\omit&\cr
\tablerule
\omit&height2pt&\omit&&\omit&&\omit&&\omit&\cr
&& 3 && 1 && \omit && $\frac12 (\frac{D-1}{D-2}) \kappa^2 \partial^{\prime}_{
\mu} \{ \, \hspace{-.1cm} \not{\hspace{-.1cm}\partial} \, i\Delta_{\rm cf}(x;x)
\, i[S](x;x') \gamma^{\mu} \}$ & \cr
\omit&height2pt&\omit&&\omit&&\omit&&\omit&\cr
\tablerule
\omit&height2pt&\omit&&\omit&&\omit&&\omit&\cr
&& 3 && 2 && a && $- \frac1{4(D-2)} \kappa^2 \partial^{\prime}_{\mu}
\{ \, \hspace{-.1cm} \not{\hspace{-.1cm} \partial} \, i\Delta_{\rm cf}(x;x) \,
i[S](x;x') \gamma^{\mu} \}$ & \cr
\omit&height2pt&\omit&&\omit&&\omit&&\omit&\cr
\tablerule
\omit&height2pt&\omit&&\omit&&\omit&&\omit&\cr
&& 3 && 2 && b && $-(\frac{D-2}8) \kappa^2 \partial^{\prime}_{\mu} \{ 
i[S](x;x') \, \partial^{\mu} i\Delta_{\rm cf}(x;x) \}$ & \cr
\omit&height2pt&\omit&&\omit&&\omit&&\omit&\cr
\tablerule
\omit&height2pt&\omit&&\omit&&\omit&&\omit&\cr
&& 3 && 2 && c && $- \frac18 \kappa^2 \hspace{-.1cm} \not{\hspace{-.1cm} 
\partial}' \{ i[S](x;x') \, \hspace{-.1cm} \not{\hspace{-.1cm} \partial} \, 
i\Delta_{\rm cf}(x;x) \}$ & \cr
\omit&height2pt&\omit&&\omit&&\omit&&\omit&\cr
\tablerule
\omit&height2pt&\omit&&\omit&&\omit&&\omit&\cr
&& 3 && 3 && a && $(\frac{D-2}{16}) \kappa^2 i[S](x;x') \partial \!\cdot\!
\partial' i\Delta_{\rm cf}(x;x')$ & \cr
\omit&height2pt&\omit&&\omit&&\omit&&\omit&\cr
\tablerule
\omit&height2pt&\omit&&\omit&&\omit&&\omit&\cr
&& 3 && 3 && b && $-\frac18 (\frac{2D-3}{D-2}) \kappa^2 \gamma^{\mu} 
i[S](x;x') \partial_{\mu} \hspace{-.1cm} \not{\hspace{-.1cm} \partial}' \,
i\Delta_{\rm cf}(x;x)$ & \cr
\omit&height2pt&\omit&&\omit&&\omit&&\omit&\cr
\tablerule
\omit&height2pt&\omit&&\omit&&\omit&&\omit&\cr
&& 3 && 3 && c && $+ \frac1{16} \kappa^2 \gamma^{\mu} i[S](x;x') 
\partial^{\prime}_{\mu} \hspace{-.1cm} \not{\hspace{-.1cm} \partial} \, 
i\Delta_{\rm cf}(x;x)$ & \cr
\omit&height2pt&\omit&&\omit&&\omit&&\omit&\cr
\tablerule}}

\caption{Contractions from the $i\Delta_{\rm cf}$ part of the Graviton 
Propagator.}

\label{Dcfcon}

\end{table}

In the second stage we substitute the fermion and conformal propagators,
\begin{eqnarray}
i\Bigl[S\Bigr](x;x') & = & -\frac{i \Gamma(\frac{D}2)}{2 \pi^{\frac{D}2}} 
\frac{\gamma^{\mu} \Delta x_{\mu}}{\Delta x^D} \; , \\
i\Delta_{\rm cf}(x;x') & = & \frac{\Gamma(\frac{D}2 \!-\!1)}{4 \pi^{\frac{D}2}}
\frac{(a a')^{1-\frac{D}2}}{\Delta x^{D-2}} \; .
\end{eqnarray}
At this stage we take advantage of the curious consequence of the automatic 
subtraction of dimension regularization that any dimension-dependent power
of zero is discarded,
\begin{equation}
\lim_{x' \rightarrow x} i\Delta_{\rm cf}(x;x') = 0 \qquad {\rm and} \qquad
\lim_{x' \rightarrow x} \partial^{\prime}_{\mu} i\Delta_{\rm cf}(x;x') = 0 \; .
\end{equation}

In the final stage we act the derivatives. These can act upon the conformal 
coordinate separation $\Delta x^{\mu} \equiv x^{\mu} \!-\! x^{\prime \mu}$, 
or upon the factor of $(a a')^{1-\frac{D}2}$ from the conformal propagator. 
We quote separate results for the cases where all derivatives act upon the 
conformal coordinate separation (Table~\ref{Dcfmost}) and the case where 
one or more of the derivatives acts upon the scale factors 
(Table~\ref{Dcfless}). In the former case the final result must in each
case take the form of a pure number times the universal factor,
\begin{equation}
\frac{(a a')^{1-\frac{D}2} \gamma^{\mu} \Delta x_{\mu}}{\Delta x^{2D}} \; .
\end{equation}

\begin{table}

\vbox{\tabskip=0pt \offinterlineskip
\def\tablerule{\noalign{\hrule}}
\halign to390pt {\strut#& \vrule#\tabskip=1em plus2em& 
\hfil#\hfil& \vrule#& \hfil#\hfil& \vrule#& \hfil#\hfil& \vrule#& \hfil#\hfil& 
\vrule#\tabskip=0pt\cr
\tablerule
\omit&height4pt&\omit&&\omit&&\omit&&\omit&\cr
&&$\!\!\!\!{\rm I}\!\!\!\!$ && $\!\!\!\!{\rm J}\!\!\!\!$ && $\!\!\!\!{\rm sub}
\!\!\!\!$ && $\!\!\!\!{\rm Coefficient\ of} \; 
\frac{\gamma^{\mu} \Delta x_{\mu}}{\Delta x^{2D}} \!\!\!\!$ & \cr
\omit&height4pt&\omit&&\omit&&\omit&&\omit&\cr
\tablerule
\omit&height2pt&\omit&&\omit&&\omit&&\omit&\cr
&& 2 && 1 && \omit && $0$ & \cr
\omit&height2pt&\omit&&\omit&&\omit&&\omit&\cr
\tablerule
\omit&height2pt&\omit&&\omit&&\omit&&\omit&\cr
&& 2 && 2 && a && $0$ & \cr
\omit&height2pt&\omit&&\omit&&\omit&&\omit&\cr
\tablerule
\omit&height2pt&\omit&&\omit&&\omit&&\omit&\cr
&& 2 && 2 && b && $-\frac14 (D\!-\!2)^2 (D\!-\!1)$ & \cr
\omit&height2pt&\omit&&\omit&&\omit&&\omit&\cr
\tablerule
\omit&height2pt&\omit&&\omit&&\omit&&\omit&\cr
&& 2 && 3 && a && $0$ & \cr
\omit&height2pt&\omit&&\omit&&\omit&&\omit&\cr
\tablerule
\omit&height2pt&\omit&&\omit&&\omit&&\omit&\cr
&& 2 && 3 && b && $\frac18 (D\!-\!2)^2 (D\!-\!1)$ & \cr
\omit&height2pt&\omit&&\omit&&\omit&&\omit&\cr
\tablerule
\omit&height2pt&\omit&&\omit&&\omit&&\omit&\cr
&& 3 && 1 && \omit && $-(D\!-\!1)^2$ & \cr
\omit&height2pt&\omit&&\omit&&\omit&&\omit&\cr
\tablerule
\omit&height2pt&\omit&&\omit&&\omit&&\omit&\cr
&& 3 && 2 && a && $\frac12 (D\!-\!1)$ & \cr
\omit&height2pt&\omit&&\omit&&\omit&&\omit&\cr
\tablerule
\omit&height2pt&\omit&&\omit&&\omit&&\omit&\cr
&& 3 && 2 && b && $-\frac18 (D\!-\!2)^2 (D\!-\!1)$ & \cr
\omit&height2pt&\omit&&\omit&&\omit&&\omit&\cr
\tablerule
\omit&height2pt&\omit&&\omit&&\omit&&\omit&\cr
&& 3 && 2 && c && $\frac14 (D\!-\!2) (D\!-\!1)$ & \cr
\omit&height2pt&\omit&&\omit&&\omit&&\omit&\cr
\tablerule
\omit&height2pt&\omit&&\omit&&\omit&&\omit&\cr
&& 3 && 3 && a && $0$ & \cr
\omit&height2pt&\omit&&\omit&&\omit&&\omit&\cr
\tablerule
\omit&height2pt&\omit&&\omit&&\omit&&\omit&\cr
&& 3 && 3 && b && $\frac14 (2D\!-\!3) (D\!-\!1)$ & \cr
\omit&height2pt&\omit&&\omit&&\omit&&\omit&\cr
\tablerule
\omit&height2pt&\omit&&\omit&&\omit&&\omit&\cr
&& 3 && 3 && c && $-\frac18 (D\!-\!2) (D\!-\!1)$ & \cr
\omit&height2pt&\omit&&\omit&&\omit&&\omit&\cr
\tablerule}}

\caption{$i\Delta_{\rm cf}$ terms in which all derivatives act upon 
$\Delta x^2(x;x')$. All contributions are multiplied by 
$\frac{i \kappa^2}{8 \pi^D} \, \Gamma(\frac{D}2) \Gamma(\frac{D}2 \!-\! 1) 
(a a')^{1-\frac{D}2}$.}

\label{Dcfmost}

\end{table}

The sum of all terms in Table~\ref{Dcfmost} is,
\begin{equation}
-i \Bigl[ \Sigma^{T\ref{Dcfmost}}\Bigr](x;x') = \frac{i \kappa^2}{2^6 \pi^D} 
\Gamma\Bigl(\frac{D}2\Bigr) \Gamma\Bigl(\frac{D}2 \!-\! 1\Bigr) (-2 D^2 \!+\! 
5D \!-\! 4) (D\!-\!1) (a a')^{1-\frac{D}2} \frac{\gamma^{\mu} 
\Delta x_{\mu}}{\Delta x^{2D}} \; . \label{most}
\end{equation}
If one simply omits the factor of $(a a')^{1-\frac{D}2}$ the result is 
the same as in flat space. Although (\ref{most}) is well defined for $x^{
\prime \mu} \!\neq\! x^{\mu}$ we must remember that $[\Sigma](x;x')$ will be 
used inside an integral in the quantum-corrected Dirac equation 
(\ref{Diraceq}). For that purpose the singularity at $x^{\prime \mu} \!=\! 
x^{\mu}$ is cubicly divergent in $D\!=\!4$ dimensions. To renormalize this 
divergence we extract derivatives with respect to the coordinate $x^{\mu}$, 
which can of course be taken outside the integral in (\ref{Diraceq}) to 
give a less singular integrand,
\begin{eqnarray}
\lefteqn{\frac{\gamma^{\mu} \Delta x_{\mu}}{\Delta x^{2D}} = \frac{
- \hspace{-.1cm} \not{\hspace{-.1cm}\partial}}{2(D\!-\!1)} \, 
\Biggl\{\frac1{\Delta x^{2D-2}} \Biggr\} \; , } \\
& & = \frac{- \hspace{-.1cm} \not{\hspace{-.1cm}\partial} \, \partial^2}{
4(D\!-\!1) (D\!-\!2)^2} \, \Bigl(\frac1{\Delta x^{2D-4}} \Bigr) \; , \\
& & = \frac{- \hspace{-.1cm} \not{\hspace{-.1cm}\partial} \, \partial^4}{
8(D\!-\!1) (D\!-\!2)^2 (D\!-\!3) (D\!-\!4)} \, \Bigl(\frac1{\Delta x^{2D-6}} 
\Bigr) \; . \label{ds}
\end{eqnarray}

Expression (\ref{ds}) is integrable in four dimensions and we could take 
$D\!=\!4$ except for the explicit factor of $1/(D\!-\!4)$. Of course that
is how ultraviolet divergences manifest in dimensional regularization. We
can segregate the divergence on a local term by employing a simple
representation for a delta function,
\begin{eqnarray}
\lefteqn{\frac{\partial^2}{D\!-\!4} \, \Bigl(\frac1{\Delta x^{2D-6}}\Bigr) 
= \frac{\partial^2}{D\!-\!4} \, \Biggl\{\frac1{\Delta x^{2D-6}} \!-\!
\frac{\mu^{D-4}}{\Delta x^{D-2}} \Biggr\} \!+\! \frac{i 4 \pi^{\frac{D}2}
\mu^{D-4}}{\Gamma(\frac{D}2 \!-\! 1)} \, \frac{\delta^D(x\!-\!x')}{D\!-\!4} 
\; , } \\
& & \hspace{1.8cm} = -\frac{\partial^2}2 \Biggl\{ \frac{\ln(\mu^2 \Delta x^2)}{
\Delta x^2} \!+\! O(D\!-\!4) \Biggr\} + \frac{i 4 \pi^{\frac{D}2} \mu^{D-4}}{
\Gamma(\frac{D}2 \!-\! 1)} \, \frac{\delta^D(x\!-\!x')}{D\!-\!4} \; . \qquad
\end{eqnarray}
The final result for Table~\ref{Dcfmost} is,
\begin{eqnarray}
\lefteqn{-i \Bigl[ \Sigma^{T\ref{Dcfmost}}\Bigr](x;x') = -\frac{i \kappa^2}{2^8
\pi^4} \frac1{a a'} \, \hspace{-.1cm} \not{\hspace{-.1cm}\partial} \,\partial^4
\Biggl\{ \frac{\ln(\mu^2 \Delta x^2)}{\Delta x^2} \Biggl\} + \, O(D\!-\!4) } 
\nonumber \\
& & \hspace{1.7cm} - \frac{\kappa^2 \mu^{D-4}}{2^8 \pi^{\frac{D}2}} 
\Gamma\Bigl(\frac{D}2\!-\!1 \Bigr) \frac{(2 D^2 \!-\! 5D \!+\! 4) 
(a a')^{1-\frac{D}2}}{(D\!-\!2) (D\!-\!3) (D\!-\!4)} \hspace{-.1cm} 
\not{\hspace{-.1cm} \partial} \, \partial^2 \delta^D(x\!-\!x') \; . \qquad
\label{2ndcon}
\end{eqnarray}

\begin{table}

\vbox{\tabskip=0pt \offinterlineskip
\def\tablerule{\noalign{\hrule}}
\halign to390pt {\strut#& \vrule#\tabskip=1em plus2em& 
\hfil#\hfil& \vrule#& \hfil#\hfil& \vrule#& \hfil#\hfil& \vrule#& 
\hfil#\hfil& \vrule#& \hfil#\hfil& \vrule#& \hfil#\hfil& \vrule#\tabskip=0pt\cr
\tablerule
\omit&height4pt&\omit&&\omit&&\omit&&\omit&&\omit&&\omit&\cr
&&$\!\!\!\!{\rm I}\!\!\!\!$ && $\!\!\!\!{\rm J}\!\!\!\!$ && $\!\!\!\!{\rm sub}
\!\!\!\!$ && $\!\!\!\!\frac{a a' H^2\gamma^{\mu} \Delta x_{\mu}}{\Delta 
x^{2D-2}}\!\!\!\!$ && $\!\!\!\!\frac{H \gamma^0}{\Delta x^{2D-2}}\!\!\!\!$
&& $\!\!\!\!\frac{H \Delta \eta \, \gamma^{\mu} \Delta x_{\mu}}{\Delta x^{2D}}
\!\!\!\!$ & \cr
\omit&height4pt&\omit&&\omit&&\omit&&\omit&&\omit&&\omit&\cr
\tablerule
\omit&height2pt&\omit&&\omit&&\omit&&\omit&&\omit&&\omit&\cr
&& 2 && 1 && \omit && $0$ && $0$ && $0$ & \cr
\omit&height2pt&\omit&&\omit&&\omit&&\omit&&\omit&&\omit&\cr
\tablerule
\omit&height2pt&\omit&&\omit&&\omit&&\omit&&\omit&&\omit&\cr
&& 2 && 2 && a && $0$ && $0$ && $0$ & \cr
\omit&height2pt&\omit&&\omit&&\omit&&\omit&&\omit&&\omit&\cr
\tablerule
\omit&height2pt&\omit&&\omit&&\omit&&\omit&&\omit&&\omit&\cr
&& 2 && 2 && b && $0$ && $-\frac12 (D\!-\!2) a'$ && $\frac12 (D\!-\!2) D 
a'$ & \cr
\omit&height2pt&\omit&&\omit&&\omit&&\omit&&\omit&&\omit&\cr
\tablerule
\omit&height2pt&\omit&&\omit&&\omit&&\omit&&\omit&&\omit&\cr
&& 2 && 3 && a && $0$ && $0$ && $0$ & \cr
\omit&height2pt&\omit&&\omit&&\omit&&\omit&&\omit&&\omit&\cr
\tablerule
\omit&height2pt&\omit&&\omit&&\omit&&\omit&&\omit&&\omit&\cr
&& 2 && 3 && b && $0$ && $\frac14 (D\!-\!2) a'$ && $-\frac14 (D\!-\!2) D
a'$ & \cr
\omit&height2pt&\omit&&\omit&&\omit&&\omit&&\omit&&\omit&\cr
\tablerule
\omit&height2pt&\omit&&\omit&&\omit&&\omit&&\omit&&\omit&\cr
&& 3 && 1 && \omit && $\frac12 (D\!-\!1)$ && $0$ && $0$ 
& \cr
\omit&height2pt&\omit&&\omit&&\omit&&\omit&&\omit&&\omit&\cr
\tablerule
\omit&height2pt&\omit&&\omit&&\omit&&\omit&&\omit&&\omit&\cr
&& 3 && 2 && a && $-\frac14$ && $0$ && $0$ & \cr
\omit&height2pt&\omit&&\omit&&\omit&&\omit&&\omit&&\omit&\cr
\tablerule
\omit&height2pt&\omit&&\omit&&\omit&&\omit&&\omit&&\omit&\cr
&& 3 && 2 && b && $-\frac18 (D\!-\!2)^2$ && $\frac14 (D\!-\!2) a$ &&
$\frac14 (D\!-\!2)^2 a'$ & \cr
\omit&height4pt&\omit&&\omit&&\omit&&\omit&&\omit&&\omit&\cr
&& \omit && \omit && \omit && \omit && \omit && $\!\!\!\!-\frac12 
(D\!-\!2) (D\!-\!1) a\!\!\!\!$ & \cr
\omit&height2pt&\omit&&\omit&&\omit&&\omit&&\omit&&\omit&\cr
\tablerule
\omit&height2pt&\omit&&\omit&&\omit&&\omit&&\omit&&\omit&\cr
&& 3 && 2 && c && $-\frac18 (D\!-\!2)$ && $0$ && $0$ & \cr
\omit&height2pt&\omit&&\omit&&\omit&&\omit&&\omit&&\omit&\cr
\tablerule
\omit&height2pt&\omit&&\omit&&\omit&&\omit&&\omit&&\omit&\cr
&& 3 && 3 && a && $\frac1{16} (D\!-\!2)^2$ && $0$ && $\frac18 (D\!-\!2)^2 
(a \!-\! a')$ & \cr
\omit&height2pt&\omit&&\omit&&\omit&&\omit&&\omit&&\omit&\cr
\tablerule
\omit&height2pt&\omit&&\omit&&\omit&&\omit&&\omit&&\omit&\cr
&& 3 && 3 && b && $-\frac18 (2D\!-\!3)$ && $0$ && $0$ & \cr
\omit&height2pt&\omit&&\omit&&\omit&&\omit&&\omit&&\omit&\cr
\tablerule
\omit&height2pt&\omit&&\omit&&\omit&&\omit&&\omit&&\omit&\cr
&& 3 && 3 && c && $\frac1{16} (D\!-\!2)$ && $0$ && $0$ & \cr
\omit&height2pt&\omit&&\omit&&\omit&&\omit&&\omit&&\omit&\cr
\tablerule}}

\caption{$i\Delta_{\rm cf}$ terms in which some derivatives act upon 
scale factors. All contributions are multiplied by 
$\frac{i \kappa^2}{16 \pi^D} \, \Gamma^2(\frac{D}2) (a a')^{1-\frac{D}2}$.}

\label{Dcfless}

\end{table}

When one or more derivative acts upon the scale factors a bewildering
variety of spacetime and gamma matrix structures result. For example,
the $[3\!\!-\!\!2]_b$ term gives,
\begin{eqnarray}
\lefteqn{-\Bigl(\frac{D\!-\!2}8\Bigr) \kappa^2 \partial_{\mu}' \Biggl\{ 
i\Bigl[S\Bigr](x;x') \partial^{\mu} i\Delta_{\rm cf}(x;x') \Biggr\} } 
\nonumber \\
& & \hspace{-.3cm} = \frac{i \kappa^2}{32 \pi^D} \Gamma^2\Bigl(\frac{D}2\Bigr) 
\partial_{\mu}' \Biggl\{ \frac{\gamma^{\nu} \Delta x_{\nu}}{\Delta x^D} 
(a a')^{1-\frac{D}2} \Biggl[ -\frac{(D\!-\!2) \Delta x^{\mu}}{\Delta x^D} 
+ \frac{(D\!-\!2) H a \delta^{\mu}_0}{2 \Delta x^{D-2}} \Biggr] \Biggr\} , 
\qquad \\
& & \hspace{-.3cm} = \frac{i \kappa^2}{32 \pi^D} \Gamma^2\Bigl(\frac{D}2\Bigr) 
(a a')^{1-\frac{D}2} \Biggl\{ -\frac{(D\!-\!1)(D\!-\!2) \gamma^{\mu} \Delta
x_{\mu}}{\Delta x^{2D}} + \frac{(D\!-\!2) H a \gamma^0}{2 \Delta x^{2D-2}}
\qquad \nonumber \\
& & \hspace{2cm} + \frac{(D\!-\!2)^2 a' H \Delta \eta \gamma^{\mu} 
\Delta x_{\mu}}{2 \Delta x^{2D}} - \frac{(D\!-\!1) (D\!-\!2) a H \Delta 
\eta \gamma^{\mu} \Delta x_{\mu}}{\Delta x^{2D}} \nonumber \\
& & \hspace{7.5cm} - \frac{(D\!-\!2)^2 a a' H^2 \gamma^{\mu} \Delta x_{\mu}}{4 
\Delta x^{2D-2}} \Biggr\} . \qquad \label{32b}
\end{eqnarray}
The first term of (\ref{32b}) originates from both derivatives acting on 
the conformal coordinate separation. It belongs in Table~\ref{Dcfmost}. 
The next three terms come from a single derivative acting on a scale factor, 
and the final term in (\ref{32b}) derives from both derivatives acting upon
scale factors. These last four terms belong in Table~\ref{Dcfless}. They can
be expressed as dimensionless functions of $D$, $a$ and $a'$ times three
basic terms,
\begin{eqnarray}
\lefteqn{\frac{i \kappa^2}{16 \pi^D} \Gamma^2\Bigl(\frac{D}2\Bigr) (a a')^{1
-\frac{D}2} \Biggl\{ -\frac18 (D\!-\!2)^2 \times \frac{a a' H^2 \gamma^{\mu}
\Delta x_{\mu}}{\Delta x^{2D-2}} + \frac14 (D\!-\!2) a \times \frac{H 
\gamma^0}{\Delta x^{2D-2}} } \nonumber \\
& & \hspace{2.8cm} + \Bigl[ \frac14 (D\!-\!2)^2 a' \!-\! \frac12 (D\!-\!1) 
(D\!-\!2) a \Bigr] \times \frac{H \Delta \eta \gamma^{\mu} \Delta x_{\mu}}{
\Delta x^{2D}} \Biggr\} . \qquad
\end{eqnarray}

These three terms turn out to be all we need, although intermediate 
expressions sometimes show other kinds. An example is the $[3\!\!-\!\!1]$ 
term,
\begin{eqnarray}
\lefteqn{\frac12 \Bigl(\frac{D\!-\!1}{D\!-\!2}\Bigr) \kappa^2 \partial_{\mu}' 
\Biggl\{\; \hspace{-.1cm} \not{\hspace{-.1cm}\partial} \, i\Delta_{\rm cf}(x;x)
\, i\Bigl[S\Bigr](x;x') \gamma^{\mu} \Biggr\} } \nonumber \\
& & \hspace{-.3cm} = \frac{i \kappa^2}{8 \pi^D} \Gamma^2\Bigl(\frac{D}2\Bigr) 
\Bigl(\frac{D\!-\!1}{D\!-\!2}\Bigr) \partial_{\mu}' \Biggl\{ (a a')^{1-
\frac{D}2} \Biggl[ \frac{\gamma^{\alpha} \Delta x_{\alpha}}{\Delta x^D}
\!+\! \frac{a H \gamma^0}{2 \Delta x^{D-2}} \Biggr] 
\frac{\gamma^{\beta} \Delta x_{\beta}}{\Delta x^D} \gamma^{\mu} \Biggr\}
, \qquad \\
& & \hspace{-.3cm} = \frac{i \kappa^2}{8 \pi^D} \Gamma^2\Bigl(\frac{D}2\Bigr) 
(a a')^{1-\frac{D}2} \Biggl\{ -2 \frac{(D\!-\!1)^2}{(D\!-\!2)} 
\frac{\gamma^{\mu} \Delta x_{\mu}}{\Delta x^{2D}} \!-\! \frac12 (D\!-\!1)
\frac{a H \gamma^0}{\Delta x^{2D-2}} \nonumber \\
& & \hspace{3cm} + \frac12 (D\!-\!1) \frac{a' H \gamma^0}{\Delta x^{2D-2}}
-\frac14 (D\!-\!1) \frac{a a' H^2 \gamma^0 \gamma^{\mu} \Delta x_{\mu}
\gamma^0}{\Delta x^{2D-2}} \Biggr\} . \qquad \label{31}
\end{eqnarray}
As before, the first term in (\ref{31}) belongs in Table~{\ref{Dcfmost}. The
second and third terms are of a type we encountered in (\ref{32b}) but the
final term is not. However, it is simple to bring this term to standard form 
by anti-commuting the $\gamma^{\mu}$ through either $\gamma^0$,
\begin{eqnarray}
a a' H^2 \gamma^0 \gamma^{\mu} \Delta x_{\mu} \gamma^0 & = &
-a a' H^2 \gamma^{\mu} \Delta x_{\mu} - 2 a a' H^2 \Delta \eta \gamma^0 \; ,\\
& = & - a a' H^2 \gamma^{\mu} \Delta x_{\mu} - 2 (a \!-\! a') H \gamma^0 \; .
\end{eqnarray}
Note our use of the identity $(a \!-\! a') = a a' H \Delta \eta$.

When all terms in Table~\ref{Dcfless} are summed it emerges that a
factor of $ H^2 a a'$ can be extracted,
\begin{eqnarray}
\lefteqn{ -i \Bigl[\Sigma^{T\ref{Dcfless}} \Bigr](x;x') = \frac{i \kappa^2}{
16 \pi^D} \Gamma^2\Bigl(\frac{D}2\Bigr) (a a')^{1-\frac{D}2} \Biggl\{
-\frac1{16} (D^2 \!-\! 7D \!+\! 8) \!\times\! \frac{a a' H^2 \gamma^{\mu} 
\Delta x_{\mu}}{\Delta x^{2D-2}} } \nonumber \\
& & \hspace{-.4cm} + \frac14 (D\!-\!2) (a \!-\! a') \!\times\! \frac{H 
\gamma^0}{\Delta x^{2D-2}} \!-\! \frac18 (D\!-\!2) (3D\!-\!2) (a \!-\! a') 
\!\times\! \frac{H \!\Delta \eta \gamma^{\mu} \Delta x_{\mu}}{\Delta x^{2D}}\!
\Biggr\} , \qquad \\
& & \hspace{-.7cm} = \frac{i \kappa^2 H^2}{16 \pi^D} \Gamma^2\Bigl(\frac{D}2
\Bigr) (a a')^{2-\frac{D}2} \Biggl\{-\frac1{16} (D^2 \!-\! 7D \!+\! 8) \times 
\frac{\gamma^{\mu} \Delta x_{\mu}}{\Delta x^{2D-2}} \nonumber \\
& & \hspace{1.5cm} + \frac14 (D\!-\!2) \times \frac{\gamma^0 \Delta \eta}{
\Delta x^{2D-2}} \!-\! \frac18 (D\!-\!2) (3D\!-\!2) \times \frac{\Delta \eta^2 
\gamma^{\mu} \Delta x_{\mu}}{\Delta x^{2D}} \Biggr\} . \label{T8fin}
\end{eqnarray}
Note the fact that this expression is odd under interchange of $x^{\mu}$ and
$x^{\prime \mu}$. Although individual contributions to the last two columns 
of Table~\ref{Dcfless} are not odd under interchange, their sum always 
produces a factor of $a \!-\! a' \!=\! a a' H \Delta \eta$ which makes 
(\ref{T8fin}) odd.

Expression (\ref{T8fin}) can be simplified using the differential identities,
\begin{eqnarray}
\frac{\Delta \eta^2 \gamma^{\mu} \Delta x_{\mu}}{\Delta x^{2D}} & = & 
\frac{\partial_0^2}{4 (D\!-\!2) (D\!-\!1)} \Bigl( \frac{\gamma^{\mu}
\Delta x_{\mu}}{\Delta x^{2D-4}}\Bigr) \nonumber \\
& & \hspace{2cm} - \frac1{2(D\!-\!1)} \, \frac{\gamma^{\mu} \Delta x_{\mu}}{
\Delta x^{2D-2}} + \frac1{D\!-\!1} \, \frac{\gamma^0 \Delta \eta}{\Delta 
x^{2D-2}} \; , \qquad \\
\frac{\gamma^0\Delta \eta}{\Delta x^{2D-2}} & = & \frac{\gamma^0 \partial_0}{2 
(D\!-\!2)} \, \Bigl(\frac1{\Delta x^{2D-4}}\Bigr) \; .
\end{eqnarray}
The result is,
\begin{eqnarray}
\lefteqn{ -i \Bigl[\Sigma^{T\ref{Dcfless}} \Bigr](x;x') = \frac{i \kappa^2 H^2
}{16 \pi^D} \Gamma^2\Bigl(\frac{D}2\Bigr) (a a')^{2-\frac{D}2} \Biggl\{-
\frac{(D^3 \!-\! 11 D^2 \!+\! 23D \!-\! 12)}{16 (D\!-\!1)} \, 
\frac{\gamma^{\mu} \Delta x_{\mu}}{\Delta x^{2D-2}} } \nonumber \\
& & \hspace{1.8cm} - \frac{D}{16 (D\!-\!1)} \, \gamma^0 \partial_0 
\Bigl(\frac1{\Delta x^{2D-4}}\Bigr) - \frac1{32} \Bigl(\frac{3D \!-\! 2}{
D \!-\!1}\Bigr) \partial_0^2 \Bigl(\frac{\gamma^{\mu} \Delta x_{\mu}}{\Delta 
x^{2D-4}}\Bigr) \Biggr\} . \qquad \label{T8b}
\end{eqnarray}
We now exploit partial integration identities of the same type as those
previously used for Table~\ref{Dcfmost},
\begin{eqnarray}
\lefteqn{\frac{\gamma^{\mu} \Delta x_{\mu}}{\Delta x^{2D-4}} = \frac{-
\hspace{-.1cm} \not{\hspace{-.1cm}\partial}}{2 (D\!-\!3)} \, \Bigl(\frac1{
\Delta x^{2D-6}}\Bigr) = -\frac{\hspace{-.1cm} \not{\hspace{-.1cm}\partial}}2 
\Bigl( \frac1{\Delta x^2} \Bigr) + O(D\!-\!4) \; , } \\
\lefteqn{\frac{\gamma^{\mu} \Delta x_{\mu}}{\Delta x^{2D-2}} = \frac{-
\hspace{-.1cm} \not{\hspace{-.1cm}\partial} \, \partial^2}{4 (D\!-\!2)
(D\!-\!3) (D\!-\!4)} \, \Bigl(\frac1{\Delta x^{2D-6}}\Bigr) \; , } \nonumber \\
& & = \frac{\hspace{-.1cm} \not{\hspace{-.1cm}\partial} \, \partial^2}{16}
\Biggl\{ \frac{\ln(\mu^2 \Delta x^2)}{\Delta x^2} \Biggr\} + O(D\!-\!4) 
- \frac{i \pi^{\frac{D}2} \mu^{D-4}}{2 \Gamma(\frac{D}2)} \,
\frac{\hspace{-.1cm} \not{\hspace{-.1cm}\partial} \, \delta^D(x\!-\!x')}{
(D\!-\!3) (D\!-\!4)} \; , \\
\lefteqn{\frac1{\Delta x^{2D-4}} = \frac{\partial^2}{2 (D\!-\!3) (D\!-\!4)} 
\Bigl(\frac1{\Delta x^{2D-6}}\Bigr) \; ,} \nonumber \\
& & = -\frac{\partial^2}4 \Biggl\{\frac{\ln(\mu^2 \Delta x^2)}{\Delta x^2}
\Biggr\} \!+\! O(D\!-\!4) \!+\! \frac{i 2 \pi^{\frac{D}2} \mu^{D-4}}{
\Gamma(\frac{D}2 \!-\!1)} \, \frac{\delta^D(x\!-\!x')}{(D\!-\!3)(D\!-\!4)} 
\; . \qquad
\end{eqnarray}
It is also useful to convert temporal derivatives to spatial ones using,
\begin{equation}
\gamma^0 \partial_0 = \; \hspace{-.1cm} \not{\hspace{-.1cm} \partial} - 
\hspace{-.1cm} \overline{\not{\hspace{-.1cm} \partial} } \qquad {\rm and}
\qquad \partial_0^2 = \nabla^2 - \partial^2 \; .
\end{equation}
Substituting these relations in (\ref{T8b}) gives,
\begin{eqnarray}
\lefteqn{ -i \Bigl[\Sigma^{T\ref{Dcfless}}\Bigr](x;x') = 
\frac{\kappa^2 H^2 \mu^{D-4} \Gamma(\frac{D}2)\, (a a')^{2-\frac{D}2}
}{2^9 \pi^{\frac{D}2} (D\!-\!1) (D\!-\!3) (D\!-\!4)} \Biggl\{-\Bigl(D^3 \!-\! 
13 D^2 \!+\! 27 D \!-\! 12 \Bigr) \hspace{-.1cm} \not{\hspace{-.1cm} \partial}}
\nonumber \\
& & \hspace{1cm} -2 D (D\!-\!2) \, \hspace{-.1cm} \overline{\not{\hspace{-.1cm}
\partial} } \Biggr\} \delta^D(x\!-\!x') + \frac{i \kappa^2 H^2}{2^9 \!\cdot\! 
3 \!\cdot\! \pi^4} \Biggl\{ \Bigl[6 \, \hspace{-.1cm} \not{\hspace{-.1cm} 
\partial} \, \partial^2 \!-\! 2 \; \hspace{-.1cm} \overline{\not{\hspace{-.1cm}
\partial} } \, \partial^2 \Bigr] \Bigl( \frac{\ln(\mu^2 \Delta x^2)}{\Delta 
x^2} \Bigr) \nonumber \\
& & \hspace{5cm} + 5 \! \not{\hspace{-.1cm}\partial} \, (\nabla^2 \!-\! 
\partial^2) \Bigl(\frac1{\Delta x^2} \Bigr) \Biggr\} \!+\! O(D\!-\!4) \; . 
\qquad \label{3rdcon}
\end{eqnarray}

\subsection{Sub-Leading Contributions from $i{\delta \! \Delta}_A$}

\begin{table}

\vbox{\tabskip=0pt \offinterlineskip
\def\tablerule{\noalign{\hrule}}
\halign to390pt {\strut#& \vrule#\tabskip=1em plus2em&
\hfil#& \vrule#& \hfil#& \vrule#& \hfil#& \vrule#& \hfil#\hfil&
\vrule#\tabskip=0pt\cr
\tablerule
\omit&height4pt&\omit&&\omit&&\omit&&\omit&\cr
&&\hidewidth {\rm I} &&\hidewidth {\rm J} \hidewidth&& 
\hidewidth {\rm sub} \hidewidth&& 
\hidewidth $iV_I^{\alpha\beta}(x) \, i[S](x;x') \, i V_J^{\rho\sigma}(x') 
\, [\mbox{}_{\alpha\beta} T^A_{\rho\sigma}] \, i\delta\!\Delta_A(x;x')$ 
\hidewidth&\cr
\omit&height4pt&\omit&&\omit&&\omit&&\omit&\cr
\tablerule
\omit&height2pt&\omit&&\omit&&\omit&&\omit&\cr
&& 2 && 1 && \omit && $-\f{1}{(D-3)}\ka^2\del^{'}_{\mu}\{
\hspace{-.1cm}\not{\hspace{-.1cm}\bar{\del}}
i[S](x;x')\g^{\mu}i\d\!\D_{A}(x;x')\}$ & \cr
\omit&height2pt&\omit&&\omit&&\omit&&\omit&\cr
\tablerule
\omit&height2pt&\omit&&\omit&&\omit&&\omit&\cr
&& 2 && 2 && a && $ \f{1}{4}\ka^2\hspace{-.1cm}
\not{\hspace{-.1cm}\bar{\del}}\{\del_{k}
i[S](x;x')\g_{k}i\d\!\D_{A}(x;x')\}$ & \cr
\omit&height2pt&\omit&&\omit&&\omit&&\omit&\cr
\tablerule
\omit&height2pt&\omit&&\omit&&\omit&&\omit&\cr
&& 2 && 2 && b && $ + \f{1}{4}\ka^2\del_{\ell}\{\g_{k}\del_{\ell}
i[S](x;x')\g_{k}i\d\!\D_{A}(x;x') \}$ & \cr
\omit&height2pt&\omit&&\omit&&\omit&&\omit&\cr
\tablerule
\omit&height2pt&\omit&&\omit&&\omit&&\omit&\cr
&& 2 && 2 && c && $- \f{1}{2 (D-3)}\ka^2\del_{k}\{
\hspace{-.1cm}\not{\hspace{-.1cm}\bar{\del}}
i[S](x;x')\g_{k}i\d\!\D_{A}(x;x')\}$ & \cr
\omit&height2pt&\omit&&\omit&&\omit&&\omit&\cr
\tablerule
\omit&height2pt&\omit&&\omit&&\omit&&\omit&\cr
&& 2 && 3 && a && $\f1{2 (D-3)}\ka^2
\hspace{-.1cm}\not{\hspace{-.1cm}\bar{\del}}
i[S](x;x') \, {\hspace{-.1cm}\not{\hspace{-.1cm}\del}}'
i\d\!\D_{A}(x;x') $ & \cr
\omit&height2pt&\omit&&\omit&&\omit&&\omit&\cr
\tablerule
\omit&height2pt&\omit&&\omit&&\omit&&\omit&\cr
&& 2 && 3 && b && $-\f{1}{4}\ka^2\g_{k}\del_{\ell}i[S](x;x')
\g_{(k}\del_{\ell)}i\d\!\D_{A}(x;x') $ & \cr
\omit&height2pt&\omit&&\omit&&\omit&&\omit&\cr
\tablerule
\omit&height2pt&\omit&&\omit&&\omit&&\omit&\cr
&& 2 && 3 && c && $ + \f{1}{4 (D-3)}\ka^2
\hspace{-.1cm}\not{\hspace{-.1cm}\bar{\del}}i[S](x;x')
\hspace{-.1cm}\not{\hspace{-.1cm}\bar{\del}}i\d\!\D_{A}(x;x')$&\cr
\omit&height2pt&\omit&&\omit&&\omit&&\omit&\cr
\tablerule
\omit&height2pt&\omit&&\omit&&\omit&&\omit&\cr
&& 3 && 1 && a && $\f{1}{2} (\f{D-1}{D-3}) \ka^2\del^{'}_{\mu}
\{\hspace{-.1cm}\not{\hspace{-.1cm}\del} 
i\d\!\D_{A}(x;x')i[S](x;x')\g^{\mu}\} $ &\cr
\omit&height2pt&\omit&&\omit&&\omit&&\omit&\cr
\tablerule
\omit&height2pt&\omit&&\omit&&\omit&&\omit&\cr
&& 3 && 1 && b && $-\f{1}{2 (D-3)}\ka^2\del^{'}_{\mu}
\{\hspace{-.1cm}\not{\hspace{-.1cm}\bar{\del}}
i\d\!\D_{A}(x;x')i[S](x;x')\g^{\mu}\} $ & \cr
\omit&height2pt&\omit&&\omit&&\omit&&\omit&\cr
\tablerule
\omit&height2pt&\omit&&\omit&&\omit&&\omit&\cr
&& 3 && 2 && a && $\f{1}{2 (D-3)}\ka^2\del_{k}
\{\hspace{-.1cm}\not{\hspace{-.1cm}\del}
i\d\!\D_{A}(x;x')i[S](x;x')\g_{k}\}$ & \cr
\omit&height2pt&\omit&&\omit&&\omit&&\omit&\cr
\tablerule
\omit&height2pt&\omit&&\omit&&\omit&&\omit&\cr
&& 3 && 2 && b && $-\f{1}{4 (D-3)}\ka^2\del_{k}
\{\hspace{-.1cm}\not{\hspace{-.1cm}\bar{\del}}
i\d\!\D_{A}(x;x')i[S](x;x')\g_{k}\} $ & \cr
\omit&height2pt&\omit&&\omit&&\omit&&\omit&\cr
\tablerule
\omit&height2pt&\omit&&\omit&&\omit&&\omit&\cr
&& 3 && 2 && c && $ + \f{1}{8}\ka^2\hspace{-.1cm}
\not{\hspace{-.1cm}\bar{\del}}\{i[S](x;x')
\hspace{-.1cm}\not{\hspace{-.1cm}\bar{\del}}
i\d\!\D_{A}(x;x')\} $ & \cr
\omit&height2pt&\omit&&\omit&&\omit&&\omit&\cr
\tablerule
\omit&height2pt&\omit&&\omit&&\omit&&\omit&\cr
&& 3 && 2 && d && $ + \f{1}{8}\ka^2\del_{k}\{\g_{\ell}
i[S](x;x')\g_{\ell}\del_{k}i\d\!\D_{A}(x;x')\} $ & \cr
\omit&height2pt&\omit&&\omit&&\omit&&\omit&\cr
\tablerule
\omit&height2pt&\omit&&\omit&&\omit&&\omit&\cr
&& 3 && 3 && a && $-\f{1}{4} (\f{D-1}{D-3}) \ka^2\g^{\mu}
i[S](x;x')\del_{\mu} \, {\hspace{-.1cm}
\not{\hspace{-.1cm}\del}}' i\d\!\D_{A}(x;x') $ &\cr
\omit&height2pt&\omit&&\omit&&\omit&&\omit&\cr
\tablerule
\omit&height2pt&\omit&&\omit&&\omit&&\omit&\cr
&& 3 && 3 && b && $-\f{1}{4 (D-3)}\ka^2\g^{\mu}
i[S](x;x')\del_{\mu}\hspace{-.1cm}
\not{\hspace{-.1cm}\bar{\del}}i\d\!\D_{A}(x;x') $ &\cr
\omit&height2pt&\omit&&\omit&&\omit&&\omit&\cr
\tablerule
\omit&height2pt&\omit&&\omit&&\omit&&\omit&\cr
&& 3 && 3 && c && $ +\f{1}{4 (D-3)}\ka^2\g_{k}
i[S](x;x')\del_{k}\hspace{-.1cm}
\not{\hspace{-.1cm}\del}' i\d\!\D_{A}(x;x') $ &\cr
\omit&height2pt&\omit&&\omit&&\omit&&\omit&\cr
\tablerule
\omit&height2pt&\omit&&\omit&&\omit&&\omit&\cr
&& 3 && 3 && d && $-\f{1}{16} (\f{D-5}{D-3}) \ka^2\g_{k}
i[S](x;x')\del_{k}\hspace{-.1cm}\not{\hspace{-.1cm}\bar{\del}}
i\d\!\D_{A}(x;x') $ & \cr
\omit&height2pt&\omit&&\omit&&\omit&&\omit&\cr
\tablerule
\omit&height2pt&\omit&&\omit&&\omit&&\omit&\cr
&& 3 && 3 && e && $-\f{1}{16}\ka^2\g_{k}i[S](x;x')\g_{k}
\nabla^2 i\d\!\D_{A}(x;x') $ & \cr
\omit&height2pt&\omit&&\omit&&\omit&&\omit&\cr
\tablerule}}
\caption{Contractions from the $i\d\!\D_{A}$ part of the graviton
propagator}

\label{DAcon}

\end{table}

In this subsection we work out the contribution from substituting the 
residual $A$-type part of the graviton propagator in Table~\ref{gen3},
\begin{equation}
i\Bigl[{}_{\alpha\beta} \Delta_{\rho\sigma}\Bigr](x;x') \longrightarrow \Bigl[
\overline{\eta}_{\alpha \rho} \overline{\eta}_{\sigma \beta} \!+\!
\overline{\eta}_{\alpha \sigma} \overline{\eta}_{\rho \beta} \!-\!
\frac2{D\!-\!3} \overline{\eta}_{\alpha\beta} \overline{\eta}_{\rho\sigma}
\Bigr] i\delta\!\Delta_A(x;x') \; . \label{DApart}
\end{equation}
As with the conformal contributions of the previous section we first 
make the requisite contractions and then act the derivatives. The result
of this first step is summarized in Table~\ref{DAcon}. We have sometimes
broken the result for a single vertex pair into as many as five terms 
because the three different tensors in (\ref{DApart}) can make distinct 
contributions, and because distinct contributions also come from breaking 
up factors of $\gamma^{\alpha} J^{\beta \mu}$. These distinct contributions 
are labeled by subscripts $a$, $b$, $c$, etc. We have tried to arrange them 
so that terms closer to the beginning of the alphabet have fewer purely 
spatial derivatives.

The next step is to act the derivatives and it is of course necessary 
to have an expression for $i\delta\!\Delta_A(x;x')$ at this stage. From
(\ref{DeltaA}) one can infer,
\begin{eqnarray}
\lefteqn{i\delta\!\Delta_A(x;x') = } \nonumber \\
& & \hspace{-.5cm} \frac{H^2}{16 \pi^{\frac{D}2}} \frac{\Gamma(\frac{D}2 \!+\! 
1)}{\frac{D}2 \!-\! 2} \frac{(a a')^{2- \frac{D}2}}{\Delta x^{D-4}} 
+ \frac{H^{D-2}}{(4\pi)^\frac{D}2} \frac{\Gamma(D\!-\!1)}{\Gamma( \frac{D}2
)} \Biggl\{- \pi\cot\Bigl(\frac{\pi}2 D\Bigr) + \ln(aa') \Biggr\} \nonumber \\
& & \hspace{-.5cm} + \frac{H^{D-2}}{(4\pi)^{\frac{D}2}} \! \sum_{n=1}^{\infty}
\! \left\{\!\frac1{n} \frac{\Gamma(n \!+\!D\!-\! 1)}{\Gamma(n \!+\! \frac{D}2)}
\Bigl(\frac{y}4 \Bigr)^n \!\!\!\! - \frac1{n \!-\! \frac{D}2 \!+\! 2}
\frac{\Gamma(n \!+\!  \frac{D}2 \!+\! 1)}{\Gamma(n \!+\! 2)} \Bigl(\frac{y}4
\Bigr)^{n - \frac{D}2 +2} \!\right\} \! . \quad \label{dA}
\end{eqnarray}
In $D\!=\!4$ the most singular contributions to (\ref{3ptloop}) have the
form, $i\delta\!\Delta_A/{\Delta x}^5$. Because the infinite series 
terms in (\ref{dA}) go like positive powers of $\Delta x^2$ these terms make 
integrable contributions to the quantum-corrected Dirac equation 
(\ref{Diraceq}). We can therefore take $D\!=\!4$ for those terms, at which 
point all the infinite series terms drop. Hence it is only necessary to keep 
the first line of (\ref{dA}) and that is all we shall ever use. 

\begin{table}
\vbox{\tabskip=0pt \offinterlineskip
\def\tablerule{\noalign{\hrule}}
\halign to390pt {\strut#& \vrule#\tabskip=1em plus2em&
\hfil#\hfil& \vrule#& \hfil#\hfil& \vrule#& \hfil#\hfil&
\vrule#\tabskip=0pt\cr
\tablerule
\omit&height4pt&\omit&&\omit&&\omit&\cr
\omit&height2pt&\omit&&\omit&&\omit&\cr
&&\omit\hidewidth {\rm Function} \hidewidth &&\omit\hidewidth 
{\rm Vertex\ Pair\ 2-1} \hidewidth && {\rm Vertex\ Pair\ 2-2} & \cr
\omit&height4pt&\omit&&\omit&&\omit&\cr
\tablerule
\omit&height2pt&\omit&&\omit&&\omit&\cr
&& $A_1 \partial^2 \hspace{-.1cm} \not{\hspace{-.1cm}\del} 
(\frac1{\Delta x^{2D-6}})$ && $\frac{(D-1)}{(D-2)(D-3)^2(D-4)}$ && $0$ & \cr
\omit&height2pt&\omit&&\omit&&\omit&\cr
\tablerule
\omit&height2pt&\omit&&\omit&&\omit&\cr
&& $A_1 \partial^2 \; \hspace{-.1cm} \overline{\not{\hspace{-.1cm}\del} }
(\frac1{\Delta x^{2D-6}})$ && $\frac{-D}{(D-2)(D-3)^2(D-4)}$ && $\frac{-1}{
(D-2)(D-3)^2(D-4)}$ & \cr
\omit&height2pt&\omit&&\omit&&\omit&\cr
\tablerule
\omit&height2pt&\omit&&\omit&&\omit&\cr
&& $A_2 \partial^2 \; \hspace{-.1cm} \overline{\not{\hspace{-.1cm}\del} }
(\frac1{\Delta x^{D-2}})$ && $\frac{-2}{D-3}$ && $0$ & \cr
\omit&height2pt&\omit&&\omit&&\omit&\cr
\tablerule
\omit&height2pt&\omit&&\omit&&\omit&\cr
&& $A_1 \nabla^2 \hspace{-.1cm} \not{\hspace{-.1cm}\del}
(\frac1{\Delta x^{2D-6}})$ && $0$ && $\frac{D (D^2 -3D -2)}{4(D-2)(D-3)^2
(D-4)}$ & \cr
\omit&height2pt&\omit&&\omit&&\omit&\cr
\tablerule
\omit&height2pt&\omit&&\omit&&\omit&\cr
&& $A_2 \nabla^2 \hspace{-.1cm} \not{\hspace{-.1cm}\del} 
(\frac1{\Delta x^{D-2}})$ && $0$ && $\frac{(D^2-3D-2)}{2(D-3)}$ & \cr
\omit&height2pt&\omit&&\omit&&\omit&\cr
\tablerule
\omit&height2pt&\omit&&\omit&&\omit&\cr
&& $A_1 \nabla^2 \; \hspace{-.1cm} \overline{\not{\hspace{-.1cm}\del}}
(\frac1{\Delta x^{2D-6}})$ && $0$ && $\frac{-D}{(D-2)(D-3)^2}$ & \cr
\omit&height2pt&\omit&&\omit&&\omit&\cr
\tablerule
\omit&height2pt&\omit&&\omit&&\omit&\cr
&& $A_2 \nabla^2 \; \hspace{-.1cm} \overline{\not{\hspace{-.1cm}\del}}
(\frac1{\Delta x^{D-2}})$ && $0$ && $-2(\frac{D-4}{D-3})$ & \cr
\omit&height2pt&\omit&&\omit&&\omit&\cr
\tablerule}}

\caption{$i\d\!\D_{A}$ terms giving both powers of $\Delta x^2$. The 
two coefficients are $A_1 \equiv \frac{i \kappa^2 H^2}{2^6 \pi^D} 
\Gamma(\frac{D}2 {\scriptstyle +1}) \Gamma(\frac{D}2) (a a')^{2- \frac{D}2}$ 
and $A_2 \equiv \frac{i \kappa^2 H^{D-2}}{2^{D+2} \pi^D} \Gamma({\scriptstyle 
D-2}) [{\scriptstyle \ln(a a') - \pi \cot}(\frac{D\pi}2)]$.}

\label{DAmostb}

\end{table}

The contributions from $i\delta\!\Delta_A$ are more complicated than those
from $i\Delta_{\rm cf}$ for several reasons. The fact that there is a second 
series in (\ref{dA}) occasions our Table~\ref{DAmostb}. These contributions 
are distinguished by all derivatives acting upon the conformal coordinate 
separation {\it and} by both series making nonzero contributions. Because 
these terms are special we shall explicitly carry out the reduction of the 
$2\!\!-\!\!2$ contribution. All three $2\!\!-\!\!2$ contractions on 
Table~\ref{DAcon} can be expressed as a certain tensor contracted into a 
generic form,
\begin{equation}
\Bigl[\delta_{ij} \delta_{k\ell} \!+\! \delta_{ik} \delta_{j\ell}
\!-\! \frac2{D\!-\!3} \delta_{i\ell} \delta_{jk} \Bigr] \times
\frac{\kappa^2}4 \partial_i \gamma_j \Bigl\{ i\delta\!\Delta_A(x;x')
\partial_k i[S](x;x') \gamma_{\ell} \Bigr\} \; . \label{generic}
\end{equation}
So we may as well work out the generic term and then do the contractions
at the end. Substituting the fermion propagator brings this generic term 
to the form,
\begin{eqnarray}
{\rm Generic} & \equiv & \frac{\kappa^2}4 \partial_i \gamma_j \Bigl\{ 
i\delta\!\Delta_A(x;x') \partial_k i[S](x;x') \gamma_{\ell} \Bigr\} \; , \\
& = & -\frac{i \kappa^2 \Gamma(\frac{D}2)}{8 \pi^{\frac{D}2}} 
\partial_i \gamma_j \Bigl\{ i\delta\!\Delta_A(x;x') \partial_k \Bigl( 
\frac{\gamma^{\mu} \Delta x_{\mu}}{\Delta x^D} \Bigr) \gamma_{\ell} \Bigr\}\; .
\end{eqnarray}

Now recall that there are two sorts of terms in the only part of
$i\delta\!\Delta_A(x;x')$ that can make a nonzero contribution for $D\!=\!4$,
\begin{eqnarray}
i\delta\!\Delta_{A1}(x;x') & \equiv & \frac{H^2}{16 \pi^{\frac{D}2}}
\frac{\Gamma(\frac{D}2\!+\!1)}{\frac{D}2\!-\!2} \frac{(a a')^{2-\frac{D}2}}{
\Delta x^{D-4}} \; , \\
i\delta\!\Delta_{A2}(x;x') & \equiv & \frac{H^{D-2}}{(4 \pi)^{\frac{D}2}}
\frac{\Gamma(D\!-\!1)}{\Gamma(\frac{D}2)} \Bigl\{-\pi\cot\Bigl(\frac{\pi}2 D
\Bigr) + \ln(a a')\Bigr\} \; .
\end{eqnarray}
Because all the derivatives are spatial we can pass the scale factors 
outside to obtain,
\begin{eqnarray}
\lefteqn{{\rm Generic}_1 } \nonumber \\
& & = -\frac{i \kappa^2 H^2}{2^6 \pi^D} \frac{\Gamma(\frac{D}2)
\Gamma(\frac{D}2\!+\!1)}{(D\!-\!4)} (a a')^{2-\frac{D}2} \partial_i \gamma_j
\Bigl\{ \frac1{\Delta x^{D-4}} \partial_k \Bigl(\frac{\gamma^{\mu} \Delta 
x_{\mu}}{\Delta x^D}\Bigr) \gamma_{\ell} \Bigr\} \; , \qquad \\
\lefteqn{{\rm Generic}_2 } \nonumber \\
& & = -\frac{i \kappa^2 H^{D-2}}{2^{D+2} \pi^D} \Gamma(D\!-\!1)
\Bigl\{-\pi\cot\Bigl(\frac{\pi}2 D \Bigr) \!+\! \ln(a a')\Bigr\} \partial_i
\gamma_j \partial_k \Bigl(\frac{\gamma^{\mu} \Delta x_{\mu}}{\Delta x^D}\Bigr) 
\gamma_{\ell} \; , \qquad \\
& & = \frac{i \kappa^2 H^{D-2}}{2^{D+2} \pi^D} \Gamma(D\!-\!2)
\Bigl\{-\pi\cot\Bigl(\frac{\pi}2 D \Bigr) \!+\! \ln(a a')\Bigr\} \partial_i
\gamma_j \partial_k \hspace{-.1cm}\not{\hspace{-.05cm} \partial} 
\gamma_{\ell} \Bigl(\frac1{\Delta x^{D-2}}\Bigr) \; . \qquad
\end{eqnarray}
To complete the reduction of the first generic term we note,
\begin{eqnarray}
\frac1{\Delta x^{D-4}} \partial_k \Bigl(\frac{\gamma^{\mu} \Delta x_{\mu}}{
\Delta x^D} \Bigr) & = & \frac{\gamma_k}{\Delta x^{2D-4}} \!-\! \frac{D 
\gamma^{\mu} \Delta x_{\mu} \Delta x_k}{\Delta x^{2D-2}} \; , \\
& = & \frac12 \Bigl(\frac{D\!-\!4}{D\!-\!2}\Bigr) \frac{\gamma_k}{\Delta 
x^{2D-4}} \!+\! \frac{D}{2(D\!-\!2)} \partial_k \Bigl(\frac{\gamma^{\mu}
\Delta x_{\mu}}{\Delta x^{2D-4}}\Bigr) \; , \\
& = & \frac1{4 (D\!-\!3)(D\!-\!2)} \Bigl\{ \gamma_k \partial^2 
- D \partial_k \hspace{-.1cm}\not{\hspace{-.05cm} \partial} \Bigr\} 
\frac1{\Delta x^{2D-6}} \; . \qquad
\end{eqnarray}
Hence the first generic term is,
\begin{eqnarray}
\lefteqn{{\rm Generic}_1 = \frac{i \kappa^2 H^2}{2^8 \pi^D} \frac{\Gamma(
\frac{D}2) \Gamma(\frac{D}2\!+\!1)}{(D\!-\!4)(D\!-\!3)(D\!-\!2)} (a a')^{
2-\frac{D}2} } \nonumber \\
& & \hspace{4cm} \times \Bigl\{D \partial_i \gamma_j \partial_k 
\hspace{-.1cm}\not{\hspace{-.05cm} \partial} \gamma_{\ell} - \partial^2 
\partial_i \gamma_j \gamma_k \gamma_{\ell} \Bigr\} \frac1{\Delta x^{2D-6}} 
\; . \qquad
\end{eqnarray}

Now we contract the tensor prefactor of (\ref{generic}) into the appropriate
spinor-differential operators. For the first generic term this is,
\begin{eqnarray}
\lefteqn{\Bigl[\delta_{ij} \delta_{k\ell} \!+\! \delta_{ik} \delta_{j\ell}
\!-\! \frac2{D\!-\!3} \delta_{i\ell} \delta_{jk} \Bigr] \times \Bigl\{D 
\partial_i \gamma_j \partial_k \hspace{-.1cm}\not{\hspace{-.05cm} \partial} 
\gamma_{\ell} - \partial^2 \partial_i \gamma_j \gamma_k \gamma_{\ell} \Bigr\} }
\nonumber \\
& & \hspace{-.5cm} = D \Bigl(\frac{D\!-\!5}{D\!-\!3}\Bigr) 
\hspace{-.1cm}\not{\hspace{-.05cm} \overline{\partial}} 
\hspace{-.1cm}\not{\hspace{-.05cm} \partial} 
\hspace{-.1cm}\not{\hspace{-.05cm} \overline{\partial}} \!+\! D \nabla^2
\gamma_i \hspace{-.1cm}\not{\hspace{-.05cm} \partial} \gamma_i \!-\!
\partial^2 \hspace{-.1cm}\not{\hspace{-.05cm} \overline{\partial}} 
\gamma_i \gamma_i \!-\! \partial^2 \gamma_i \hspace{-.1cm}\not{\hspace{-.05cm} 
\overline{\partial}} \gamma_i \!+\! \frac2{D\!-\!3} \partial^2 \gamma_i 
\gamma_i \hspace{-.1cm}\not{\hspace{-.05cm} \overline{\partial}} \; . \qquad
\end{eqnarray}
This term can be simplified using the identities,
\begin{eqnarray}
\hspace{-.1cm}\not{\hspace{-.05cm} \overline{\partial}} 
\hspace{-.1cm}\not{\hspace{-.05cm} \partial} 
\hspace{-.1cm}\not{\hspace{-.05cm} \overline{\partial}} & = & -
\hspace{-.1cm}\not{\hspace{-.05cm} \overline{\partial}} 
\hspace{-.1cm}\not{\hspace{-.05cm} \overline{\partial}} 
\hspace{-.1cm}\not{\hspace{-.05cm} \partial} - 2
\hspace{-.1cm}\not{\hspace{-.05cm} \overline{\partial}} \nabla^2 = \nabla^2
\hspace{-.1cm}\not{\hspace{-.05cm} \partial} - 2
\hspace{-.1cm}\not{\hspace{-.05cm} \overline{\partial}} \nabla^2 =
-\nabla^2 \hspace{-.1cm}\not{\hspace{-.05cm} \partial} \!+\! 2 \nabla^2
\gamma^0 \partial_0 \; , \\
\gamma_i \hspace{-.1cm}\not{\hspace{-.05cm} \partial} \gamma_i & = & 
-\gamma_i \gamma_i \hspace{-.1cm}\not{\hspace{-.05cm} \partial} -2
\hspace{-.1cm}\not{\hspace{-.05cm} \overline{\partial}} = (D\!-\!1)
\hspace{-.1cm}\not{\hspace{-.05cm} \partial} - 2
\hspace{-.1cm}\not{\hspace{-.05cm} \overline{\partial}} = (D\!-\!3)
\hspace{-.1cm}\not{\hspace{-.05cm} \partial} + 2 \gamma^0 \partial_0 \; , \\
\hspace{-.1cm}\not{\hspace{-.05cm} \overline{\partial}} \gamma_i \gamma_i & = & 
-(D\!-\!1) \hspace{-.1cm}\not{\hspace{-.05cm} \overline{\partial}} =
\gamma_i \gamma_i \hspace{-.1cm}\not{\hspace{-.05cm} \overline{\partial}} 
\; , \\
\gamma_i \hspace{-.1cm}\not{\hspace{-.05cm} \overline{\partial}} \gamma_i & = & 
-\gamma_i \gamma_i \hspace{-.1cm}\not{\hspace{-.05cm} \overline{\partial}} -2
\hspace{-.1cm}\not{\hspace{-.05cm} \overline{\partial}} = (D\!-\!3)
\hspace{-.1cm}\not{\hspace{-.05cm} \overline{\partial}} \; .
\end{eqnarray}
Applying these identities gives,
\begin{eqnarray}
\lefteqn{\Bigl[\delta_{ij} \delta_{k\ell} \!+\! \delta_{ik} \delta_{j\ell}
\!-\! \frac2{D\!-\!3} \delta_{i\ell} \delta_{jk} \Bigr] \times \Bigl\{D 
\partial_i \gamma_j \partial_k \hspace{-.1cm}\not{\hspace{-.05cm} \partial} 
\gamma_{\ell} - \partial^2 \partial_i \gamma_j \gamma_k \gamma_{\ell} \Bigr\} }
\nonumber \\
& & \hspace{2.7cm} = \Bigl(D^2 \!-\! \frac{2D}{D\!-\!3}\Bigr) \nabla^2
\hspace{-.1cm}\not{\hspace{-.05cm} \partial} \!-\! 4D \Bigl(\frac{D\!-\!4}{D\!-
\!3}\Bigr) \nabla^2  \hspace{-.1cm}\not{\hspace{-.05cm} \overline{\partial}} 
\!-\! \frac4{D\!-\!3} \partial^2 \hspace{-.1cm}\not{\hspace{-.05cm} 
\overline{\partial}} \; . \qquad
\end{eqnarray}
For the second generic term the relevant contraction is,
\begin{eqnarray}
\lefteqn{\Bigl[\delta_{ij} \delta_{k\ell} \!+\! \delta_{ik} \delta_{j\ell}
\!-\! \frac2{D\!-\!3} \delta_{i\ell} \delta_{jk} \Bigr] \times \partial_i
\gamma_j \partial_k \hspace{-.1cm}\not{\hspace{-.05cm} \partial} \gamma_{\ell}}
\nonumber \\
& & \hspace{4cm} = \Bigl(\frac{D\!-\!5}{D\!-\!3}\Bigr) 
\hspace{-.1cm}\not{\hspace{-.05cm} \overline{\partial}} 
\hspace{-.1cm}\not{\hspace{-.05cm} \partial} 
\hspace{-.1cm}\not{\hspace{-.05cm} \overline{\partial}} \!+\! \nabla^2
\gamma_i \hspace{-.1cm}\not{\hspace{-.05cm} \partial} \gamma_i \; , \\
& & \hspace{4cm} = \Bigl(D \!-\! \frac{2}{D\!-\!3}\Bigr) \nabla^2
\hspace{-.1cm}\not{\hspace{-.05cm} \partial} \!-\! 4 \Bigl(\frac{D\!-\!4}{D\!-
\!3}\Bigr) \nabla^2  \hspace{-.1cm}\not{\hspace{-.05cm} \overline{\partial}} 
\; .
\end{eqnarray}

\begin{table}

\vbox{\tabskip=0pt \offinterlineskip
\def\tablerule{\noalign{\hrule}}
\halign to390pt {\strut#& \vrule#\tabskip=1em plus2em&
\hfil#\hfil& \vrule#& \hfil#\hfil& \vrule#& \hfil#\hfil& \vrule#& \hfil#\hfil& 
\vrule#& \hfil#\hfil& \vrule#& \hfil#\hfil& \vrule#& \hfil#\hfil&
\vrule#\tabskip=0pt\cr
\tablerule
\omit&height4pt&\omit&&\omit&&\omit&&\omit&&\omit&&\omit&&\omit&\cr
&& $\!\!\!\!{\rm I}\!\!\!\!\!\!$ && $\!\!\!\!{\rm J}\!\!\!\!\!\!$ && 
$\!\!\!\!{\rm sub}\!\!\!\!\!\!$ && 
$\!\!\!\!\frac{\gamma^{\mu} \Delta x_{\mu}}{\Delta x^{2D-2}}\!\!\!\!$ &&
$\!\!\!\!\frac{\gamma^i \Delta x_i}{\Delta x^{2D-2}}\!\!\!\!$ &&
$\!\!\!\frac{\Vert \Delta \vec{x}\Vert^2 \gamma^{\mu} \Delta x_{\mu}}{\Delta 
x^{2D}}\!\!\!\!$ &&
$\!\!\!\!\frac{\Vert \Delta \vec{x}\Vert^2 \gamma^i \Delta x_i}{\Delta 
x^{2D}}\!\!\!\!$ &\cr
\omit&height4pt&\omit&&\omit&&\omit&&\omit&&\omit&&\omit&&\omit&\cr
\tablerule
\omit&height2pt&\omit&&\omit&&\omit&&\omit&&\omit&&\omit&&\omit&\cr
&& 2 && 3 && a && $2(\frac{D-1}{D-3})$ && $\!\!\!\!-\frac{2D}{D-3}\!\!\!$ 
&& $0$ && $0$ & \cr
\omit&height2pt&\omit&&\omit&&\omit&&\omit&&\omit&&\omit&&\omit&\cr
\tablerule
\omit&height2pt&\omit&&\omit&&\omit&&\omit&&\omit&&\omit&&\omit&\cr
&& 2 && 3 && b && $0$ && $1$ && $\frac{D^2}2$ && $-2 {\scriptstyle D}$ & \cr
\omit&height2pt&\omit&&\omit&&\omit&&\omit&&\omit&&\omit&&\omit&\cr
\tablerule
\omit&height2pt&\omit&&\omit&&\omit&&\omit&&\omit&&\omit&&\omit&\cr
&& 2 && 3 && c && $0$ && $\!\!\!\!-(\frac{D-1}{D-3})\!\!\!\!$ && 
$-\frac{D}{D-3}$ && $\frac{2D}{D-3}$ & \cr
\omit&height2pt&\omit&&\omit&&\omit&&\omit&&\omit&&\omit&&\omit&\cr
\tablerule
\omit&height2pt&\omit&&\omit&&\omit&&\omit&&\omit&&\omit&&\omit&\cr
&& 3 && 1 && a && $\!\!\!\!-\frac{4(D-1)(D-2)}{D-3}\!\!\!\!$ && $0$ && 
$0$ && $0$ & \cr
\omit&height2pt&\omit&&\omit&&\omit&&\omit&&\omit&&\omit&&\omit&\cr
\tablerule
\omit&height2pt&\omit&&\omit&&\omit&&\omit&&\omit&&\omit&&\omit&\cr
&& 3 && 1 && b && $2(\frac{D-1}{D-3})$ && $\!\!\!\!2(\frac{D-4}{D-3})\!\!\!\!$ 
&& $0$ && $0$ & \cr
\omit&height2pt&\omit&&\omit&&\omit&&\omit&&\omit&&\omit&&\omit&\cr
\tablerule
\omit&height2pt&\omit&&\omit&&\omit&&\omit&&\omit&&\omit&&\omit&\cr
&& 3 && 2 && a && $0$ && $\!\!\!\! 4 (\frac{D-2}{D-3})\!\!\!\!$ && $0$
&& $0$ & \cr
\omit&height2pt&\omit&&\omit&&\omit&&\omit&&\omit&&\omit&&\omit&\cr
\tablerule
\omit&height2pt&\omit&&\omit&&\omit&&\omit&&\omit&&\omit&&\omit&\cr
&& 3 && 2 && b && $-(\frac{D-1}{D-3})$ && $\!\!\!\! (\frac{D+1}{D-3})
\!\!\!\!$ && $\!\!\!\!2 (\frac{D-1}{D-3})\!\!\!\!$ && $-4 (\frac{D-1}{D-3})$ 
& \cr
\omit&height2pt&\omit&&\omit&&\omit&&\omit&&\omit&&\omit&&\omit&\cr
\tablerule
\omit&height2pt&\omit&&\omit&&\omit&&\omit&&\omit&&\omit&&\omit&\cr
&& 3 && 2 && c && $\frac12 ({\scriptstyle D-1})$ && $\!\!\!\! -\frac12 
({\scriptstyle D+1})\!\!\!\!$ && $\!\!\!\!- ({\scriptstyle D-1})\!\!\!\!$ 
&& $2 ({\scriptstyle D-1})$ & \cr
\omit&height2pt&\omit&&\omit&&\omit&&\omit&&\omit&&\omit&&\omit&\cr
\tablerule
\omit&height2pt&\omit&&\omit&&\omit&&\omit&&\omit&&\omit&&\omit&\cr
&& 3 && 2 && d && $\frac12 ({\scriptstyle D-1})^2$ && $\!\!\!\! -\frac12 
({\scriptstyle D+1})\!\!\!\!$ && $\!\!\!\!- ({\scriptstyle D-1})^2\!\!\!\!$ 
&& $2 ({\scriptstyle D-1})$ & \cr
\omit&height2pt&\omit&&\omit&&\omit&&\omit&&\omit&&\omit&&\omit&\cr
\tablerule
\omit&height2pt&\omit&&\omit&&\omit&&\omit&&\omit&&\omit&&\omit&\cr
&& 3 && 3 && a && $\!\!\!\!2\frac{(D-1)(D-2)}{(D-3)}\!\!\!\!$ && $0$
&& $0$ && $0$ & \cr
\omit&height2pt&\omit&&\omit&&\omit&&\omit&&\omit&&\omit&&\omit&\cr
\tablerule
\omit&height2pt&\omit&&\omit&&\omit&&\omit&&\omit&&\omit&&\omit&\cr
&& 3 && 3 && b && $-(\frac{D-1}{D-3})$ && $\!\!\!\! -(\frac{D-4}{D-3})\!\!\!\!$
&& $0$ && $0$ & \cr
\omit&height2pt&\omit&&\omit&&\omit&&\omit&&\omit&&\omit&&\omit&\cr
\tablerule
\omit&height2pt&\omit&&\omit&&\omit&&\omit&&\omit&&\omit&&\omit&\cr
&& 3 && 3 && c && $-(\frac{D-1}{D-3})$ && $\!\!\!\! -(\frac{D-4}{D-3})\!\!\!\!$
&& $0$ && $0$ & \cr
\omit&height2pt&\omit&&\omit&&\omit&&\omit&&\omit&&\omit&&\omit&\cr
\tablerule
\omit&height2pt&\omit&&\omit&&\omit&&\omit&&\omit&&\omit&&\omit&\cr
&& 3 && 3 && d && $\!\!\!\!-\frac{(D-1)(D-5)}{4 (D-3)}\!\!\!\!$ && $\!\!\!\!
\frac12 (\frac{D-5}{D-3})\!\!\!\!$ && $\frac{(D-5)(D-2)}{4 (D-3)}$ && 
$\!\!\!\!-\frac{(D-5)(D-2)}{2(D-3)}\!\!\!\!$ & \cr
\omit&height2pt&\omit&&\omit&&\omit&&\omit&&\omit&&\omit&&\omit&\cr
\tablerule
\omit&height2pt&\omit&&\omit&&\omit&&\omit&&\omit&&\omit&&\omit&\cr
&& 3 && 3 && e && $-\frac14 ({\scriptstyle D-1})^2$ && $\!\!\!\! \frac12 
({\scriptstyle D-1})\!\!\!\!$ && $\!\!\!\!\frac14 ({\scriptstyle D-2})
({\scriptstyle D-1})\!\!\!\!$ && $-\frac12 ({\scriptstyle D-2})$ & \cr
\omit&height2pt&\omit&&\omit&&\omit&&\omit&&\omit&&\omit&&\omit&\cr
\tablerule}}

\caption{$i\d\!\D_{A}$ terms in which all derivatives act upon 
$\Delta x^2(x;x')$. All contributions are multiplied by $\frac{i \kappa^2 
H^2}{2^6 \pi^D} \Gamma(\frac{D}2 {\scriptstyle +1}) \Gamma(\frac{D}2) 
(a a')^{2- \frac{D}2}$. }

\label{DAmostc}

\end{table}

In summing the contributions from Table~\ref{DAmostb} it is best to take
advantage of cancellations between $A_1$ and $A_2$ terms. These occur between
the 2nd and 3rd terms in the second column, the 4th and 5th terms of the 3rd
column, and the 6th and 7th terms of the 3rd column. In each of these cases
the result is finite; and it actually vanishes in the final case! Only the
first term of column 2 and the 2nd term of column 3 contribute divergences.
The result for the three contributions from $[2\!\!-\!\!1]$ in 
Table~\ref{DAmostb} is,
\begin{eqnarray}
\lefteqn{-\frac{\kappa^2 H^2 \mu^{D-4}}{2^5 \pi^{\frac{D}2}} \frac{(D\!-\!1) 
\Gamma(\frac{D}2\!+\!1)}{(D\!-\!3)^2 (D\!-\!4)} \, (a a')^{2 -\frac{D}2} 
\hspace{-.1cm} \not{\hspace{-.1cm} \partial} \, \delta^D(x\!-\!x') } 
\nonumber \\
& & + \frac{i \kappa^2 H^2}{2^6 \pi^4} \Biggl\{\! -\frac32 \partial^2
\hspace{-.1cm} \not{\hspace{-.1cm} \partial} \Bigl[ \frac{\ln(\mu^2 
\Delta x^2)}{\Delta x^2}\Bigr] \!+\! \partial^2 \,\hspace{-.1cm} 
\overline{\not{\hspace{-.1cm} \partial}} \Bigl[ \frac{4 \!+\! 2 
\ln(\frac14 H^2 \Delta x^2)}{\Delta x^2}\Bigr] \!\Biggr\} \!+\! O(D\!-\!4) . 
\qquad
\end{eqnarray}
The result for the five contributions from $[2\!\!-\!\!2]$ in 
Table~\ref{DAmostb} is,
\begin{eqnarray}
\lefteqn{\frac{\kappa^2 H^2 \mu^{D-4}}{2^5 \pi^{\frac{D}2}} \frac{
\Gamma(\frac{D}2\!+\!1)}{(D\!-\!3)^2 (D\!-\!4)} \, (a a')^{2 -\frac{D}2} \;
\hspace{-.1cm} \overline{\not{\hspace{-.1cm} \partial}} \, \delta^D(x\!-\!x') } 
\nonumber \\
& & + \frac{i \kappa^2 H^2}{2^6 \pi^4} \Biggl\{\! \frac12 \partial^2 \,
\hspace{-.1cm} \overline{\not{\hspace{-.1cm} \partial}} \Bigl[ \frac{\ln(\mu^2 
\Delta x^2)}{\Delta x^2}\Bigr] \!-\! \nabla^2 \hspace{-.1cm} 
\not{\hspace{-.1cm} \partial} \Bigl[ \frac{2 \!+\! \ln(\frac14 H^2 \Delta x^2)
}{\Delta x^2}\Bigr] \!\Biggr\} + O(D\!-\!4) . \qquad
\end{eqnarray}
As might be expected from the similarities in their reductions, these
two terms combine together nicely in the total for Table~\ref{DAmostb},
\begin{eqnarray}
\lefteqn{-i \Bigl[\Sigma^{T\ref{DAmostb}}\Bigr](x;x') = \frac{\kappa^2 H^2
\mu^{D-4}}{2^5 \pi^{\frac{D}2}} \frac{\Gamma(\frac{D}2\!+\!1) (a a')^{2-
\frac{D}2}}{(D\!-\!3)^2 (D\!-\!4)} \Bigl[-(D\!-\!1) \hspace{-.1cm} 
\not{\hspace{-.1cm} \partial} + \hspace{-.1cm} \overline{\not{\hspace{-.1cm} 
\partial}} \, \Bigr] \delta^D(x\!-\!x') } \nonumber \\
& & \hspace{3cm} + \frac{i \kappa^2 H^2}{2^6 \pi^4} \Biggl\{\Bigl( -\frac32
\hspace{-.1cm} \not{\hspace{-.1cm} \partial} \partial^2 \!+\! \frac12 \;
\hspace{-.1cm} \overline{\not{\hspace{-.1cm} \partial}} \, \partial^2 \Bigl)
\Bigl[ \frac{\ln(\mu^2 \Delta x^2)}{\Delta x^2}\Bigr] \nonumber \\
& & \hspace{3.5cm} + \Bigl(2 \; \hspace{-.1cm} \overline{\not{\hspace{-.1cm} 
\partial}} \, \partial^2 \!-\! \hspace{-.1cm} \not{\hspace{-.1cm} \partial} \,
\nabla^2 \Bigr) \Bigl[ \frac{2 \!+\! \ln(\frac14 H^2 \Delta x^2)}{\Delta x^2}
\Bigr] \Biggr\} + O(D\!-\!4) . \qquad \label{4thcon}
\end{eqnarray}

The next class is comprised of terms in which only the first series of
$i\delta\!\Delta_A$ makes a nonzero contribution when all derivatives
act upon the conformal coordinate separation. The results for this class
of terms are summarized in Table~\ref{DAmostc}. In reducing these terms
the following derivatives occur many times,
\begin{eqnarray}
\partial_i i \delta\!\Delta_A(x;x') & = & -\frac{H^2}{8 \pi^{\frac{D}2}} \,
\Gamma\Bigl(\frac{D}2\!+\!1\Bigr) (a a')^{2-\frac{D}2} \, \frac{\Delta x^i}{
\Delta x^{D-2}} = -\partial_i' i \delta\!\Delta_A(x;x') \; ,  \\
\partial_0 i \delta\!\Delta_A(x;x') & = & \frac{H^2}{8 \pi^{\frac{D}2}}
\, \Gamma\Bigl(\frac{D}2\!+\!1\Bigr) (a a')^{2-\frac{D}2} \Biggl\{
\frac{\Delta \eta}{\Delta x^{D-2}} \!-\! \frac{a H}{2 \Delta x^{D-4}} \Biggr\} 
\nonumber \\
& & \hspace{5cm} + \frac{H^{D-2}}{2^D \pi^{\frac{D}2}} \frac{\Gamma(
D\!-\!1)}{\Gamma(\frac{D}2)} \, a H \; , \qquad \\
\partial_0' i \delta\!\Delta_A(x;x') & = & \frac{H^2}{8 \pi^{\frac{D}2}}
\, \Gamma\Bigl(\frac{D}2\!+\!1\Bigr) (a a')^{2-\frac{D}2} \Biggl\{
-\frac{\Delta \eta}{\Delta x^{D-2}} \!-\! \frac{a' H}{2 \Delta x^{D-4}} \Biggr\}
\nonumber \\
& & \hspace{5cm} + \frac{H^{D-2}}{2^D \pi^{\frac{D}2}} \frac{\Gamma(D\!-\!1)}{
\Gamma(\frac{D}2)} \, a' H \; . \qquad
\end{eqnarray}
We also make use of a number of gamma matrix identities,
\begin{eqnarray}
\gamma^{\mu} \gamma_{\mu} & = & - D \quad {\rm and} \quad \gamma^i \gamma^i =
-(D\!-\!1) \; , \label{gammafirst} \\
\gamma^{\mu} \gamma^{\nu} \gamma_{\mu} & = & (D\!-\!2) \gamma^{\nu} \quad 
{\rm and} \quad \gamma^i \gamma^{\nu} \gamma^i = (D\!-\!1) \gamma^{\nu} -
2 \overline{\gamma}^{\nu} \; , \qquad \\
(\gamma^{\mu} \Delta x_{\mu})^2 & = & - \Delta x^2 \quad {\rm and} \quad
(\gamma^i \Delta x^i)^2 = -\Vert \Delta \vec{x} \Vert^2 \; , \\
\gamma^i \gamma^{\mu} \Delta x_{\mu} \gamma^i & = & (D\!-\!1) \gamma^{\mu}
\Delta x_{\mu} - 2 \gamma^i \Delta x^i \; , \\
\gamma^i \Delta x^i \gamma^{\mu} \Delta x_{\mu} \gamma^j \Delta x^j & = & 
\Vert \Delta \vec{x} \Vert^2 \gamma^{\mu} \Delta x_{\mu} - 2 \Vert \Delta 
\vec{x} \Vert^2 \gamma^i \Delta x^i \; . \label{gammalast}
\end{eqnarray}

In summing the many terms of Table~\ref{DAmostc} the constant $K \equiv
D - 2/(D\!-\!3)$ occurs suspiciously often,
\begin{eqnarray}
\lefteqn{-i \Bigl[\Sigma^{T\ref{DAmostc}}\Bigr](x;x') = \frac{i \kappa^2 H^2}{
2^6 \pi^D} \Gamma\Bigl(\frac{D}2 \!+\! 1\Bigr) \Gamma\Bigl(\frac{D}2\Bigr) 
(a a')^{2-\frac{D}2} } \nonumber \\
& & \times \Biggl\{ \Bigl[-2 (D\!-\!1) \!+\! \Bigl(\frac{D\!-\!1}4 \Bigr) K 
\Bigr] \frac{\gamma^{\mu} \Delta x_{\mu}}{\Delta x^{2D-2}} + \Bigl[-(D\!-\!2) 
\!+\! \frac{K}2 \Bigr] \frac{\gamma^i \Delta x_i}{\Delta x^{2D-2}} \nonumber \\
& & \hspace{1.5cm} - \Bigl(\frac{D\!-\!2}4 \Bigr) K \frac{\Vert \Delta \vec{x} 
\Vert^2 \gamma^{\mu} \Delta x_{\mu}}{\Delta x^{2D}} + \frac{(D\!-\!2)(D\!-\!4)
}{ (D\!-\!3)} \frac{\Vert \Delta \vec{x} \Vert^2 \gamma^i \Delta x_i}{
\Delta x^{2D}} \Biggr\} . \qquad \label{S12-1}
\end{eqnarray}
The last two terms can be reduced using the identities,
\begin{eqnarray}
\frac{\Vert \Delta \vec{x} \Vert^2 \gamma^{\mu} \Delta x_{\mu}}{\Delta x^{2D}} 
&\!\!\!\!=\!\!\!\!& \frac12 \frac{\gamma^{\mu} \Delta x_{\mu}}{\Delta x^{2D-2}} 
\!+\!  \frac1{D\!-\!1} \frac{\gamma^i \Delta x_i}{\Delta x^{2D-2}} \!+\! 
\frac{\nabla^2}{4(D\!-\!2) (D\!-\!1)} \Bigl(\frac{\gamma^{\mu} 
\Delta x_{\mu}}{\Delta x^{2D-4}} \Bigr) , \qquad \label{difID1} \\
\frac{\Vert \Delta \vec{x} \Vert^2 \gamma^i \Delta x_i}{\Delta x^{2D}} 
&\!\!\!\!= \!\!\!\!&\frac12 \Bigl(\frac{D\!+\!1}{D\!-\!1} \Bigr) 
\frac{\gamma^i \Delta x_i}{\Delta x^{2D-2}} + \frac{\nabla^2}{4(D\!-\!2) 
(D\!-\!1)} \Bigl(\frac{\gamma^i \Delta x_i}{\Delta x^{2D-4}} \Bigr) . \qquad
\label{difID2}
\end{eqnarray}
Substituting these in (\ref{S12-1}) gives,
\begin{eqnarray}
\lefteqn{-i \Bigl[\Sigma^{T\ref{DAmostc}}\Bigr](x;x') = \frac{i \kappa^2 H^2}{
2^6 \pi^D} \Gamma\Bigl(\frac{D}2 \!+\! 1\Bigr) \Gamma\Bigl(\frac{D}2\Bigr) 
(a a')^{2-\frac{D}2} \Biggl\{ \Bigl[-2 (D\!-\!1) \!+\! \frac{D K}8 \Bigr] } 
\nonumber \\
& & \hspace{1cm} \times \frac{\gamma^{\mu} \Delta x_{\mu}}{\Delta x^{2D-2}}
+ \Bigl[-\frac{(D\!-\!2)(D^2 \!-\! 5D \!+\! 10)}{2 (D\!-\!1)(D\!-\!3)} \!+\! 
\frac{D K}{4 (D\!-\!1)} \Bigr] \frac{\gamma^i \Delta x_i}{\Delta x^{2D-2}} 
\nonumber \\
& & \hspace{2.5cm} - \frac{K \nabla^2}{16 (D\!-\!1)} \Bigl(\frac{\gamma^{\mu} 
\Delta x_{\mu}}{\Delta x^{2D-4}} \Bigr) + \frac{(D\!-\!4) \nabla^2 }{4
(D\!-\!1) (D\!-\!3)} \frac{\gamma^i \Delta x_i}{\Delta x^{2D-4}} \Biggr\} . 
\qquad \label{S12-2}
\end{eqnarray}
We then apply the same formalism as in the previous sub-section to partially 
integrate, extract the local divergences and take $D\!=\!4$ for the remaining, 
integrable and ultraviolet finite nonlocal terms,
\newpage
\begin{eqnarray}
\lefteqn{-i \Bigl[\Sigma^{T\ref{DAmostc}}\Bigr](x;x') = \frac{\kappa^2 H^2
\mu^{D-4}}{2^7 \pi^{\frac{D}2}} \frac{\Gamma(\frac{D}2 \!+\! 1) (a a')^{2 -
\frac{D}2}}{(D\!-\!3) (D\!-\!4)} } \nonumber \\
& & \hspace{-.5cm} \times \Biggl\{ \Bigl[\frac{D K}8 \!-\! 2(D\!-\!1)\Bigr] 
\hspace{-.1cm} \not{\hspace{-.1cm} \partial} \!+\! \Bigl[\frac{D K}{4(D\!-\!1)} 
\!-\! \frac{(D\!-\!2)(D^2 \!-\! 5D \!+\! 10)}{2 (D\!-\! 1) (D\!-\!3)} \Bigr] 
\; \hspace{-.1cm} \overline{\not{\hspace{-.1cm} \partial} } \Biggr\} 
\delta^D(x\!-\!x') \nonumber \\
& & \hspace{-.5cm} +\frac{i \kappa^2 H^2}{2^9 \!\cdot\! 3 \!\cdot\! \pi^4} 
\Biggl\{\!\Bigl[-15 \hspace{-.1cm} \not{\hspace{-.1cm} \partial} \, \partial^2
\!-\! 4 \; \hspace{-.1cm} \overline{\not{\hspace{-.1cm} \partial}} \,\partial^2
\Bigr] \Bigl(\frac{\ln(\mu^2 \Delta x^2)}{\Delta x^2} \Bigr) \!+\! \hspace{
-.1cm} \not{\hspace{-.1cm} \partial} \, \nabla^2 \Bigl(\frac1{\Delta x^2} 
\Bigr)\!\Biggr\} \!+\! O(D\!-\!4) . \qquad \label{5thcon}
\end{eqnarray}

\begin{table}

\vbox{\tabskip=0pt \offinterlineskip
\def\tablerule{\noalign{\hrule}}
\halign to390pt {\strut#& \vrule#\tabskip=1em plus2em&
\hfil#\hfil& \vrule#& \hfil#\hfil& \vrule#& \hfil#\hfil& \vrule#& \hfil#\hfil& 
\vrule#& \hfil#\hfil& \vrule#& \hfil#\hfil&
\vrule#\tabskip=0pt\cr
\tablerule
\omit&height4pt&\omit&&\omit&&\omit&&\omit&&\omit&&\omit&\cr
&& $\!\!\!\!{\rm I}\!\!\!\!\!\!$ && $\!\!\!\!{\rm J}\!\!\!\!\!\!$ && 
$\!\!\!\!{\rm sub}\!\!\!\!\!\!$ && 
$\!\!\!\!\frac{H \gamma^0}{\Delta x^{2D-4}}\!\!\!\!$ &&
$\!\!\!\!\frac{H \gamma^i \Delta x_i \gamma^{\mu} \Delta x_{\mu} \gamma^0}{
\Delta x^{2D-2}}\!\!\!\!$ && $\!\!\!\frac{H^2 a a' \gamma^{\mu} \Delta x_{\mu}
}{\Delta x^{2D-4}}\!\!\!\!$ &\cr
\omit&height4pt&\omit&&\omit&&\omit&&\omit&&\omit&&\omit&\cr
\tablerule
\omit&height2pt&\omit&&\omit&&\omit&&\omit&&\omit&&\omit&\cr
&& 2 && 1 && \omit && $\!\!\!\!2(\frac{D-1}{D-3}) a'\!\!\!\!$ && 
$\!\!\!\!(\frac{2 D}{D-3}) a' \!\!\!\!$ && $0$ & \cr
\omit&height2pt&\omit&&\omit&&\omit&&\omit&&\omit&&\omit&\cr
\tablerule
\omit&height2pt&\omit&&\omit&&\omit&&\omit&&\omit&&\omit&\cr
&& 2 && 3 && a && $\!\!\!\!-(\frac{D-1}{D-3}) a'\!\!\!\!$ && 
$\!\!\!\!(\frac{-D}{D-3}) a' \!\!\!\!$ && $0$ & \cr
\omit&height2pt&\omit&&\omit&&\omit&&\omit&&\omit&&\omit&\cr
\tablerule
\omit&height2pt&\omit&&\omit&&\omit&&\omit&&\omit&&\omit&\cr
&& 3 && 1 && a && $0$ && $0$ && 
$\!\!\!\!\frac{(D-1)(D-4)}{2 (D-3)}\!\!\!\!$ & \cr
\omit&height2pt&\omit&&\omit&&\omit&&\omit&&\omit&&\omit&\cr
\tablerule
\omit&height2pt&\omit&&\omit&&\omit&&\omit&&\omit&&\omit&\cr
&& 3 && 1 && b && $0$ && $\!\!\!\!(\frac{D-4}{D-3}) a'\!\!\!\!$ 
&& $0$ & \cr
\omit&height2pt&\omit&&\omit&&\omit&&\omit&&\omit&&\omit&\cr
\tablerule
\omit&height2pt&\omit&&\omit&&\omit&&\omit&&\omit&&\omit&\cr
&& 3 && 2 && a && $\!\!\!\!-(\frac{D-1}{D-3}) a\!\!\!\!$ && 
$\!\!\!\! -2 (\frac{D-2}{D-3}) a\!\!\!\!$ && $0$ & \cr
\omit&height2pt&\omit&&\omit&&\omit&&\omit&&\omit&&\omit&\cr
\tablerule
\omit&height2pt&\omit&&\omit&&\omit&&\omit&&\omit&&\omit&\cr
&& 3 && 3 && a && $0$ && $0$ && $\!\!\!\!-\frac{(D-1)(D-4)}{4 (D-3)}\!\!\!\!$ 
& \cr
\omit&height2pt&\omit&&\omit&&\omit&&\omit&&\omit&&\omit&\cr
\tablerule
\omit&height2pt&\omit&&\omit&&\omit&&\omit&&\omit&&\omit&\cr
&& 3 && 3 && b && $0$ && $\!\!\!\! \frac12 (\frac{D-4}{D-3}) a\!\!\!\!$
&& $0$ & \cr
\omit&height2pt&\omit&&\omit&&\omit&&\omit&&\omit&&\omit&\cr
\tablerule
\omit&height2pt&\omit&&\omit&&\omit&&\omit&&\omit&&\omit&\cr
&& 3 && 3 && c && $0$ && $\!\!\!\! -\frac12 (\frac{D-4}{D-3}) a'\!\!\!\!$
&& $0$ & \cr
\omit&height2pt&\omit&&\omit&&\omit&&\omit&&\omit&&\omit&\cr
\tablerule}}

\caption{$i\d\!\D_{A}$ terms in which some derivatives act upon the
scale factors of the first series. The factor 
$\frac{i \kappa^2 H^2}{2^6 \pi^D} \Gamma(\frac{D}2 {\scriptstyle +1}) 
\Gamma(\frac{D}2) (a a')^{2- \frac{D}2}$ multiplies all contributions.}

\label{DAlessa}

\end{table}

\begin{table}

\vbox{\tabskip=0pt \offinterlineskip
\def\tablerule{\noalign{\hrule}}
\halign to390pt {\strut#& \vrule#\tabskip=1em plus2em&
\hfil#\hfil& \vrule#& \hfil#\hfil& \vrule#& \hfil#\hfil& 
\vrule#& \hfil#\hfil& \vrule#& \hfil#\hfil& \vrule#& \hfil#\hfil& 
\vrule#\tabskip=0pt\cr
\tablerule
\omit&height4pt&\omit&&\omit&&\omit&&\omit&&\omit&&\omit&\cr
&& $\!\!\!\!{\rm I}\!\!\!\!\!\!$ && $\!\!\!\!{\rm J}\!\!\!\!\!\!$ && 
$\!\!\!\!{\rm sub}\!\!\!\!\!\!$ && $\!\!\!\frac{H \gamma^0}{\Delta x^{D}}
\!\!\!\!$ && $\!\!\!\!\frac{H \gamma^i \Delta x_i \gamma^{\mu} \Delta x_{\mu} 
\gamma^0}{\Delta x^{D+2}}\!\!\!\!$ && $\!\!\!\! \partial^2 
(\frac{H \gamma^0}{\Delta x^{D-2}}) \!\!\!\!$ &\cr
\omit&height4pt&\omit&&\omit&&\omit&&\omit&&\omit&&\omit&\cr
\tablerule
\omit&height2pt&\omit&&\omit&&\omit&&\omit&&\omit&&\omit&\cr
&& 2 && 1 && \omit && $\!\!\!\!-2(\frac{D-1}{D-3}) a'\!\!\!\!$ 
&& $\!\!\!\!-(\frac{2D}{D-3}) a'\!\!\!\!$ && $0$ & \cr
\omit&height2pt&\omit&&\omit&&\omit&&\omit&&\omit&&\omit&\cr
\tablerule
\omit&height2pt&\omit&&\omit&&\omit&&\omit&&\omit&&\omit&\cr
&& 2 && 3 && a && $(\frac{D-1}{D-3}) a'$ && $(\frac{D}{D-3}) a'$ && $0$ & \cr
\omit&height2pt&\omit&&\omit&&\omit&&\omit&&\omit&&\omit&\cr
\tablerule
\omit&height2pt&\omit&&\omit&&\omit&&\omit&&\omit&&\omit&\cr
&& 3 && 1 && a && $0$ && $0$ && $\!\!\!\!\frac{(D-1) \, a}{(D-2)(D-3)}\!\!\!\!$ 
&\cr
\omit&height2pt&\omit&&\omit&&\omit&&\omit&&\omit&&\omit&\cr
\tablerule
\omit&height2pt&\omit&&\omit&&\omit&&\omit&&\omit&&\omit&\cr
&& 3 && 2 && a && $(\frac{D-1}{D-3}) a$ && $\!\!\!\!(\frac{D}{D-3}) a\!\!\!\!$ 
&& $0$ & \cr
\omit&height2pt&\omit&&\omit&&\omit&&\omit&&\omit&&\omit&\cr
\tablerule}}

\caption{$i\d\!\D_{A}$ terms in which some derivatives act upon the
scale factors of the second series. All contributions are multiplied by 
$\frac{i \kappa^2 H^{D-2}}{2^{D+2} \pi^D} \Gamma({\scriptstyle D-1})$.}

\label{DAlessb}

\end{table}

The final class is comprised of terms in which one or more derivatives
act upon a scale factor. Within this class we report contributions
from the first series in Table~\ref{DAlessa} and contributions from the
second series in Table~\ref{DAlessb}. Each nonzero entry in the 4th and 
5th columns of Table~\ref{DAlessa} diverges logarithmically like 
$1/\Delta x^{2D-4}$. However, the sum in each case results in an
additional factor of $a\!-\!a' \!=\! a a' H \Delta \eta$ which makes 
the contribution from Table~\ref{DAlessa} integrable,
\begin{eqnarray}
\lefteqn{-i\Bigl[\Sigma^{T\ref{DAlessa}}\Bigr](x;x') = 
\frac{i \kappa^2 H^4}{2^6 \pi^D} \Gamma\Bigl(\frac{D}2 \!+\! 1\Bigr)
\Gamma\Bigl(\frac{D}2\Bigr) (a a')^{3-\frac{D}2} \Biggl\{ -\Bigl(\frac{D\!-\!1
}{D\!-\!3}\Bigr) \frac{\gamma^0 \Delta \eta}{\Delta x^{2D-4}} } \nonumber \\
& & \hspace{1.5cm} -\frac12 \Bigl(\frac{3D\!-\!4}{D\!-\!3}\Bigr)
\frac{\gamma^i \Delta x_i \gamma^{\mu} \Delta x_{\mu} \gamma^0 \Delta \eta}{
\Delta x^{2D-2}} + \frac{(D\!-\!1)(D\!-\!4)}{4(D\!-\!3)}
\frac{\gamma^{\mu} \Delta x_{\mu}}{\Delta x^{2D-4}} \Biggr\} . \qquad
\end{eqnarray}
This is another example of the fact that the self-energy is odd under
interchange of $x^{\mu}$ and $x^{\prime \mu}$. 

The same thing happens with the contribution from Table~\ref{DAlessb},
\begin{eqnarray}
\lefteqn{-i\Bigl[\Sigma^{T\ref{DAlessb}}\Bigr](x;x') = 
\frac{i \kappa^2 H^D}{2^{D+2} \pi^D} \Gamma(D-1) a a' \Biggl\{ 
\Bigl(\frac{D\!-\!1}{D\!-\!3}\Bigr) \frac{\gamma^0 \Delta \eta}{\Delta x^{D}} }
\nonumber \\
& & \hspace{1.3cm} + \Bigl(\frac{D}{D\!-\!3}\Bigr) \frac{\gamma^i \Delta x_i 
\gamma^{\mu} \Delta x_{\mu} \gamma^0 \Delta \eta}{\Delta x^{D+2}} + 
\gamma^0 \Bigl(\frac{D\!-\!1}{D\!-\!3}\Bigr) \frac{i 2 \pi^{\frac{D}2}}{
\Gamma(\frac{D}2)} \frac{\delta^D(x\!-\!x')}{H a} \Biggr\} . \qquad
\end{eqnarray}
We can therefore set $D\!=\!4$, at which point the two Tables cancel except
for the delta function term,
\begin{equation}
-i\Bigl[\Sigma^{T\ref{DAlessa}+\ref{DAlessb}}\Bigr](x;x') =
\frac{\kappa^2 H^{D-2}}{(4\pi)^{\frac{D}2}} \frac{\Gamma(D\!-\!1)}{
\Gamma(\frac{D}2)} \times -\frac12 \Bigl(\frac{D\!-\!1}{D\!-\!3}\Bigr) 
a H \gamma^0 \delta^D(x\!-\!x') \!+\! O(D\!-\!4) . \label{6thcon}
\end{equation}
It is worth commenting that this term violates the reflection symmetry
(\ref{refl}). In $D\!=\!4$ it cancels the similar term in (\ref{1stcon}).

\subsection{Sub-Leading Contributions from $i{\delta \! \Delta}_B$}

In this subsection we work out the contribution from substituting the 
residual $B$-type part of the graviton propagator in Table~\ref{gen3},
\begin{equation}
i\Bigl[{}_{\alpha\beta} \Delta_{\rho\sigma}\Bigr] \longrightarrow -\Bigl[
\delta^0_{\alpha} \delta^0_{\sigma} \overline{\eta}_{\beta \rho} +
\delta^0_{\alpha} \delta^0_{\rho} \overline{\eta}_{\beta \sigma} +
\delta^0_{\beta} \delta^0_{\sigma} \overline{\eta}_{\alpha \rho} +
\delta^0_{\beta} \delta^0_{\rho} \overline{\eta}_{\alpha \sigma} \Bigr] 
i\delta\!\Delta_B \; . \label{DBpart}
\end{equation}
As in the two previous sub-sections we first make the requisite contractions 
and then act the derivatives. The result of this first step is summarized in 
Table~\ref{DBcon}. We have sometimes broken the result for a single vertex 
pair into parts because the four different tensors in (\ref{DBpart}) can make 
distinct contributions, and because distinct contributions also come from 
breaking up factors of $\gamma^{\alpha} J^{\beta \mu}$. These distinct 
contributions are labeled by subscripts $a$, $b$, $c$, etc. 

\begin{table}

\vbox{\tabskip=0pt \offinterlineskip
\def\tablerule{\noalign{\hrule}}
\halign to390pt {\strut#& \vrule#\tabskip=1em plus2em&
\hfil#\hfil& \vrule#& \hfil#\hfil& \vrule#& \hfil#\hfil& \vrule#& \hfil#\hfil&
\vrule#\tabskip=0pt\cr
\tablerule
\omit&height4pt&\omit&&\omit&&\omit&&\omit&\cr
&&$\!\!\!\!{\rm I}\!\!\!\!$ && $\!\!\!\!{\rm J} \!\!\!\!$ && 
$\!\!\!\! {\rm sub} \!\!\!\!$ &&
$iV_I^{\alpha\beta}(x) \, i[S](x;x') \, i V_J^{\rho\sigma}(x') 
\, [\mbox{}_{\alpha\beta} T^B_{\rho\sigma}] \, i\delta\!\Delta_B(x;x')$ &\cr
\omit&height4pt&\omit&&\omit&&\omit&&\omit&\cr
\tablerule
\omit&height2pt&\omit&&\omit&&\omit&&\omit&\cr
&& 2 && 1 && \omit && $0$ & \cr
\omit&height2pt&\omit&&\omit&&\omit&&\omit&\cr
\tablerule
\omit&height2pt&\omit&&\omit&&\omit&&\omit&\cr
&& 2 && 2 && a && $-\f{1}{2}\ka^2\del^{'}_0 \{
 \g^{(0}\del^{k)} i[S](x;x')\g_{k} i\d\!\D_{B}(x;x') \} $ & \cr
\omit&height2pt&\omit&&\omit&&\omit&&\omit&\cr
\tablerule
\omit&height2pt&\omit&&\omit&&\omit&&\omit&\cr
&& 2 && 2 && b && $-\f{1}{2}\ka^2\del_k \{ \g^{(0}\del^{k)}
i[S](x;x')\g^{0} i\d\!\D_{B}(x;x')\}$ & \cr
\omit&height2pt&\omit&&\omit&&\omit&&\omit&\cr
\tablerule
\omit&height2pt&\omit&&\omit&&\omit&&\omit&\cr
&& 2 && 3 && a && $-\f{1}{8}\ka^2\g_k\del_{0}i[S](x;x')
\g^k \, \del^{'}_{0}i\d\!\D_{B}(x;x')$ & \cr
\omit&height2pt&\omit&&\omit&&\omit&&\omit&\cr
\tablerule
\omit&height2pt&\omit&&\omit&&\omit&&\omit&\cr
&& 2 && 3 && b && $\f{1}{8}\ka^2\g ^0\del^{'}_{0} 
i\d\!\D_{B}(x;x') \, \del_{k}i[S](x;x')\g^{k} $ & \cr
\omit&height2pt&\omit&&\omit&&\omit&&\omit&\cr
\tablerule
\omit&height2pt&\omit&&\omit&&\omit&&\omit&\cr
&& 2 && 3 && c && $-\f{1}{8}\ka^2\g^k\del_{k} 
i\d\!\D_{B}(x;x') \, \del_{0}i[S](x;x')\g^{0} $ & \cr
\omit&height2pt&\omit&&\omit&&\omit&&\omit&\cr
\tablerule
\omit&height2pt&\omit&&\omit&&\omit&&\omit&\cr
&& 2 && 3 && d && $\f{1}{8}\ka^2\g^0\del^{k}i[S](x;x')
\g^0 \, \del_{k}i\d\!\D_{B}(x;x') $ & \cr
\omit&height2pt&\omit&&\omit&&\omit&&\omit&\cr
\tablerule
\omit&height2pt&\omit&&\omit&&\omit&&\omit&\cr
&& 3 && 1 && \omit && $0$ & \cr
\omit&height2pt&\omit&&\omit&&\omit&&\omit&\cr
\tablerule
\omit&height2pt&\omit&&\omit&&\omit&&\omit&\cr
&& 3 && 2 && a && $\f{1}{8}\ka^2\del^{'}_{0}\{\g^{k} i[S](x;x')
\g_k \, \del_{0}i\d\!\D_{B}(x;x')\}$ & \cr
\omit&height2pt&\omit&&\omit&&\omit&&\omit&\cr
\tablerule
\omit&height2pt&\omit&&\omit&&\omit&&\omit&\cr
&& 3 && 2 && b && $\f{1}{8}\ka^2\g^k\del_{k}\{ i[S](x;x')
\g^0 \, \del_{0}i\d\!\D_{B}(x;x')\} $ & \cr
\omit&height2pt&\omit&&\omit&&\omit&&\omit&\cr
\tablerule
\omit&height2pt&\omit&&\omit&&\omit&&\omit&\cr
&& 3 && 2 && c && $-\f{1}{8}\ka^2\g^0\del^{'}_{0}\{ i[S](x;x')
\g^k \, \del_{k}i\d\!\D_{B}(x;x')\} $ & \cr
\omit&height2pt&\omit&&\omit&&\omit&&\omit&\cr
\tablerule
\omit&height2pt&\omit&&\omit&&\omit&&\omit&\cr
&& 3 && 2 && d && $-\f{1}{8}\ka^2\del_{k}\{\g^{0} i[S](x;x')
\g^0 \, \del^{k}i\d\!\D_{B}(x;x')\}$ & \cr
\omit&height2pt&\omit&&\omit&&\omit&&\omit&\cr
\tablerule
\omit&height2pt&\omit&&\omit&&\omit&&\omit&\cr
&& 3 && 3 && a && $-\f{1}{16}\ka^2\g_{k}i[S](x;x')\g^{k}
\del_0\del^{'}_{0}i\d\!\D_{B}(x;x') $ & \cr
\omit&height2pt&\omit&&\omit&&\omit&&\omit&\cr
\tablerule
\omit&height2pt&\omit&&\omit&&\omit&&\omit&\cr
&& 3 && 3 && b && $\f{1}{16}\ka^2\g^{0}i[S](x;x')\g^k
\, \del_k\del^{'}_{0}i\d\!\D_{B}(x;x') $ & \cr
\omit&height2pt&\omit&&\omit&&\omit&&\omit&\cr
\tablerule
\omit&height2pt&\omit&&\omit&&\omit&&\omit&\cr
&& 3 && 3 && c && $-\f{1}{16}\ka^2\g^{k}i[S](x;x')\g^0
\, \del_0\del_{k}i\d\!\D_{B}(x;x') $ & \cr
\omit&height2pt&\omit&&\omit&&\omit&&\omit&\cr
\tablerule
\omit&height2pt&\omit&&\omit&&\omit&&\omit&\cr
&& 3 && 3 && d && $\f{1}{16}\ka^2\g^{0}i[S](x;x')\g^0
\nabla^2 i\d\!\D_{B}(x;x') $ & \cr
\omit&height2pt&\omit&&\omit&&\omit&&\omit&\cr
\tablerule}}

\caption{Contractions from the $i\d\!\D_B$ part of the graviton
propagator.}

\label{DBcon}

\end{table}

$i\delta\!\Delta_B(x;x')$ is the residual of the $B$-type propagator 
(\ref{DeltaB}) after the conformal contribution has
been subtracted,
\begin{eqnarray}
\lefteqn{i\delta\!\Delta_B(x;x') = \frac{H^2 \Gamma(\frac{D}2)}{16 
\pi^{\frac{D}2}} \frac{(a a')^{2-\frac{D}2}}{\Delta x^{D-4}} 
-\frac{H^{D-2}}{(4\pi)^\frac{D}2} \frac{\Gamma(D\!-\!2)}{\Gamma\Bigl(\frac{D}2 
\Bigr)} } \nonumber \\
& & \hspace{2cm} + \frac{H^{D-2}}{(4 \pi)^{\frac{D}2}} \sum_{n=1}^{\infty} 
\left\{ \frac{\Gamma(n \!+\!  \frac{D}2)}{\Gamma(n \!+\! 2)} \Bigl( \frac{y}4 
\Bigr)^{n - \frac{D}2 +2} - \frac{\Gamma(n \!+\! D \!-\! 2)}{\Gamma(n \!+\! 
\frac{D}2)} \Bigl(\frac{y}4 \Bigr)^n \right\} \; . \qquad \label{dB}
\end{eqnarray}
As was the case for the $i\delta\!\Delta_A(x;x')$ contributions considered 
in the previous sub-section, this diagram is not sufficiently singular for
the infinite series terms from $i\delta\!\Delta_B(x;x')$ to make a nonzero
contribution in the $D\!=\!4$ limit. Unlike $i\delta\!\Delta_A(x;x')$, even
the $n\!=\!0$ terms of $i\delta\!\Delta_B(x;x')$ vanish for $D\!=\!4$. This
means they can only contribute when multiplied by a divergence.

\begin{table}

\vbox{\tabskip=0pt \offinterlineskip
\def\tablerule{\noalign{\hrule}}
\halign to390pt {\strut#& \vrule#\tabskip=1em plus2em&
\hfil#\hfil& \vrule#& \hfil#\hfil& \vrule#& \hfil#\hfil& \vrule#& \hfil#\hfil& 
\vrule#& \hfil#\hfil& \vrule#& \hfil#\hfil& \vrule#& \hfil#\hfil&
\vrule#\tabskip=0pt\cr
\tablerule
\omit&height4pt&\omit&&\omit&&\omit&&\omit&&\omit&&\omit&&\omit&\cr
&& $\!\!\!\!{\rm I}\!\!\!\!\!\!$ && $\!\!\!\!{\rm J}\!\!\!\!\!\!$ && 
$\!\!\!\!{\rm sub}\!\!\!\!\!\!$ && 
$\!\!\!\!\frac{\gamma^{\mu} \Delta x_{\mu}}{\Delta x^{2D-2}}\!\!\!\!$ &&
$\!\!\!\!\frac{\gamma^i \Delta x_i}{\Delta x^{2D-2}}\!\!\!\!$ &&
$\!\!\!\frac{\Vert \Delta \vec{x}\Vert^2 \gamma^{\mu} \Delta x_{\mu}}{\Delta 
x^{2D}}\!\!\!\!$ &&
$\!\!\!\!\frac{\Vert \Delta \vec{x}\Vert^2 \gamma^i \Delta x_i}{\Delta 
x^{2D}}\!\!\!\!$ &\cr
\omit&height4pt&\omit&&\omit&&\omit&&\omit&&\omit&&\omit&&\omit&\cr
\tablerule
\omit&height2pt&\omit&&\omit&&\omit&&\omit&&\omit&&\omit&&\omit&\cr
&& 2 && 3 && a && ${\scriptstyle (D-1)^2}$ && ${\scriptstyle -(D+1)}$ 
&& ${\scriptstyle -D (D-1)}$ && ${\scriptstyle 2D}$ & \cr
\omit&height2pt&\omit&&\omit&&\omit&&\omit&&\omit&&\omit&&\omit&\cr
\tablerule
\omit&height2pt&\omit&&\omit&&\omit&&\omit&&\omit&&\omit&&\omit&\cr
&& 2 && 3 && b && ${\scriptstyle (D-1)}$ && ${\scriptstyle -2D + 1}$ 
&& ${\scriptstyle -D}$ && ${\scriptstyle 2D}$ & \cr
\omit&height2pt&\omit&&\omit&&\omit&&\omit&&\omit&&\omit&&\omit&\cr
\tablerule
\omit&height2pt&\omit&&\omit&&\omit&&\omit&&\omit&&\omit&&\omit&\cr
&& 2 && 3 && c && ${\scriptstyle 0}$ && ${\scriptstyle -(D-1)}$ && 
${\scriptstyle -D}$ && ${\scriptstyle 2D}$ & \cr
\omit&height2pt&\omit&&\omit&&\omit&&\omit&&\omit&&\omit&&\omit&\cr
\tablerule
\omit&height2pt&\omit&&\omit&&\omit&&\omit&&\omit&&\omit&&\omit&\cr
&& 2 && 3 && d && ${\scriptstyle 0}$ && ${\scriptstyle -1}$ && 
${\scriptstyle -D}$ && ${\scriptstyle 2D}$ & \cr
\omit&height2pt&\omit&&\omit&&\omit&&\omit&&\omit&&\omit&&\omit&\cr
\tablerule
\omit&height2pt&\omit&&\omit&&\omit&&\omit&&\omit&&\omit&&\omit&\cr
&& 3 && 2 && a && $\!\!\!\!{\scriptstyle -2(D-1)(D-2)}\!\!\!\!$ && 
${\scriptstyle 3D-5}$ && ${\scriptstyle 2 (D-1)^2}$ && 
$\!\!\!\!{\scriptstyle -4(D-1)}$ & \cr
\omit&height2pt&\omit&&\omit&&\omit&&\omit&&\omit&&\omit&&\omit&\cr
\tablerule
\omit&height2pt&\omit&&\omit&&\omit&&\omit&&\omit&&\omit&&\omit&\cr
&& 3 && 2 && b && $\!\!\!\!{\scriptstyle -(D-1)}\!\!\!\!$ && ${\scriptstyle 
3 (D-1)}$ && ${\scriptstyle 2 (D-1)}$ && $\!\!\!\!{\scriptstyle -4(D-1)}$ & \cr
\omit&height2pt&\omit&&\omit&&\omit&&\omit&&\omit&&\omit&&\omit&\cr
\tablerule
\omit&height2pt&\omit&&\omit&&\omit&&\omit&&\omit&&\omit&&\omit&\cr
&& 3 && 2 && c && $\!\!\!\!{\scriptstyle 0}\!\!\!\!$ && ${\scriptstyle 
2D -3}$ && ${\scriptstyle 2 (D-1)}$ && $\!\!\!\!{\scriptstyle -4(D-1)}$ & \cr
\omit&height2pt&\omit&&\omit&&\omit&&\omit&&\omit&&\omit&&\omit&\cr
\tablerule
\omit&height2pt&\omit&&\omit&&\omit&&\omit&&\omit&&\omit&&\omit&\cr
&& 3 && 2 && d && $\!\!\!\!{\scriptstyle -(D-1)}\!\!\!\!$ && ${\scriptstyle 
2D -1}$ && $\!\!\!\!{\scriptstyle 2 (D-1)}\!\!\!\!$ && 
$\!\!\!\!{\scriptstyle -4(D-1)}$ & \cr
\omit&height2pt&\omit&&\omit&&\omit&&\omit&&\omit&&\omit&&\omit&\cr
\tablerule
\omit&height2pt&\omit&&\omit&&\omit&&\omit&&\omit&&\omit&&\omit&\cr
&& 3 && 3 && a && $\!\!\!\!{\scriptstyle \frac12 (D-1)(D-3)}\!\!\!\!$ && 
$\!\!\!\!{\scriptstyle -(D-3)}\!\!\!\!$ && $\!\!\!\!{\scriptstyle -\frac12
(D-1)(D-2)}\!\!\!\!$ && $\!\!\!\!{\scriptstyle (D-2)}\!\!\!\!$ & \cr
\omit&height2pt&\omit&&\omit&&\omit&&\omit&&\omit&&\omit&&\omit&\cr
\tablerule
\omit&height2pt&\omit&&\omit&&\omit&&\omit&&\omit&&\omit&&\omit&\cr
&& 3 && 3 && b && $\!\!\!\!{\scriptstyle 0}\!\!\!\!$ && $\!\!\!\!
{\scriptstyle -\frac12 (D-2)}\!\!\!\!$ && $\!\!\!\!{\scriptstyle -\frac12
(D-2)}\!\!\!\!$ && $\!\!\!\!{\scriptstyle (D-2)}\!\!\!\!$ & \cr
\omit&height2pt&\omit&&\omit&&\omit&&\omit&&\omit&&\omit&&\omit&\cr
\tablerule
\omit&height2pt&\omit&&\omit&&\omit&&\omit&&\omit&&\omit&&\omit&\cr
&& 3 && 3 && c && $\!\!\!\!{\scriptstyle 0}\!\!\!\!$ && $\!\!\!\!
{\scriptstyle -\frac12 (D-2)}\!\!\!\!$ && $\!\!\!\!{\scriptstyle -\frac12
(D-2)}\!\!\!\!$ && $\!\!\!\!{\scriptstyle (D-2)}\!\!\!\!$ & \cr
\omit&height2pt&\omit&&\omit&&\omit&&\omit&&\omit&&\omit&&\omit&\cr
\tablerule
\omit&height2pt&\omit&&\omit&&\omit&&\omit&&\omit&&\omit&&\omit&\cr
&& 3 && 3 && d && ${\scriptstyle \frac12 (D-1)}$ && ${\scriptstyle -(D-1)}$
&& ${\scriptstyle -\frac12 (D-2)}$ && ${\scriptstyle (D-2)}$ & \cr
\omit&height2pt&\omit&&\omit&&\omit&&\omit&&\omit&&\omit&&\omit&\cr
\tablerule
\omit&height2pt&\omit&&\omit&&\omit&&\omit&&\omit&&\omit&&\omit&\cr
\tablerule
\omit&height2pt&\omit&&\omit&&\omit&&\omit&&\omit&&\omit&&\omit&\cr
&& $\!\!\!\!{\rm Total}\!\!\!\!$ && \omit && \omit && $\!\!\!\!{\scriptstyle 
-\frac12 (D-1)(D-2)} \!\!\!\!$ && $\!\!\!\!{\scriptstyle 3 (D-2)}\!\!\!\!$ 
&& $\!\!\!\! {\scriptstyle \frac12 (D+2)(D-2)}\!\!\!\!$ && $\!\!\!\!
{\scriptstyle -4(D-2)}\!\!\!\!$ &\cr
\omit&height2pt&\omit&&\omit&&\omit&&\omit&&\omit&&\omit&&\omit&\cr
\tablerule}}

\caption{$i\delta\!\Delta_B$ terms in which all derivatives act upon 
$\Delta x^2(x;x')$. All contributions are multiplied by $\frac{i \kappa^2 
H^2}{2^8 \pi^D} \Gamma^2(\frac{D}2) (D\!-\!4) (a a')^{2- \frac{D}2}$. }

\label{DBmost}

\end{table}

Contributions from the $[2\!\!-\!\!2]$ vertex pair require special treatment
to take advantage of the cancellation between the two series. We will work
out the ``a'' term from Table~\ref{DBcon},
\begin{eqnarray}
\lefteqn{\Bigl[2\!\!-\!\!2\Bigr]_a = -\frac{i \kappa^2 \Gamma(\frac{D}2\!-\!1)
}{16 \pi^{\frac{D}2}} \partial_0' \Biggl\{ i\delta\!\Delta_B(x;x') \Bigl(
\gamma^0 \hspace{-.1cm} \not{\hspace{-.1cm} \partial} \hspace{-.1cm} \; 
\overline{\not{\hspace{-.1cm} \partial}} \!-\! \gamma^i \hspace{-.1cm} 
\not{\hspace{-.1cm} \partial} \gamma^i \partial_0\Bigr) \Bigl[\frac1{\Delta 
x^{D-2}} \Bigr] \Biggr\} , } \\
& & \hspace{-.5cm} = \frac{i \kappa^2 \Gamma(\frac{D}2\!-\!1)}{16 
\pi^{\frac{D}2}} \partial_0' \Biggl\{ i\delta\!\Delta_B(x;x') \Bigl(-3 
\partial_0 \; \hspace{-.1cm} \overline{\not{\hspace{-.1cm} \partial}} \!+\! 
\gamma^0 \nabla^2 \!+\! (D\!-\!1) \hspace{-.1cm} \not{\hspace{-.1cm} \partial} 
\partial_0\Bigr) \Bigl[\frac1{\Delta x^{D-2}} \Bigr] \Biggr\} . \qquad
\end{eqnarray}
A key identity for reducing the $[2\!-\!2]$ terms involves commuting two
derivatives through $1/\Delta x^{D-4}$,
\begin{equation}
\frac1{\Delta x^{D-4}} \partial_{\mu} \partial_{\nu} \Bigl[\frac1{\Delta 
x^{D-2}}\Bigr] = \frac1{4(D\!-\!3)} \Bigl( -\eta_{\mu\nu} \partial^2 \!+\!
D \partial_{\mu} \partial_{\nu} \Bigr) \Bigl[\frac1{\Delta x^{2D-6}}\Bigr] \; .
\label{keyID}
\end{equation}
This can be used to extract the derivatives from the first term of
$i\delta\!\Delta_B(x;x')$, at which point the result is integrable and we can
take $D\!=\!4$,
\begin{eqnarray}
\lefteqn{\Bigl[2\!\!-\!\!2\Bigr]^1_a = \frac{i \kappa^2 H^2}{2^8 \pi^D}
\Gamma\Bigl(\frac{D}2\Bigr) \Gamma\Bigl(\frac{D}2\!-\!1\Bigr) }
\nonumber \\
& & \hspace{2cm} \times \partial_0' \Biggl\{ \frac{(a a')^{2-\frac{D}2}}{\Delta
x^{D-4}} \Bigl(-3 \partial_0 \; \hspace{-.1cm} \overline{\not{\hspace{-.1cm} 
\partial}} \!+\! \gamma^0 \nabla^2 \!+\! (D\!-\!1) \hspace{-.1cm} \not{
\hspace{-.1cm} \partial} \partial_0\Bigr) \Bigl[\frac1{\Delta x^{D-2}} \Bigr] 
\Biggr\} , \qquad \\
& & = \frac{i \kappa^2 H^2}{2^9 \pi^D} \frac{\Gamma(\frac{D}2 \!+\!1) \Gamma(
\frac{D}2\!-\!1)}{D\!-\!3} (a a')^{2-\frac{D}2} \nonumber \\
& & \hspace{.5cm} \times \Bigl(-\partial_0 \!-\! \frac12 (D\!-\!4) H a'\Bigr) 
\Bigl(-3 \partial_0 \; \hspace{-.1cm} \overline{\not{\hspace{-.1cm} 
\partial}} \!+\! \gamma^0 \nabla^2 \!+\! (D\!-\!1) \hspace{-.1cm} \not{
\hspace{-.1cm} \partial} \partial_0\Bigr) \Bigl[\frac1{\Delta x^{2D-6}} 
\Bigr] , \qquad \\
& & = -\frac{i \kappa^2 H^2}{2^8 \pi^4} \gamma^0 \partial_0 \Bigl(3 
\partial_0^2 \!+\! \nabla^2\Bigr) \Bigl[\frac1{\Delta x^2}\Bigr] + 
O(D\!-\!4) \; .
\end{eqnarray}
Of course the second term of $i\delta\!\Delta_B$ is constant so the derivatives
are already extracted,
\begin{eqnarray}
\Bigl[2\!\!-\!\!2\Bigr]^2_a & = & \frac{i \kappa^2 H^{D-2}}{2^{D+3} \pi^D}
\frac{\Gamma(D\!-\!2)}{D\!-\!2} \partial_0 \Bigl(-3 \partial_0 \; \hspace{-.1cm}
\overline{\not{\hspace{-.1cm} \partial}} \!+\! \gamma^0 \nabla^2 \!+\! 
(D\!-\!1) \hspace{-.1cm} \not{\hspace{-.1cm} \partial} \partial_0\Bigr) 
\Bigl[\frac1{\Delta x^{D-2}} \Bigr] , \qquad \\
& = & \frac{i \kappa^2 H^2}{2^8 \pi^4} \gamma^0 \partial_0 \Bigl(3 
\partial_0^2 \!+\! \nabla^2\Bigr) \Bigl[\frac1{\Delta x^2}\Bigr] + 
O(D\!-\!4) \; .
\end{eqnarray}
Hence the total for $[2\!\!-\!\!2]_a$ is zero in $D\!=\!4$ dimensions!

The analogous result for the initial reduction of the other $[2\!\!-\!\!2]$ 
term is,
\begin{eqnarray}
\lefteqn{\Bigl[2\!\!-\!\!2\Bigr]_b = \frac{i \kappa^2 \Gamma(\frac{D}2\!-\!1)}{
16 \pi^{\frac{D}2}} } \nonumber \\
& & \times \partial_k \Biggl\{ i\delta\!\Delta_B(x;x') \Bigl(-\gamma^0 
\partial_0 \partial_k \!+\! \hspace{-.1cm} \overline{\not{\hspace{-.1cm} 
\partial}} \, \partial_k \!+\! \gamma_k \partial_0^2 \!+\! \gamma_k \; 
\hspace{-.1cm} \overline{\not{\hspace{-.1cm} \partial}} \gamma^0 \partial_0 
\Bigr) \Bigl[\frac1{\Delta x^{D-2}}\Bigr] \Biggr\} . \qquad
\end{eqnarray}
The results for each of the two terms of $i\delta\!\Delta_B$ are,
\newpage
\begin{eqnarray}
\Bigl[2\!\!-\!\!2\Bigr]^1_b & = & \frac{i \kappa^2 H^2}{2^9 \pi^D} 
\frac{\Gamma(\frac{D}2 \!+\!1) \Gamma(\frac{D}2\!-\!1)}{D\!-\!3} 
(a a')^{2-\frac{D}2} \nonumber \\
& & \hspace{3.5cm} \times \Bigl(-2 \gamma^0 \partial_0 \nabla^2 + 
\hspace{-.1cm} \overline{\not{\hspace{-.1cm} \partial}} \, \nabla^2 + 
\hspace{-.1cm} \overline{\not{\hspace{-.1cm} \partial}} \, \partial_0^2 \Bigr) 
\Bigl[\frac1{\Delta x^{2D-6}} \Bigr] , \qquad \\
& = & \frac{i \kappa^2 H^2}{2^8 \pi^4} \Bigl(-2 \gamma^0 \partial_0 \nabla^2 +
\hspace{-.1cm} \overline{\not{\hspace{-.1cm} \partial}} \, \nabla^2 + 
\hspace{-.1cm} \overline{\not{\hspace{-.1cm} \partial}} \, \partial_0^2 \Bigr) 
\Bigl[\frac1{\Delta x^2}\Bigr] + O(D\!-\!4) \; , \\
\Bigl[2\!\!-\!\!2\Bigr]^2_b & = & \frac{i \kappa^2 H^{D-2}}{2^{D+3} \pi^D}
\frac{\Gamma(D\!-\!2)}{D\!-\!2} \Bigl(2 \gamma^0 \partial_0 \nabla^2 -
\hspace{-.1cm} \overline{\not{\hspace{-.1cm} \partial}} \, \nabla^2 - 
\hspace{-.1cm} \overline{\not{\hspace{-.1cm} \partial}} \, \partial_0^2 \Bigr) 
\Bigl[\frac1{\Delta x^{D-2}} \Bigr] , \qquad \\
& = & \frac{i \kappa^2 H^2}{2^8 \pi^4} \Bigl(2 \gamma^0 \partial_0 \nabla^2 -
\hspace{-.1cm} \overline{\not{\hspace{-.1cm} \partial}} \, \nabla^2 - 
\hspace{-.1cm} \overline{\not{\hspace{-.1cm} \partial}} \, \partial_0^2 \Bigr) 
\Bigl[\frac1{\Delta x^2}\Bigr] + O(D\!-\!4) \; .
\end{eqnarray}
Hence the entire contribution from $[2\!\!-\!\!2]$ vanishes in $D\!=\!4$.

The lower vertex pairs all involve at least one derivative of 
$i\delta\!\Delta_B$,
\begin{eqnarray}
\partial_i i \delta\!\Delta_B(x;x') & = & -\frac{H^2 \Gamma(\frac{D}2)}{16 
\pi^{\frac{D}2}} (D\!-\!4) (a a')^{2-\frac{D}2} \, \frac{\Delta x^i}{
\Delta x^{D-2}} = -\partial_i' \delta\!\Delta_B(x;x') \; , \qquad \\
\partial_0 i \delta\!\Delta_B(x;x') & = & \frac{H^2 \Gamma(\frac{D}2)}{16 
\pi^{\frac{D}2}} (D\!-\!4) (a a')^{2-\frac{D}2} \Biggl\{ \frac{\Delta \eta}{
\Delta x^{D-2}} \!-\! \frac{a H}{2 \Delta x^{D-4}} \Biggr\} , \\
\partial_0' i \delta\!\Delta_B(x;x') & = & \frac{H^2 \Gamma(\frac{D}2)}{16
\pi^{\frac{D}2}} (D\!-\!4) (a a')^{2-\frac{D}2} \Biggl\{-\frac{\Delta \eta}{
\Delta x^{D-2}} \!-\! \frac{a' H}{2 \Delta x^{D-4}} \Biggr\} .
\end{eqnarray}
These reductions are very similar to those of the analogous $i\delta\!\Delta_A$
terms. We make use of the same gamma matrix identities 
(\ref{gammafirst}-\ref{gammalast}) that were used in the previous sub-section.
The only really new feature is that one sometimes encounters factors of
$\Delta \eta^2$ which we always resolve as,
\begin{equation}
\Delta \eta^2 = -\Delta x^2 + \Vert \Delta \vec{x} \Vert^2 \; .
\end{equation}
Table~\ref{DBmost} gives our results for the most singular contributions,
those in which all derivatives act upon the conformal coordinate separation
$\Delta x^2$.

The only really unexpected thing about Table~\ref{DBmost} is the overall
factor of $(D\!-\!2)$ common to each of the four sums,
\begin{eqnarray}
\lefteqn{-i \Bigl[\Sigma^{T\ref{DBmost}}\Bigr](x;x') = \frac{i \kappa^2 H^2}{
2^8 \pi^D} \Gamma^2\Bigl(\frac{D}2\Bigr) (D\!-\!2) (D\!-\!4) 
(a a')^{2-\frac{D}2} \Biggl\{ -\frac12 (D\!-\!1) \frac{\gamma^{\mu} 
\Delta x_{\mu}}{\Delta x^{2D-2}} } \nonumber \\
& & \hspace{2cm} + 3 \frac{\gamma^i \Delta x_i}{\Delta x^{2D-2}} + \frac12 
(D\!+\!2) \frac{\Vert \Delta \vec{x} \Vert^2 \gamma^{\mu} \Delta x_{\mu}}{
\Delta x^{2D}} - 4 \frac{\Vert \Delta \vec{x} \Vert^2 \gamma^i \Delta x_i}{
\Delta x^{2D}} \Biggr\} . \qquad
\end{eqnarray}
As with the result of Table~\ref{DAmostc}, we use the differential identities
(\ref{difID1}-\ref{difID2}) to prepare the last two terms for partial
integration,
\begin{eqnarray}
\lefteqn{-i \Bigl[\Sigma^{T\ref{DBmost}}\Bigr](x;x') = \frac{i \kappa^2 H^2}{
2^8 \pi^D} \Gamma^2\Bigl(\frac{D}2\Bigr) (D\!-\!2) (D\!-\!4) 
(a a')^{2-\frac{D}2} } \nonumber \\
& & \times \Biggl\{ -\frac14 (D\!-\!4) \frac{\gamma^{\mu} \Delta x_{\mu}}{
\Delta x^{2D-2}} + \frac12 \Bigl( \frac{3 D \!-\! 8}{D\!-\!1}\Bigr) 
\frac{\gamma^i \Delta x_i}{\Delta x^{2D-2}} \nonumber \\
& & \hspace{1cm} + \frac{(D\!+\!2) \, \nabla^2}{8 (D\!-\!1) 
(D\!-\!2)} \Bigl(\frac{\gamma^{\mu} \Delta x_{\mu}}{\Delta x^{2D -4}} \Bigr)
- \frac{\nabla^2} {(D\!-\!1)(D\!-\!2)} \Bigl(\frac{\gamma^i \Delta x_i}{
\Delta x^{2D-4}} \Bigr) \Biggr\} , \qquad \\
& & = \frac{i \kappa^2 H^2}{2^8 \pi^D} \Gamma^2\Bigl(\frac{D}2\Bigr) 
(a a')^{2-\frac{D}2} \Biggl\{ \frac1{16} \Bigl(\frac{D\!-\!4}{D\!-\!3}\Bigr)
\hspace{-.1cm} \not{\hspace{-.1cm} \partial} \partial^2 - \frac18 \frac{(3D
\!-\!8)}{(D\!-\!1) (D\!-\!3)} \; \hspace{-.1cm} \overline{\not{\hspace{-.1cm} 
\partial}} \, \partial^2 \nonumber \\
& & \hspace{1cm} - \frac1{16} \frac{(D\!+\!2) (D\!-\!4)}{(D\!-\!1) (D\!-\!3)}
\hspace{-.1cm} \not{\hspace{-.1cm} \partial} \nabla^2 + \frac12 \frac{(D
\!-\!4)}{(D\!-\!1) (D\!-\!3)} \; \hspace{-.1cm} \overline{\not{\hspace{-.1cm} 
\partial}} \, \nabla^2 \Biggr\} \frac1{\Delta x^{2D-6}} . 
\end{eqnarray}
The expression is now integrable so we can take $D\!=\!4$,
\begin{equation}
-i \Bigl[\Sigma^{T\ref{DBmost}}\Bigr](x;x') = \frac{i \kappa^2 H^2}{2^8 \pi^4}
\Bigl\{-\frac16 \; \hspace{-.1cm} \overline{\not{\hspace{-.1cm} 
\partial}} \, \partial^2 \Bigr\} \frac1{\Delta x^2} + O(D\!-\!4) \; . 
\label{7thcon}
\end{equation}

Unlike the $i\delta\!\Delta_A$ terms there is no net contribution when one or
more of the derivatives acts upon a scale factor. If both derivatives act on 
scale factors the result is integrable in $D\!=\!4$ dimensions, and vanishes 
owing to the factor of $(D\!-\!4)^2$ from differentiating both $a^{2-\frac{D}2}$
and $a^{\prime 2 - \frac{D}2}$. If a single derivative acts upon a scale 
factor, the result is a factor of either $(D\!-\!4) a$ or $(D\!-\!4) a'$ 
times a term which is logarithmically divergent and {\it even} under 
interchange of $x^{\mu}$ and $x^{\prime \mu}$. As we have by now seen many 
times, the sum of all such terms contrives to obey reflection symmetry 
(\ref{refl}) by the separate extra factors of $(D\!-\!4) a$ and $(D\!-\!4) a'$ 
combining to give,
\begin{equation}
(D\!-\!4) (a - a') = (D\!-\!4) a a' H \Delta \eta \; .
\end{equation}
Of course this makes the sum integrable in $D\!=\!4$ dimensions, at which 
point we can take $D\!=\!4$ and the result vanishes on account of the overall
factor of $(D\!-\!4)$.

\subsection{Sub-Leading Contributions from $i{\delta \! \Delta}_C$}

The point of this subsection is to compute the contribution from replacing
the graviton propagator in Table~\ref{gen3} by its residual $C$-type part,
\begin{equation}
i\Bigl[{}_{\alpha\beta} \Delta_{\rho\sigma}\Bigr] \!\rightarrow \!2 \Biggl[
\frac{\eta_{\alpha\beta} \eta_{\rho\sigma}}{(D\!-\!2)(D\!-\!3)} \!+\!
\frac{\delta^0_{\alpha} \delta^0_{\beta} \eta_{\rho\sigma} \!+\! 
\eta_{\alpha\beta} \delta^0_{\rho} \delta^0_{\sigma}}{D\!-\!3} \!+\! 
\Bigl(\frac{D\!-\!2}{D\!-\!3}\Bigr) \delta^0_{\alpha} \delta^0_{\beta} 
\delta^0_{\rho} \delta^0_{\sigma} \Biggr] i\delta\!\Delta_C . \label{DCpart}
\end{equation}
As in the previous sub-sections we first make the requisite contractions 
and then act the derivatives. The result of this first step is summarized in 
Table~\ref{DCcon}. We have sometimes broken the result for a single vertex 
pair into parts because the four different tensors in (\ref{DCpart}) can make 
distinct contributions, and because distinct contributions also come from 
breaking up factors of $\gamma^{\alpha} J^{\beta \mu}$. These distinct 
contributions are labeled by subscripts $a$, $b$, $c$, etc. 

\begin{table}

\vbox{\tabskip=0pt \offinterlineskip
\def\tablerule{\noalign{\hrule}}
\halign to390pt {\strut#& \vrule#\tabskip=1em plus2em& 
\hfil#\hfil& \vrule#& \hfil#\hfil& \vrule#& \hfil#\hfil& \vrule#& \hfil#\hfil& 
\vrule#\tabskip=0pt\cr
\tablerule
\omit&height4pt&\omit&&\omit&&\omit&&\omit&\cr
&&$\!\!\!\! {\rm I}\!\!\!\!$ && $\!\!\!\! {\rm J} \!\!\!\!$ && 
$\!\!\!\! {\rm sub} \!\!\!\!$ &&
$iV_I^{\alpha\beta}(x) \, i[S](x;x') \, i V_J^{\rho\sigma}(x') 
\, [\mbox{}_{\alpha\beta} T^C_{\rho\sigma}] \, i\delta\!\Delta_C(x;x')$ &\cr
\omit&height4pt&\omit&&\omit&&\omit&&\omit&\cr
\tablerule
\omit&height2pt&\omit&&\omit&&\omit&&\omit&\cr
&& 2 && 1 && a && $-\frac1{(D-3)(D-2)} \kappa^2 \hspace{-.1cm} \not{\hspace{
-.1cm} \partial} \, \delta^D(x\!-\!x') \, i\delta\!\Delta_C(x;x)$ & \cr
\omit&height2pt&\omit&&\omit&&\omit&&\omit&\cr
\tablerule
\omit&height2pt&\omit&&\omit&&\omit&&\omit&\cr
&& 2 && 1 && b && $-\frac1{D-3} \kappa^2 \partial'_{\mu} \{ \gamma^0
\partial_0 \, i[S](x;x') \gamma^{\mu} \, i\delta\!\Delta_C(x;x') \}$ & \cr
\omit&height2pt&\omit&&\omit&&\omit&&\omit&\cr
\tablerule
\omit&height2pt&\omit&&\omit&&\omit&&\omit&\cr
&& 2 && 2 && a && $\frac1{2(D-3)(D-2)} \kappa^2 \hspace{-.1cm} 
\not{\hspace{-.1cm} \partial} \, \delta^D(x\!-\!x') \, 
i\delta\!\Delta_C(x;x)$ & \cr
\omit&height2pt&\omit&&\omit&&\omit&&\omit&\cr
\tablerule
\omit&height2pt&\omit&&\omit&&\omit&&\omit&\cr
&& 2 && 2 && b && $-\frac1{2(D-3)} \kappa^2 \gamma^0 \partial_0 \, 
\delta^D(x\!-\!x') \, i\delta\!\Delta_C(x;x)$ & \cr
\omit&height2pt&\omit&&\omit&&\omit&&\omit&\cr
\tablerule
\omit&height2pt&\omit&&\omit&&\omit&&\omit&\cr
&& 2 && 2 && c && $+\frac1{2(D-3)} \kappa^2 \partial'_{\mu} \{ \gamma^0
\partial_0 \, i[S](x;x') \gamma^{\mu} \, i\delta\!\Delta_C(x;x') \}$ & \cr
\omit&height2pt&\omit&&\omit&&\omit&&\omit&\cr
\tablerule
\omit&height2pt&\omit&&\omit&&\omit&&\omit&\cr
&& 2 && 2 && d && $-\frac12 (\frac{D-2}{D-3}) \kappa^2 \partial'_0 \{ 
\gamma^0 \partial_0 \, i[S](x;x') \gamma^0 \, i\delta\!\Delta_C(x;x') \}$ & \cr
\omit&height2pt&\omit&&\omit&&\omit&&\omit&\cr
\tablerule
\omit&height2pt&\omit&&\omit&&\omit&&\omit&\cr
&& 2 && 3 && a && $-\frac{(D-1)}{4(D-3)(D-2)} \kappa^2 \, \delta^D(x\!-\!x')
\, \hspace{-.1cm} \not{\hspace{-.1cm} \partial}' \,i \delta\!\Delta_C(x;x')$ 
& \cr
\omit&height2pt&\omit&&\omit&&\omit&&\omit&\cr
\tablerule
\omit&height2pt&\omit&&\omit&&\omit&&\omit&\cr
&& 2 && 3 && b && $+\frac1{4(D-3)} \kappa^2 \, \delta^D(x\!-\!x') \gamma^i 
\partial_i' \, i\delta\!\Delta_C(x;x')$ & \cr
\omit&height2pt&\omit&&\omit&&\omit&&\omit&\cr
\tablerule
\omit&height2pt&\omit&&\omit&&\omit&&\omit&\cr
&& 2 && 3 && c && $+\frac14 (\frac{D-1}{D-3}) \kappa^2 \gamma^0 \partial_0
\, i[S](x;x') \, \hspace{-.1cm} \not{\hspace{-.1cm} \partial}' \,i 
\delta\!\Delta_C(x;x')$ & \cr
\omit&height2pt&\omit&&\omit&&\omit&&\omit&\cr
\tablerule
\omit&height2pt&\omit&&\omit&&\omit&&\omit&\cr
&& 2 && 3 && d && $-\frac14 (\frac{D-2}{D-3}) \kappa^2 \gamma^0 \partial_0 
\, i[S](x;x') \gamma^i \partial_i' \, i\delta\!\Delta_C(x;x')$ & \cr
\omit&height2pt&\omit&&\omit&&\omit&&\omit&\cr
\tablerule
\omit&height2pt&\omit&&\omit&&\omit&&\omit&\cr
&& 3 && 1 && a && $-\frac{(D-1)}{2(D-3)(D-2)} \kappa^2 \partial'_{\mu} \{ \,
\hspace{-.1cm} \not{\hspace{-.1cm} \partial} \, i \delta\!\Delta_C(x;x') \,
i[S](x;x') \gamma^{\mu} \}$ & \cr
\omit&height2pt&\omit&&\omit&&\omit&&\omit&\cr
\tablerule
\omit&height2pt&\omit&&\omit&&\omit&&\omit&\cr
&& 3 && 1 && b && $+\frac1{2(D-3)} \kappa^2 \partial'_{\mu} \{ \gamma^i
\partial_i \,i \delta\!\Delta_C(x;x') \, i[S](x;x') \gamma^{\mu} \}$ & \cr
\omit&height2pt&\omit&&\omit&&\omit&&\omit&\cr
\tablerule
\omit&height2pt&\omit&&\omit&&\omit&&\omit&\cr
&& 3 && 2 && a && $\frac{(D-1)}{4(D-3)(D-2)} \kappa^2 \partial'_{\mu} \{ \,
\hspace{-.1cm} \not{\hspace{-.1cm} \partial} \, i \delta\!\Delta_C(x;x') \,
i[S](x;x') \gamma^{\mu} \}$ & \cr
\omit&height2pt&\omit&&\omit&&\omit&&\omit&\cr
\tablerule
\omit&height2pt&\omit&&\omit&&\omit&&\omit&\cr
&& 3 && 2 && b && $-\frac14 (\frac{D-1}{D-3}) \kappa^2 \partial'_0 \{ \,
\hspace{-.1cm} \not{\hspace{-.1cm} \partial} \, i \delta\!\Delta_C(x;x') \,
i[S](x;x') \gamma^0 \}$ & \cr
\omit&height2pt&\omit&&\omit&&\omit&&\omit&\cr
\tablerule
\omit&height2pt&\omit&&\omit&&\omit&&\omit&\cr
&& 3 && 2 && c && $- \frac1{4(D-3)} \kappa^2 \partial'_{\mu} \{ \gamma^i
\partial_i \,i \delta\!\Delta_C(x;x') \, i[S](x;x') \gamma^{\mu} \}$ & \cr
\omit&height2pt&\omit&&\omit&&\omit&&\omit&\cr
\tablerule
\omit&height2pt&\omit&&\omit&&\omit&&\omit&\cr
&& 3 && 2 && d && $+\frac14 (\frac{D-2}{D-3}) \kappa^2 \partial'_0 \{
\gamma^i \partial_i \,i \delta\!\Delta_C(x;x') \, i[S](x;x') \gamma^0 \}$ & \cr
\omit&height2pt&\omit&&\omit&&\omit&&\omit&\cr
\tablerule
\omit&height2pt&\omit&&\omit&&\omit&&\omit&\cr
&& 3 && 3 && a && $\frac{(D-1)^2}{8(D-3)(D-2)} \kappa^2 \gamma^{\mu} 
i[S](x;x') \partial_{\mu} \hspace{-.1cm} \not{\hspace{-.1cm} \partial}' \, 
i \delta\!\Delta_C(x;x')$ & \cr
\omit&height2pt&\omit&&\omit&&\omit&&\omit&\cr
\tablerule
\omit&height2pt&\omit&&\omit&&\omit&&\omit&\cr
&& 3 && 3 && b && $-\frac18 (\frac{D-1}{D-3}) \kappa^2 \gamma^{\mu} i[S](x;x')
\partial_{\mu} \gamma^j \partial_j' \, i \delta\!\Delta_C(x;x')$ & \cr
\omit&height2pt&\omit&&\omit&&\omit&&\omit&\cr
\tablerule
\omit&height2pt&\omit&&\omit&&\omit&&\omit&\cr
&& 3 && 3 && c && $-\frac18 (\frac{D-1}{D-3}) \kappa^2 \gamma^i i[S](x;x')
\partial_i \, \hspace{-.1cm} \not{\hspace{-.1cm} \partial}' \, 
i \delta\!\Delta_C(x;x')$ & \cr
\omit&height2pt&\omit&&\omit&&\omit&&\omit&\cr
\tablerule
\omit&height2pt&\omit&&\omit&&\omit&&\omit&\cr
&& 3 && 3 && d && $+\frac18 (\frac{D-2}{D-3}) \kappa^2 \gamma^i i[S](x;x')
\partial_i \gamma^j \partial_j' \,i \delta\!\Delta_C(x;x')$ & \cr
\omit&height2pt&\omit&&\omit&&\omit&&\omit&\cr
\tablerule}}

\caption{Contractions from the $i\delta\!\Delta_C$ part of the graviton 
propagator.}

\label{DCcon}

\end{table}

Here $i\delta\!\Delta_C(x;x')$ is the residual of the $C$-type propagator 
(\ref{DeltaC}) after the conformal contribution has been subtracted,
\begin{eqnarray}
\lefteqn{i \delta\!\Delta_C(x;x') = \frac{H^2}{16 \pi^{\frac{D}2}} 
\Bigl( \frac{D}2 \!-\! 3\Bigr) \Gamma\Bigl(\frac{D}2 \!-\! 1\Bigr) 
\frac{(a a')^{2-\frac{D}2}}{\Delta x^{D-4}}+ \frac{H^{D-2}}{(4\pi)^{\frac{D}2}}
\frac{\Gamma(D \!-\! 3)}{\Gamma(\frac{D}2)} } \nonumber \\
& & \hspace{-.7cm} - \frac{H^{D-2}}{(4\pi)^{\frac{D}2}} \!\!\sum_{n=1}^{\infty} 
\!\!\left\{ \!\!\Bigl(n \!-\! \frac{D}2 \!+\! 3\Bigr) \frac{\Gamma(n \!+\!
\frac{D}2 \!-\! 1)}{\Gamma(n \!+\! 2)} \Bigl(\frac{y}4 \Bigr)^{n -\frac{D}2 +2}
\!\!\!\!\!\!\! - (n\!+\!1) \frac{\Gamma(n \!+\! D \!-\! 3)}{\Gamma(n \!+\! 
\frac{D}2)} \Bigl(\frac{y}4 \Bigr)^n \!\right\} \!. \qquad \label{dC}
\end{eqnarray}
As with the contributions from $i\delta\!\Delta_B(x;x')$ considered in the
previous sub-section, the only way $i\delta\!\Delta_C(x;x')$ can give a
nonzero contribution in $D\!=\!4$ dimensions is for it to multiply a singular
term. That means only the $n\!=\!0$ term can possibly contribute. Even for 
the $n\!=\!0$ term, both derivatives must act upon a $\Delta x^2$ to make a
nonzero contribution in $D\!=\!4$ dimensions.

Those of the $[2\!\!-\!\!1]$ and $[2\!\!-\!\!2]$ vertex pairs which are not
proportional to delta functions after the initial contraction of 
Table~\ref{DCcon} all contrive to give delta functions in the end. This 
happens through the same key identity (\ref{keyID}) which was used to
reduce the analogous terms in the previous subsection. In each case we have
finite constants times different contractions of the following tensor function,
\begin{eqnarray}
\lefteqn{ \partial_{\mu}' \Biggl\{ i\delta\!\Delta_C(x;x') \partial_{\alpha}
\partial_{\beta} \Bigl[\frac1{\Delta x^{D-2}}\Bigr] \Biggr\} = \frac{H^{D-2}}{
(4\pi)^{\frac{D}2}} \frac{\Gamma(D\!-\!3)}{\Gamma(\frac{D}2)}\, \partial_{\mu}'
\partial_{\alpha} \partial_{\beta} \Bigl[\frac1{\Delta x^{D-2}}\Bigr] }
\nonumber \\
& & \hspace{2cm} + \frac{H^2}{16 \pi^{\frac{D}2}} \Bigl(\frac{D}2 \!-\!3\Bigr) 
\Gamma\Bigl(\frac{D}2 \!-\!1\Bigr) \partial_{\mu}' \Biggl\{ \frac{(a a')^{2-
\frac{D}2}}{\Delta x^{D-4}} \, \partial_{\alpha} \partial_{\beta} \Bigl[\frac1{
\Delta x^{D-2}} \Bigr]\Biggr\} , \\
& & \hspace{-.5cm} = \frac{H^{D-2}}{(4\pi)^{\frac{D}2}} \frac{\Gamma(D\!-\!3)}{
\Gamma(\frac{D}2)} \, \partial_{\mu}' \partial_{\alpha} \partial_{\beta} 
\Bigl[\frac1{\Delta x^{D-2}}\Bigr] + \frac{H^{D-2}}{16 \pi^{\frac{D}2}} 
\Bigl(\frac{D}2 \!-\!3\Bigr) \Gamma\Bigl(\frac{D}2 \!-\!1\Bigr) 
(a a')^{2-\frac{D}2} \nonumber \\
& & \hspace{2cm} \times \Bigl(\partial_{\mu}' \!-\! \frac12 (D\!-\!4) H a'
\Bigr) \Biggl\{ \frac{D \, \partial_{\alpha} \partial_{\beta}}{4 (D\!-\!3)} 
- \frac{\eta_{\alpha\beta} \partial^2}{4(D\!-\!3)} \Biggr\} \Bigl[\frac1{\Delta
x^{2D-6}} \Bigr] , \qquad \\
& & \hspace{-.5cm} = \frac{H^2}{16 \pi^2} \partial_{\mu}' \partial_{\alpha} 
\partial_{\beta} \Bigl[\frac1{\Delta x^2}\Bigr] - \frac{H^2}{16 \pi^2} 
\partial_{\mu}' \Bigl( \partial_{\alpha} \partial_{\beta} \!-\! \frac14 
\eta_{\alpha\beta} \partial^2\Bigr) \Bigl[\frac1{\Delta x^2}\Bigr] + 
O(D\!-\!4) , \qquad \\
& & \hspace{-.5cm} = -\frac{i H^2}{16} \, \eta_{\alpha\beta} \partial_{\mu} 
\delta^4(x\!-\!x') + O(D\!-\!4) \; . \label{deltalim}
\end{eqnarray}

It remains to multiply (\ref{deltalim}) by the appropriate prefactors and take
the appropriate contraction. For example, the $[2\!\!-\!\!1]_b$ contribution 
is,
\begin{eqnarray}
\lefteqn{-\frac{\kappa^2}{D\!-\!3} \times \frac{i \Gamma(\frac{D}2 \!-\!1)}{4 
\pi^{\frac{D}2}} \times \gamma^0 \delta^{\alpha}_0 \gamma^{\beta} \gamma^{\mu} 
\times -\frac{i H^2}{16} \, \eta_{\alpha \beta} \partial_{\mu} 
\delta^4(x\!-\!x') } \nonumber \\
& & \hspace{4cm} = \frac{\kappa^2 H^2}{16 \pi^2} \times \frac14 
\hspace{-.1cm} \not{\hspace{-.1cm} \partial} \delta^4(x\!-\!x') + O(D\!-\!4) 
\; . \qquad
\end{eqnarray}
We have summarized the results in Table~\ref{DCdelta}, along with all terms 
for which the initial contractions of Table~\ref{DCcon} produced delta 
functions. The sum of all such terms is,
\begin{equation}
-i \Bigl[\Sigma^{T\ref{DCdelta}}\Bigr](x;x') = \frac{\kappa^2 H^2}{16 \pi^2}
\Bigl\{ -\frac38 \hspace{-.1cm} \not{\hspace{-.1cm} \partial} - \frac14 \;
\hspace{-.1cm} \overline{\not{\hspace{-.1cm} \partial} \, } \Bigr\}
\delta^4(x\!-\!x') + O(D\!-\!4) \; . \label{8thcon}
\end{equation}

\begin{table}

\vbox{\tabskip=0pt \offinterlineskip
\def\tablerule{\noalign{\hrule}}
\halign to390pt {\strut#& \vrule#\tabskip=1em plus2em&
\hfil#\hfil& \vrule#& \hfil#\hfil& \vrule#& \hfil#\hfil& \vrule#& 
\hfil#\hfil& \vrule#& \hfil#\hfil& \vrule#\tabskip=0pt\cr
\tablerule
\omit&height4pt&\omit&&\omit&&\omit&&\omit&&\omit&\cr
&&$\!\!\!\!{\rm I}\!\!\!\!$ && $\!\!\!\!{\rm J} \!\!\!\!$ && 
$\!\!\!\! {\rm sub} \!\!\!\!$ && $\!\!\!\! \frac{\kappa^2 H^2}{16 \pi^2}
\hspace{-.1cm} \not{\hspace{-.1cm} \partial} \, \delta^4(x\!-\!x') \!\!\!\!$ 
&& $\!\!\!\! \frac{\kappa^2 H^2}{16 \pi^2} \; \hspace{-.1cm} \overline{\not{
\hspace{-.1cm} \partial}} \, \delta^4(x\!-\!x') \!\!\!\!$ &\cr
\omit&height4pt&\omit&&\omit&&\omit&&\omit&&\omit&\cr
\tablerule
\omit&height2pt&\omit&&\omit&&\omit&&\omit&&\omit&\cr
&& 2 && 1 && a && $-\frac12$ && $0$ & \cr
\omit&height2pt&\omit&&\omit&&\omit&&\omit&&\omit&\cr
\tablerule
\omit&height2pt&\omit&&\omit&&\omit&&\omit&&\omit&\cr
&& 2 && 1 && b && $\frac14$ && $0$ & \cr
\omit&height2pt&\omit&&\omit&&\omit&&\omit&&\omit&\cr
\tablerule
\omit&height2pt&\omit&&\omit&&\omit&&\omit&&\omit&\cr
&& 2 && 2 && a && $\frac14$ && $0$ & \cr
\omit&height2pt&\omit&&\omit&&\omit&&\omit&&\omit&\cr
\tablerule
\omit&height2pt&\omit&&\omit&&\omit&&\omit&&\omit&\cr
&& 2 && 2 && b && $-\frac12$ && $\frac12$ & \cr
\omit&height2pt&\omit&&\omit&&\omit&&\omit&&\omit&\cr
\tablerule
\omit&height2pt&\omit&&\omit&&\omit&&\omit&&\omit&\cr
&& 2 && 2 && c && $-\frac18$ && $0$ & \cr
\omit&height2pt&\omit&&\omit&&\omit&&\omit&&\omit&\cr
\tablerule
\omit&height2pt&\omit&&\omit&&\omit&&\omit&&\omit&\cr
&& 2 && 2 && d && $\frac14$ && $-\frac14$ & \cr
\omit&height2pt&\omit&&\omit&&\omit&&\omit&&\omit&\cr
\tablerule
\omit&height2pt&\omit&&\omit&&\omit&&\omit&&\omit&\cr
&& 2 && 3 && a && $0$ && $0$ & \cr
\omit&height2pt&\omit&&\omit&&\omit&&\omit&&\omit&\cr
\tablerule
\omit&height2pt&\omit&&\omit&&\omit&&\omit&&\omit&\cr
&& 2 && 3 && b && $0$ && $0$ & \cr
\omit&height2pt&\omit&&\omit&&\omit&&\omit&&\omit&\cr
\tablerule
\omit&height2pt&\omit&&\omit&&\omit&&\omit&&\omit&\cr
\tablerule
\omit&height2pt&\omit&&\omit&&\omit&&\omit&&\omit&\cr
&& $\!\!\!\! {\rm Total} \!\!\!\!$ && \omit && \omit 
&& $-\frac38$ && $-\frac14$ & \cr
\omit&height2pt&\omit&&\omit&&\omit&&\omit&&\omit&\cr
\tablerule}}

\caption{Delta functions from the $i\delta\!\Delta_C$ part of the graviton 
propagator.}

\label{DCdelta}

\end{table}

\begin{table}

\vbox{\tabskip=0pt \offinterlineskip
\def\tablerule{\noalign{\hrule}}
\halign to390pt {\strut#& \vrule#\tabskip=1em plus2em&
\hfil#\hfil& \vrule#& \hfil#\hfil& \vrule#& \hfil#\hfil& \vrule#& \hfil#\hfil& 
\vrule#& \hfil#\hfil& \vrule#& \hfil#\hfil& \vrule#& \hfil#\hfil&
\vrule#\tabskip=0pt\cr
\tablerule
\omit&height4pt&\omit&&\omit&&\omit&&\omit&&\omit&&\omit&&\omit&\cr
&& $\!\!\!\!\!\!{\rm I}\!\!\!\!\!\!$ && $\!\!\!\!\!\!{\rm J}\!\!\!\!\!\!$ && 
$\!\!\!\!\!\!{\rm sub}\!\!\!\!\!\!$ && 
$\!\!\!\!\frac{\gamma^{\mu} \Delta x_{\mu}}{\Delta x^{2D-2}}\!\!\!\!$ &&
$\!\!\!\!\frac{\gamma^i \Delta x_i}{\Delta x^{2D-2}}\!\!\!\!$ &&
$\!\!\!\frac{\Vert \vec{x}\Vert^2 \gamma^{\mu} \Delta x_{\mu}}{\Delta 
x^{2D}}\!\!\!\!$ &&
$\!\!\!\!\frac{\Vert \vec{x}\Vert^2 \gamma^i \Delta x_i}{\Delta 
x^{2D}}\!\!\!\!$ &\cr
\omit&height4pt&\omit&&\omit&&\omit&&\omit&&\omit&&\omit&&\omit&\cr
\tablerule
\omit&height2pt&\omit&&\omit&&\omit&&\omit&&\omit&&\omit&&\omit&\cr
&& 2 && $\!\!\!\!3\!\!\!\!$ && $\!\!\!\!{\rm c}\!\!\!\!$
&& ${\scriptstyle -(D-1)^2}$ && ${\scriptstyle D (D-1)}$ 
&& ${\scriptstyle 0}$ && ${\scriptstyle 0}$ & \cr
\omit&height2pt&\omit&&\omit&&\omit&&\omit&&\omit&&\omit&&\omit&\cr
\tablerule
\omit&height2pt&\omit&&\omit&&\omit&&\omit&&\omit&&\omit&&\omit&\cr
&& 2 && $\!\!\!\!3\!\!\!\!$ && $\!\!\!\!{\rm d}\!\!\!\!$
&& ${\scriptstyle 0}$ && $\!\!\!\!{\scriptstyle (D-1)(D-2)}
\!\!\!\!$ && ${\scriptstyle D (D-2)}$ && $\!\!\!\!{\scriptstyle -2 D (D-2)}
\!\!\!\!$ & \cr
\omit&height2pt&\omit&&\omit&&\omit&&\omit&&\omit&&\omit&&\omit&\cr
\tablerule
\omit&height2pt&\omit&&\omit&&\omit&&\omit&&\omit&&\omit&&\omit&\cr
&& 3 && $\!\!\!\!1\!\!\!\!$ && $\!\!\!\!{\rm a}\!\!\!\!$
&& ${\scriptstyle 4 (D-1)}$ && ${\scriptstyle 0}$ && 
${\scriptstyle 0}$ && ${\scriptstyle 0}$ & \cr
\omit&height2pt&\omit&&\omit&&\omit&&\omit&&\omit&&\omit&&\omit&\cr
\tablerule
\omit&height2pt&\omit&&\omit&&\omit&&\omit&&\omit&&\omit&&\omit&\cr
&& 3 && $\!\!\!\!1\!\!\!\!$ && $\!\!\!\!{\rm b}\!\!\!\!$
&& ${\scriptstyle -2 (D-1)}$ && ${\scriptstyle -2 (D-4)}$ && 
${\scriptstyle 0}$ && ${\scriptstyle 0}$ & \cr
\omit&height2pt&\omit&&\omit&&\omit&&\omit&&\omit&&\omit&&\omit&\cr
\tablerule
\omit&height2pt&\omit&&\omit&&\omit&&\omit&&\omit&&\omit&&\omit&\cr
&& 3 && $\!\!\!\!2\!\!\!\!$ && $\!\!\!\!{\rm a}\!\!\!\!$
&& $\!\!\!\!{\scriptstyle -2(D-1)}\!\!\!\!$ && ${\scriptstyle 0}$ && 
${\scriptstyle 0}$ && $\!\!\!\!{\scriptstyle 0}\!\!\!\!$ & \cr
\omit&height2pt&\omit&&\omit&&\omit&&\omit&&\omit&&\omit&&\omit&\cr
\tablerule
\omit&height2pt&\omit&&\omit&&\omit&&\omit&&\omit&&\omit&&\omit&\cr
&& 3 && $\!\!\!\!2\!\!\!\!$ && $\!\!\!\!{\rm b}\!\!\!\!$
&& $\!\!\!\!{\scriptstyle 2 (D-1) (D-2) }\!\!\!\!$ && $\!\!\!\!
{\scriptstyle -2 (D-1) (D-2)}\!\!\!\!$ && ${\scriptstyle 0}$ && 
$\!\!\!\!{\scriptstyle 0}\!\!\!\!$ & \cr
\omit&height2pt&\omit&&\omit&&\omit&&\omit&&\omit&&\omit&&\omit&\cr
\tablerule
\omit&height2pt&\omit&&\omit&&\omit&&\omit&&\omit&&\omit&&\omit&\cr
&& 3 && $\!\!\!\!2\!\!\!\!$ && $\!\!\!\!{\rm c}\!\!\!\!$
&& $\!\!\!\!{\scriptstyle (D-1)}\!\!\!\!$ && ${\scriptstyle 
(D -4)}$ && ${\scriptstyle 0}$ && $\!\!\!\!{\scriptstyle 0}\!\!\!\!$ & \cr
\omit&height2pt&\omit&&\omit&&\omit&&\omit&&\omit&&\omit&&\omit&\cr
\tablerule
\omit&height2pt&\omit&&\omit&&\omit&&\omit&&\omit&&\omit&&\omit&\cr
&& 3 && $\!\!\!\!2\!\!\!\!$ && $\!\!\!\!{\rm d}\!\!\!\!$
&& $\!\!\!\!{\scriptstyle 0}\!\!\!\!$ && $\!\!\!\!{\scriptstyle 
-(2D-3) (D-2)}\!\!\!\!$ && $\!\!\!\!{\scriptstyle -2 (D-1) (D-2)}\!\!\!\!\!\!$ 
&& $\!\!\!\!{\scriptstyle 4 (D-1) (D-2)}\!\!\!\!$ & \cr
\omit&height2pt&\omit&&\omit&&\omit&&\omit&&\omit&&\omit&&\omit&\cr
\tablerule
\omit&height2pt&\omit&&\omit&&\omit&&\omit&&\omit&&\omit&&\omit&\cr
&& 3 && $\!\!\!\!3\!\!\!\!$ && $\!\!\!\!{\rm a}\!\!\!\!$
&& $\!\!\!\!{\scriptstyle -(D-1)^2}\!\!\!\!$ && $\!\!\!\!{\scriptstyle 0}
\!\!\!\!$ && $\!\!\!\!{\scriptstyle 0}\!\!\!\!$ && $\!\!\!\!{\scriptstyle 0}
\!\!\!\!$ & \cr
\omit&height2pt&\omit&&\omit&&\omit&&\omit&&\omit&&\omit&&\omit&\cr
\tablerule
\omit&height2pt&\omit&&\omit&&\omit&&\omit&&\omit&&\omit&&\omit&\cr
&& 3 && $\!\!\!\!3\!\!\!\!$ && $\!\!\!\!{\rm b}\!\!\!\!$
&& $\!\!\!\!{\scriptstyle \frac12 (D-1)^2}\!\!\!\!$ && $\!\!\!\!
{\scriptstyle \frac12 (D-1) (D-4)}\!\!\!\!$ && $\!\!\!\!{\scriptstyle 0}
\!\!\!\!$ && $\!\!\!\!{\scriptstyle 0}\!\!\!\!$ & \cr
\omit&height2pt&\omit&&\omit&&\omit&&\omit&&\omit&&\omit&&\omit&\cr
\tablerule
\omit&height2pt&\omit&&\omit&&\omit&&\omit&&\omit&&\omit&&\omit&\cr
&& 3 && $\!\!\!\!3\!\!\!\!$ && $\!\!\!\!{\rm c}\!\!\!\!$
&& $\!\!\!\!{\scriptstyle \frac12 (D-1)^2}\!\!\!\!$ && $\!\!\!\!
{\scriptstyle \frac12 (D-1) (D-4)}\!\!\!\!$ && $\!\!\!\!{\scriptstyle 0}
\!\!\!\!$ && $\!\!\!\!{\scriptstyle 0}\!\!\!\!$ & \cr
\omit&height2pt&\omit&&\omit&&\omit&&\omit&&\omit&&\omit&&\omit&\cr
\tablerule
\omit&height2pt&\omit&&\omit&&\omit&&\omit&&\omit&&\omit&&\omit&\cr
&& 3 && $\!\!\!\!3\!\!\!\!$ && $\!\!\!\!{\rm d}\!\!\!\!$
&& ${\scriptstyle -\frac12 (D-1) (D-2)}$ && ${\scriptstyle (D-2)}$
&& ${\scriptstyle \frac12 (D-2)^2}$ && ${\scriptstyle -(D-2)^2}$ & \cr
\omit&height2pt&\omit&&\omit&&\omit&&\omit&&\omit&&\omit&&\omit&\cr
\tablerule
\omit&height2pt&\omit&&\omit&&\omit&&\omit&&\omit&&\omit&&\omit&\cr
\tablerule
\omit&height2pt&\omit&&\omit&&\omit&&\omit&&\omit&&\omit&&\omit&\cr
&& $\!\!\!\!{\rm Total}\!\!\!\!$ && \omit && \omit && $\!\!\!\!{\scriptstyle 
\frac12 (D-1)(D-2)} \!\!\!\!$ && $\!\!\!\!{\scriptstyle 2(D-1) -D (D-2)}
\!\!\!\!$ && $\!\!\!\! {\scriptstyle -\frac12 (D-2)^2}\!\!\!\!$ && $\!\!\!\!
{\scriptstyle (D-2)^2}\!\!\!\!$ &\cr
\omit&height2pt&\omit&&\omit&&\omit&&\omit&&\omit&&\omit&&\omit&\cr
\tablerule}}

\caption{$i\delta\!\Delta_C$ terms in which all derivatives act upon 
$\Delta x^2(x;x')$. All contributions are multiplied by $\frac{i \kappa^2 
H^2}{2^8 \pi^D} \Gamma(\frac{D}2) \Gamma(\frac{D}2 \!-\!1) \frac{(D-4)(D-6)}{
D-3} (a a')^{2- \frac{D}2}$. }

\label{DCmost}

\end{table}

All the lower vertex pairs involve one or more derivatives of 
$i\delta\!\Delta_C$,
\begin{eqnarray}
\partial_i i \delta\!\Delta_C & = & -\frac{H^2 \Gamma(\frac{D}2\!-\!1)}{32
\pi^{\frac{D}2}} (D\!-\!6) (D\!-\!4) (a a')^{2-\frac{D}2} \, \frac{\Delta x^i}{
\Delta x^{D-2}} = -\partial_i' i\delta\!\Delta_C \; , \qquad \\
\partial_0 i \delta\!\Delta_C & = & \frac{H^2 \Gamma(\frac{D}2\!-\!1)}{32
\pi^{\frac{D}2}} (D\!-\!6)(D\!-\!4) (a a')^{2-\frac{D}2} \Biggl\{ \frac{\Delta 
\eta}{\Delta x^{D-2}} \!-\! \frac{a H}{2 \Delta x^{D-4}} \Biggr\} , \qquad \\
\partial_0' i \delta\!\Delta_C & = & \frac{H^2 \Gamma(\frac{D}2\!-\!1)}{32
\pi^{\frac{D}2}} (D\!-\!6)(D\!-\!4) (a a')^{2-\frac{D}2} \Biggl\{-\frac{\Delta 
\eta}{\Delta x^{D-2}} \!-\! \frac{a' H}{2 \Delta x^{D-4}} \Biggr\} . \qquad
\end{eqnarray}
Their reduction follows the same pattern as in the previous two sub-sections.
Table~\ref{DCmost} summarizes the results for the case in which all derivatives
act upon the conformal coordinate separation $\Delta x^2$.

When summed, three of the columns of Table~\ref{DCmost} reveal a factor of 
$(D\!-\!2)$ which we extract,
\begin{eqnarray}
\lefteqn{-i \Bigl[\Sigma^{T\ref{DCmost}}\Bigr](x;x') = \frac{i \kappa^2 H^2}{
2^8 \pi^D} \Gamma\Bigl(\frac{D}2\Bigr) \Gamma\Bigl(\frac{D}2 \!-\!1\Bigr)
\frac{(D\!-\!2) (D\!-\!4) (D\!-\!6)}{(D\!-\!3)} (a a')^{2-\frac{D}2} }
\nonumber \\
& & \times \Biggl\{ \frac12 (D\!-\!1) \frac{\gamma^{\mu} \Delta x_{\mu}}{
\Delta x^{2D-2}} + \Bigl[ 2\Bigl(\frac{D\!-\!1}{D\!-\!2}\Bigr) \!-\! D \Bigr]
\frac{\gamma^i \Delta x_i}{\Delta x^{2D-2}} \nonumber \\
& & \hspace{3cm} - \frac12 (D\!-\!2) \frac{\Vert \Delta \vec{x} \Vert^2 
\gamma^{\mu} \Delta x_{\mu}}{\Delta x^{2D}} + (D\!-\!2) \frac{\Vert 
\Delta \vec{x} \Vert^2 \gamma^i \Delta x_i}{\Delta x^{2D}} \Biggr\} . \qquad
\label{T18}
\end{eqnarray}
We partially integrate (\ref{T18}) with the aid of (\ref{difID1}-\ref{difID2})
and then take $D\!=\!4$, just as we did for the sum of Table~\ref{DBmost},
\begin{eqnarray}
\lefteqn{-i \Bigl[\Sigma^{T\ref{DCmost}}\Bigr](x;x') = \frac{i \kappa^2 H^2}{
2^8 \pi^D} \Gamma\Bigl(\frac{D}2\Bigr) \Gamma\Bigl(\frac{D}2 \!-\!1\Bigr)
\frac{(D\!-\!2) (D\!-\!4) (D\!-\!6)}{(D\!-\!3)} (a a')^{2-\frac{D}2} }
\nonumber \\
& & \times \Biggl\{ \frac{D}4 \frac{\gamma^{\mu} \Delta x_{\mu}}{\Delta 
x^{2D-2}} + \Bigl[2 \Bigl(\frac{D\!-\!1}{D\!-\!2}\Bigr) \!-\! \frac{D^2}{2
(D\!-\!1)} \Bigr] \frac{\gamma^i \Delta x_i}{\Delta x^{2D-2}} \nonumber \\
& & \hspace{4cm} - \frac{\nabla^2}{8 (D\!-\!1)} \Bigl(\frac{\gamma^{\mu} 
\Delta x_{\mu}}{\Delta x^{2D-4}}\Bigr) + \frac{\nabla^2}{4(D\!-\!1)} 
\Bigl(\frac{\gamma^i \Delta x_i}{\Delta x^{2D-4}}\Bigr) \Biggr\} , \qquad \\
& & \hspace{-.5cm} = \frac{i \kappa^2 H^2}{2^8 \pi^D} \Gamma\Bigl(\frac{D}2
\Bigr) \Gamma\Bigl(\frac{D}2 \!-\!1\Bigr) \frac{(D\!-\!2) (D\!-\!6)}{(D\!-\!1) 
(D\!-\!3)^2} (a a')^{2-\frac{D}2} \Biggl\{- \frac{D (D\!-\!1)}{16 (D\!-\!2)}
\hspace{-.1cm} \not{\hspace{-.1cm} \partial} \partial^2 \nonumber \\
& & + \frac{(D^3 \!-\! 6D^2 \!+\! 8 D \!-\! 4)}{8 (D\!-\!2)^2}
\; \hspace{-.1cm} \overline{\not{\hspace{ -.1cm} \partial}} \, \partial^2 
\!+\! \Bigl(\frac{D\!-\!4}{16}\Bigr) \hspace{-.1cm} \not{\hspace{-.1cm} 
\partial} \nabla^2\!-\!\Bigl(\frac{D\!-\!4}8 \Bigr) \, \hspace{-.1cm} 
\overline{\not{\hspace{-.1cm} \partial}} \, \nabla^2 \Biggr\} 
\frac1{\Delta x^{2D-6}} , \qquad \\
& & \hspace{-.5cm} = \frac{i \kappa^2 H^2}{2^8 \pi^4} \Bigl\{ \frac12
\hspace{-.1cm} \not{\hspace{-.1cm} \partial} \, \partial^2 +\frac16 \;
\hspace{-.1cm} \overline{\not{\hspace{-.1cm} \partial}} \, \partial^2 \Bigr\}
\frac1{\Delta x^2} + O(D\!-\!4) \; . \label{9thcon}
\end{eqnarray}
As already explained, terms for which one or more derivative acts upon a 
scale factor make no contribution in $D\!=\!4$ dimensions, so this is the 
final nonzero contribution.

\section{Renormalization}

The regulated result we have worked so hard to compute derives from summing 
expressions (\ref{1stcon}), (\ref{2ndcon}), (\ref{3rdcon}), (\ref{4thcon}), 
(\ref{5thcon}), (\ref{6thcon}), (\ref{7thcon}), (\ref{8thcon}) and 
(\ref{9thcon}),
\begin{eqnarray}
\lefteqn{-i\Bigl[\Sigma\Bigr](x;x') = \kappa^2 \Biggl\{\beta_1 (a a')^{1-
\frac{D}2} \hspace{-.1cm} \not{\hspace{-.1cm} \partial} \partial^2 + \beta_2 
(a a')^{2-\frac{D}2} H^2 \hspace{-.1cm} \not{\hspace{-.1cm} \partial} + 
\beta_3 (a a')^{2-\frac{D}2} H^2 \; \hspace{-.1cm} \overline{\not{
\hspace{-.1cm} \partial}} } \nonumber \\
& & \hspace{2cm} + b_2 H^2 \hspace{-.1cm} \not{\hspace{-.1cm} \partial} + 
b_3 H^2 \; \hspace{-.1cm} \overline{\not{\hspace{-.1cm} \partial}} \Biggr\} 
\delta^D(x\!-\!x') + \frac{\kappa^2 H^2}{16 \pi^2} \times -3 \ln(a) \; 
\hspace{-.1cm} \overline{\not{\hspace{-.1cm} \partial}} \delta^4(x\!-\!x') 
\nonumber \\
& & \hspace{.2cm} -\frac{i \kappa^2}{2^8 \pi^4} (a a')^{-1} \hspace{-.1cm} 
\not{\hspace{-.1cm} \partial} \partial^4 \Bigl[ \frac{\ln(\mu^2 \Delta x^2)}{
\Delta x^2} \Bigr] + \frac{i \kappa^2 H^2}{2^8 \pi^4} \Biggl\{ \Bigl(
-\frac{15}2 \hspace{-.1cm} \not{\hspace{-.1cm} \partial} \, \partial^2 +
\hspace{-.1cm} \overline{\not{\hspace{-.1cm} \partial}} \, \partial^2 \Bigr)
\Bigl[ \frac{\ln(\mu^2 \Delta x^2)}{\Delta x^2} \Bigr] \nonumber \\
& & \hspace{1cm} + \Bigl(8 \; \hspace{-.1cm} \overline{\not{\hspace{-.1cm} 
\partial}} \partial^2 \!-\! 4 \hspace{-.1cm} \not{\hspace{-.1cm} \partial} 
\nabla^2 \Bigr) \Bigl[ \frac{\ln(\frac14 H^2\Delta x^2)}{\Delta x^2} \Bigr] 
\!-\! 7 \hspace{-.1cm} \not{\hspace{-.1cm} \partial} \, \nabla^2 \Bigl[
\frac1{\Delta x^2} \Bigr]\!\Biggr\} \!+\! O(D\!-\!4) . \qquad \label{regres}
\end{eqnarray}
The various $D$-dependent constants in (\ref{regres}) are,
\begin{eqnarray}
\beta_1 & \!\!=\!\! & \frac{\mu^{D-4}}{2^8 \pi^{\frac{D}2}} \frac{\Gamma(
\frac{D}2 \!-\!1)}{(D\!-\!3) (D\!-\!4)} \Biggl\{ -2 D \!+\! 1 \!-\! 
\frac2{D\!-\!2} \Biggr\} , \\
\beta_2 &\!\! =\!\! & \frac{\mu^{D-4}}{2^9 \pi^{\frac{D}2}} 
\frac{\Gamma(\frac{D}2 \!+\! 1)}{(D\!-\!3) (D\!-\!4)} \Biggl\{\frac12 D^2 \!-\!
10 D \!+\! 15 \!-\! \frac{24}{D} \!-\! \frac6{D\!-\!1} \!-\! \frac{35}{D\!-\!3} 
\Biggr\} , \qquad \\
\beta_3 & \!\!= \!\!& \frac{\mu^{D-4}}{2^9 \pi^{\frac{D}2}} 
\frac{\Gamma(\frac{D}2 \!+\!1)}{(D\!-\!3) (D\!-\!4)} \Biggl\{-D \!+\! 3 \!+\!
\frac9{D\!-\!3} \Biggr\} , \\
b_2 &\!\!=\!\!& \frac{H^{D-4}}{(4 \pi)^{\frac{D}2}} \frac{\Gamma(D\!-\!1)}{
\Gamma(\frac{D}2)} \Biggl\{ -\frac{(D\!+\!1) (D\!-\!1) (D\!-\!4)}{2 (D\!-\!3)}
\times \frac{\pi}2 \cot\Bigl(\frac{\pi D}2\Bigr) \nonumber \\
& & \hspace{4.5cm} - \frac{(D\!-\!1)(D^3 \!-\! 8 D^2 \!+\! 23 D \!-\! 32)}{8
(D\!-\!2)^2 (D\!-\!3)^2} - \frac7{48}\Biggr\} , \qquad \\
b_3 &\!\!=\!\!& \frac{H^{D-4}}{(4 \pi)^{\frac{D}2}} \frac{\Gamma(D\!-\!1)}{
\Gamma(\frac{D}2)} \Biggl\{ \frac34 \Bigl(D \!-\! \frac2{D\!-\!3}\Bigr)
\times \frac{\pi}2 \cot\Bigl(\frac{\pi D}2\Bigr) \nonumber \\
& & \hspace{6cm} + \frac34 \frac{(D^2 \!-\! 6 D \!+\! 8)}{(D\!-\!2)^2 
(D\!-\!3)^2} - \frac52 \Biggr\} . 
\end{eqnarray}
In obtaining these expressions we have always chosen to convert finite,
$D\!=\!4$ terms with $\partial^2$ acting on $1/\Delta x^2$, into delta 
functions,
\begin{equation}
\partial^2 \Bigl[\frac1{\Delta x^2}\Bigr] = i 4 \pi^2 \delta^4(x\!-\!x') \; .
\end{equation}
All such terms have then been included in $b_2$ and $b_3$.

\vspace{-1.5cm}

\begin{center}
\begin{picture}(300,70)(0,0)
\ArrowLine(150,20)(90,20) 
\ArrowLine(210,20)(150,20)
\Vertex(150,20){3}
\Text(151,20)[]{\LARGE $\times$}
\Text(150,10)[b]{$x$}
\end{picture}
\\ {\rm Fig.~3: Contribution from counterterms.}
\end{center}

The local divergences in this expression are canceled by the BPHZ counterterms
enumerated at the end of section 3. The generic diagram topology is depicted 
in Fig.~3, and the analytic form is,
\begin{eqnarray}
\lefteqn{-i\Bigl[\Sigma^{\rm ctm}\Bigr](x;x') = \sum_{I=1}^3 i C_{Iij} \,
\delta^D(x-x') \; , } \\
& & = -\kappa^2 \Bigl\{\alpha_1 (a a')^{-1} \hspace{-.1cm} \not{\hspace{-.1cm}
\partial} \partial^2 + \alpha_2 D (D\!-\!1) H^2 \hspace{-.1cm} \not{\hspace{
-.1cm} \partial} + \alpha_3 H^2 \; \hspace{-.1cm} \overline{\not{\hspace{-.1cm} 
\partial}} \Bigr\} \delta^D(x\!-\!x') \; . \label{genctm}
\end{eqnarray}
In comparing (\ref{regres}) and (\ref{genctm}) it would seem that the simplest
choice for the coefficients $\alpha_i$ is,
\begin{equation}
\alpha_1 = \beta_1 \quad , \quad \alpha_2 = \frac{\beta_2 \!+\! b_2}{D 
(D\!-\!1)} \quad {\rm and} \quad \alpha_3 = \beta_3 \!+\! b_3 \; . \label{sim}
\end{equation}
This choice absorbs all local constants but one is of course left with
time dependent terms proportional to $\ln(a a')$,
\begin{eqnarray}
\beta_1 (a a')^{1-\frac{D}2} - \alpha_1 (a a')^{-1} & = & +\frac1{2^6 \pi^2}
\frac{\ln(a a')}{a a'} + O(D\!-\!4) \; , \\
\beta_2 (a a')^{2-\frac{D}2} + b_2 - D (D\!-\! 1) \alpha_2 & = & 
+\frac{7.5}{2^6 \pi^2} \ln(a a') + O(D\!-\!4) \; , \\
\beta_3 (a a')^{2-\frac{D}2} + b_3 - \alpha_3 & = & -\frac1{2^6 \pi^2} 
\ln(a a') + O(D\!-\!4) \; .
\end{eqnarray}
Our final result for the renormalized self-energy is,
\begin{eqnarray}
\lefteqn{-i\Bigl[\Sigma^{\rm ren}\Bigr](x;x') \!=\!\!\frac{\kappa^2}{2^6 \pi^2}
\Biggl\{\!\frac{\ln(a a')}{a a'} \hspace{-.1cm} \not{\hspace{-.1cm} \partial} 
\partial^2 \!+\! \frac{15}2 \ln(a a') H^2\!\hspace{-.1cm} \not{\hspace{-.1cm} 
\partial} \!-\! 7 \ln(a a') H^2 \; \hspace{-.1cm} \overline{\not{\hspace{-.1cm}
\partial}} \!\Biggr\} \delta^4(x \!-\! x') } \nonumber \\
& & \hspace{.2cm} - \frac{i \kappa^2}{2^8 \pi^4} (a a')^{-1} \hspace{-.1cm} 
\not{\hspace{-.1cm} \partial} \partial^4 \Bigl[ \frac{\ln(\mu^2 \Delta x^2)}{
\Delta x^2} \Bigr] + \frac{i \kappa^2 H^2}{2^8 \pi^4} \Biggl\{\Bigl(-\frac{15}2
\hspace{-.1cm} \not{\hspace{-.1cm} \partial} \, \partial^2 +
\hspace{-.1cm} \overline{\not{\hspace{-.1cm} \partial}} \, \partial^2 \Bigr)
\Bigl[ \frac{\ln(\mu^2 \Delta x^2)}{\Delta x^2} \Bigr] \nonumber \\
& & \hspace{3cm} + \Bigl(8 \; \hspace{-.1cm} \overline{\not{\hspace{-.1cm} 
\partial}} \partial^2 \!-\! 4 \hspace{-.1cm} \not{\hspace{-.1cm} \partial} 
\nabla^2 \Bigr) \Bigl[ \frac{\ln(\frac14 H^2\Delta x^2)}{\Delta x^2} \Bigr] 
\!-\! 7 \hspace{-.1cm} \not{\hspace{-.1cm} \partial} \, \nabla^2 \Bigl[
\frac1{\Delta x^2} \Bigr]\!\Biggr\} . \qquad \label{ren}
\end{eqnarray}

\section{Discussion}

We have used dimensional regularization to compute quantum gravitational
corrections to the fermion self-energy at one loop order in a locally de Sitter
background. Our regulated result is (\ref{regres}). Although Dirac $+$ Einstein
is not perturbatively renormalizable \cite{DVN} we obtained a finite result
(\ref{ren}) by absorbing the divergences with BPHZ counterterms. 

For this 1PI function, and at one loop order, only three counterterms are 
necessary. None of them represents redefinitions of terms in the Lagrangian
of Dirac $+$ Einstein. Two of the required counterterms (\ref{invctms}) are 
generally coordinate invariant fermion bilinears of dimension six. The third 
counterterm (\ref{nictm}) is the only other fermion bilinear of dimension six
which respects the symmetries (\ref{homot}-\ref{dilx}) of our de Sitter 
noninvariant gauge (\ref{GR}) and also obeys the reflection property 
(\ref{refl}) of the self-energy for massless fermions.

Although our renormalized result could be changed by altering the finite 
parts of the three BPHZ counterterms, this does not affect its leading behavior
in the far infrared. It is simple to be quantitative about this. Were we to
make finite shifts $\Delta \alpha_i$ in our counterterms (\ref{sim}) the 
induced change in the renormalized self-energy would be,
\begin{equation}
-i \Bigl[\Delta \Sigma^{\rm ren}\Bigr](x;x') = -\kappa^2 \Biggl\{
\frac{\Delta \alpha_1}{a a'} \hspace{-.1cm} \not{\hspace{-.1cm} \partial} 
\partial^2 + 12 \Delta \alpha_2 H^2 \hspace{-.1cm} \not{\hspace{-.1cm} 
\partial} + \Delta \alpha_3 H^2 \; \hspace{-.1cm} \overline{\not{\hspace{-.1cm} 
\partial}} \Biggr\} \delta^4(x\!-\!x') \; . \label{arb}
\end{equation}
No physical principle seems to fix the $\Delta \alpha_i$ so any result that
derives from their values is arbitrary. This is why BPHZ renormalization does
not yield a complete theory. However, at late times (which accesses the far 
infrared because all momenta are redshifted by $a(t) = e^{Ht}$) the local part 
of the renormalized self-energy (\ref{ren}) is dominated by the large 
logarithms,
\begin{equation}
\frac{\kappa^2}{2^6 \pi^2} \Biggl\{\frac{\ln(a a')}{a a'} \hspace{-.1cm} 
\not{\hspace{-.1cm} \partial} \partial^2 + \frac{15}2 \ln(a a') H^2
\hspace{-.1cm} \not{\hspace{-.1cm} \partial} - 7 \ln(a a') H^2 \; 
\hspace{-.1cm} \overline{\not{\hspace{-.1cm} \partial}} \Biggr\} \delta^4(x 
\!-\! x') \; . \label{fixed}
\end{equation}
The coefficients of these logarithms are finite and completely fixed by our
calculation. As long as the shifts $\Delta \alpha_i$ are finite, their impact
(\ref{arb}) must eventually be dwarfed by the large logarithms (\ref{fixed}).

None of this should seem surprising, although it does with disturbing
regularity. The comparison we have just made is a standard feature of low 
energy effective field theory and has a very old and distinguished pedigree
\cite{BN,SW,FS,HS,CDH,CD,DMC1,DL,JFD1,JFD2,MV,HL,ABS,KK1,KK2}. Loops of 
massless particles make finite, nonanalytic contributions which cannot be 
changed by local counterterms and which dominate the far infrared. Further, 
these effects must occur as well, {\it with precisely the same numerical 
values}, in whatever fundamental theory ultimately resolves the ultraviolet 
problem of quantum gravity. That is why Feinberg and Sucher got exactly the 
same long range force from the exchange of massless neutrinos using Fermi 
theory \cite{FS,HS} as one would get from the Standard Model \cite{HS}.

So we can use (\ref{ren}) reliably in the far infrared. Our motivation for 
undertaking this exercise was to search for a gravitational analogue of what 
Yukawa-coupling a massless, minimally coupled scalar does to massless 
ferm\-i\-ons 
during inflation \cite{PW}. Obtaining (\ref{ren}) completes the first part in 
that program. What remains is to use our result to solve the quantum-corrected 
Dirac equation (\ref{Diraceq}). We shall undertake that in a subsequent paper. 
However, it seems clear that the dominant effect must come from the following 
six terms,
\begin{eqnarray}
\lefteqn{-i \Bigl[\Sigma^{\rm ren}\Bigr](x;x') \longrightarrow \frac{\kappa^2 
H^2}{2^6 \pi^2} \Biggl\{\frac{15}2 \ln(a a') \hspace{-.1cm} \not{\hspace{-.1cm} 
\partial} - 7 \ln(a a') \; \hspace{-.1cm} \overline{\not{\hspace{-.1cm} 
\partial}} \Biggr\} \delta^4(x \!-\! x') + \frac{i \kappa^2 H^2}{2^8 \pi^4} }
\nonumber \\
& & \hspace{-.5cm} \times \Biggl\{\Bigl(-\frac{15}2 \hspace{-.1cm}\not{\hspace{
-.1cm} \partial} \, \partial^2 + \hspace{-.1cm} \overline{\not{\hspace{-.1cm} 
\partial}} \, \partial^2 \Bigr) \Bigl[ \frac{\ln(\mu^2 \Delta x^2)}{\Delta x^2}
\Bigr] + \Bigl(8 \; \hspace{-.1cm} \overline{\not{\hspace{-.1cm} \partial}} 
\partial^2 \!-\! 4 \hspace{-.1cm} \not{\hspace{-.1cm} \partial} \nabla^2 \Bigr)
\Bigl[ \frac{\ln(\frac14 H^2\Delta x^2)}{\Delta x^2} \Bigr] \Biggr\} . \qquad
\label{big}
\end{eqnarray}
As adumbrated in the Introduction, these terms are only enhanced by a factor
of $\ln(a)$ relative to the classical part of the Dirac equation 
(\ref{Diraceq}). That is much weaker than the $a \ln(a)$ enhancement 
engendered by a Yukawa scalar \cite{PW} but it can still lead to interesting 
effects. Note that any such effect will be independent of assumptions about 
the existence and couplings of light scalars during inflation.

We have already commented on the importance of the logarithm terms 
(\ref{big}). During inflation these infrared logarithms are ubiquitous
in quantum corrections from massless, minimally coupled scalars and
gravitons. A heroic recent analysis of scalar-driven inflation at
arbitrarily high loop order was not able to exclude the possibility 
that they might even contaminate the power spectra of cosmological density
perturbations \cite{SW2}! The proportional correction they make in that
case must be small because the logarithms would only start to grow at
horizon crossing, and must cease growing when the mode reenters the
horizon after inflation. So the largest enhancement for a currently
observable mode would be $\ln(a) \ltwid 100$. This must be set against
the tiny loop counting parameter of $G H^2 \ltwid 10^{-12}$.

The more significant corrections would be to modes which are still
enormously super-horizon. These are also down by the constant $G H^2$,
but the time-dependent enhancement factor $\ln(a)$ could be arbitrarily 
big. That is what we shall study in our follow-up work. Of course 
loops of such modes can also engender large corrections to effective 
interactions of low dimension. These corrections can grow so large that 
perturbation theory eventually breaks down. Starobinski\u{\i} has 
advocated gaining quantitative control over this regime by summing the 
leading infrared logarithms at each order \cite{AAS}. With Yokoyama he 
has given a complete solution for the case of a minimally coupled scalar 
with arbitrary potential which is a spectator to de Sitter inflation 
\cite{SY}. 

The asymptotic late time effect is small in the simple scalar models 
for which the leading logarithm expansion has been summed. However, it 
is by no means clear that the same must be true for more complicated 
theories that also show infrared logarithms such as quantum gravity 
\cite{TW4,TW5,TW6}, scalar QED \cite{PTW,KW} or Yukawa theory \cite{PW,DW}.
Another application of our result (\ref{ren}) is to serve as ``data'' in 
checking a leading logarithm formulation of Dirac $+$ Einstein during 
inflation \cite{RPW3,TW8}. Because the fermion does not itself engender 
infrared logarithms it may serve as a spectator for what is going on in 
the gravitational sector. In the leading logarithm limit one could obtain 
an explicit operator expression for the fermion, in terms of its own free 
field and the infrared part of the metric. Although the infrared part of 
the metric would not be known to all orders, it is known to lowest order, 
and that would suffice to compare with one loop results such as (\ref{ren}). 
This might serve as an important intermediate point in the difficult task 
of generalizing Starobinski\u{\i}'s techniques to full blown quantum gravity.

It is well to close with a comment on accuracy. Although parts of this 
computation are quite intricate we have good confidence that (\ref{ren}) is 
correct for three reasons. First, there is the flat space limit of taking 
$H$ to zero while taking the conformal time to be $\eta = -e^{-H t}/H$ with
$t$ held fixed. This checks the leading conformal contributions. Our second
reason for confidence is the fact that all divergences can be absorbed using 
just the three counterterms we have inferred in section 3 on the basis of
symmetry. This was by no means the case for individual terms; many separate 
pieces must be added to eliminate other divergences. The final check comes 
from the fact that the self-energy of a massless fermion must be odd under 
interchange of its two coordinates. This was again not true for separate 
contributions, yet it emerged when terms were summed.

\vskip 1cm

\centerline{\bf Acknowledgements}

This work was partially supported by NSF grant PHY-0244714 and by 
the Institute for Fundamental Theory at the University of Florida.

\end{document}